\documentclass[twocolumn,times,trackchanges]{aastex63}
\usepackage{xcolor}
\usepackage{threeparttable,booktabs}

\DeclareUnicodeCharacter{2212}{-}
\usepackage{graphicx} 
\usepackage[mathscr]{euscript}
\DeclareSymbolFont{rsfs}{U}{rsfs}{m}{n}
\DeclareSymbolFontAlphabet{\mathscrsfs}{rsfs}
\usepackage{makecell}

\shorttitle{ALMA LEGA-C Galaxies}
\shortauthors{Woodrum et al. 2022}

\graphicspath{{./}{Figures/}}

\begin{document}

\title{Molecular Gas Reservoirs in Massive Quiescent Galaxies at $\mathrm{z\sim0.7}$ Linked to Late Time Star Formation}

\author[0000-0001-5962-7260]{Charity Woodrum}
\affiliation{Steward Observatory, University of Arizona, 933 North Cherry Avenue, Tucson, AZ 85721, USA}

\author[0000-0003-2919-7495]{Christina C. Williams}
\affiliation{Steward Observatory, University of Arizona, 933 North Cherry Avenue, Tucson, AZ 85721, USA}
\affiliation{NSF’s National Optical-Infrared Astronomy Research Laboratory, 950 North Cherry Avenue, Tucson, AZ 85719, USA }

\author[0000-0002-7893-6170]{Marcia Rieke}
\affiliation{Steward Observatory, University of Arizona, 933 North Cherry Avenue, Tucson, AZ 85721, USA}

\author[0000-0001-6755-1315]{Joel Leja}
\affiliation{Department of Astronomy \& Astrophysics, The Pennsylvania State University, University Park, PA 16802, USA}
\affiliation{Institute for Computational \& Data Sciences, The Pennsylvania State University, University Park, PA, USA}
\affiliation{Institute for Gravitation and the Cosmos, The Pennsylvania State University, University Park, PA 16802, USA}

\author[0000-0002-9280-7594]{Benjamin D. Johnson}
\affiliation{Center for Astrophysics - Harvard \& Smithsonian, 60 Garden Street, Cambridge, MA 02138, USA}

\author[0000-0001-5063-8254]{Rachel Bezanson}
\affiliation{Department of Physics and Astronomy and PITT PACC, University of Pittsburgh, Pittsburgh, PA 15260, USA}

\author[0000-0001-5448-1821]{Robert Kennicutt}
\affiliation{Steward Observatory, University of Arizona, 933 North Cherry Avenue, Tucson, AZ 85721, USA}
\affiliation{Department of Physics and Astronomy and George P. and Cynthia Woods Mitchell Institute for Fundamental Physics and Astronomy, Texas A\&M University, 4242 TAMU, College Station, TX 77843-4242, USA}

\author[0000-0003-3256-5615]{Justin Spilker}
\affiliation{Department of Physics and Astronomy and George P. and Cynthia Woods Mitchell Institute for Fundamental Physics and Astronomy, Texas A\&M University, 4242 TAMU, College Station, TX 77843-4242, USA}

\author[0000-0002-8224-4505]{Sandro Tacchella}
\affiliation{Kavli Institute for Cosmology, University of Cambridge, Madingley Road, Cambridge, CB3 0HA, UK}
\affiliation{Cavendish Laboratory, University of Cambridge, 19 JJ Thomson Avenue, Cambridge, CB3 0HE, UK}

\begin{abstract}

We explore how the presence of detectable molecular gas depends on the inferred star formation histories (SFHs) in eight massive, quiescent galaxies at $\mathrm{z\sim0.7}$. Half of the sample have clear detections of molecular gas, traced by CO(2-1). We find that the molecular gas content is unrelated to the rate of star formation decline prior to the most recent 1 Gyr, suggesting that the gas reservoirs are not left over from their primary star formation epoch. However, the recent SFHs of CO-detected galaxies demonstrate evidence for secondary bursts of star formation in their last Gyr. The fraction of stellar mass formed in these  secondary bursts ranges from $\mathrm{f_{burst}\approx0.3-6\%}$, and ended between $\mathrm{t_{end\mbox{-}burst}\approx0-330~Myr}$ ago. The CO-detected galaxies form a higher fraction of mass in the last Gyr ($\mathrm{f_{M_{1Gyr}}=2.6\pm1.8\%}$) compared to the CO-undetected galaxies ($\mathrm{f_{M_{1Gyr}}=0.2\pm0.1\%}$). The galaxies with  gas reservoirs have enhanced late-time star formation, highlighting this as a contributing factor to the observed heterogeneity in the gas reservoirs in high-redshift quiescent galaxies. We find that the amount of gas and star formation driven by these secondary bursts are inconsistent with that expected from dry minor mergers, and instead are likely driven by recently accreted gas i.e., gas-rich minor mergers. This conclusion would not have been made based on $\mathrm{SFR_{UV+IR}}$ measurements alone, highlighting the power of detailed SFH modeling in the interpretation of gas reservoirs. Larger samples are needed to understand the frequency of low-level rejuvenation among quiescent galaxies at intermediate redshifts, and to what extent this drives the diversity of molecular gas reservoirs.

\end{abstract}


\section{Introduction} \label{sec:intro}

The most massive galaxies in the nearby universe (log $\mathrm{M_*>11\ M_{\odot}}$) formed the majority of their stars rapidly, during the first few billion years of the universe's history 
\citep[e.g.][]{Thomas2010}. As cosmic time passes, massive galaxies rapidly truncate or ``quench" their star formation, and by $z\sim1$ the majority of massive galaxies have transitioned to a quiescent, passive phase \citep[e.g.,][]{Bell2004, Faber2007, Muzzin2013b, Tomczak2014, Davidzon2017}. Decades of study have revealed independent classes of mechanisms by which galaxies might quench, with internal physics linked to mass and external effects due to environment apparently both contributing independently to the growth of the red sequence \citep[e.g.,][]{Peng2010}. Despite quenching being a well-established phenomena, one of the most challenging problems in galaxy evolution is understanding the physical processes driving it. 

To provide insights into the physical mechanisms responsible for quenching galaxies, we can study the different timescales over which they form and quench \citep{Wu2018, Belli2019, Tacchella2022a, Akhshik2022}. For example, it has been proposed that a slow path ($\approx 1$Gyr) to quiescence is from the cessation of cosmic gas supply and subsequent slow consumption of the remaining gas \citep{Feldmann2015}. On the other hand, a rapid quenching ($\approx 10^8$yr) could be due to gas-rich processes such as major mergers \citep[e.g.,][]{Hopkins2006} or disk instabilities 
\citep{Yang2008, Dekel2009, Dekel2014, Zolotov2015}. To further complicate things, multiple mechanisms could be at play simultaneously. For example, the growth of a stellar bulge may stabilize cold gas against collapse, decreasing star formation efficiency \citep{Martig2009}, but to quench it may require additional processes such as the inability to accrete gas because of shock-heating at the virial radius of massive dark matter halos \citep{Birnboim2003, Keres2005, Dekel2006, Faucher2011}.

Different evolutionary pathways may be accompanied by changes in galactic structure. Studies have found that compact quiescent galaxies are older than more extended, normal-sized quiescent galaxies \citep{Tacchella2017, Williams2017, Lee2018, Wu2018, Estrada-Carpenter2020} and that quiescent galaxies at high redshift are on average more compact than quiescent galaxies at lower redshifts \citep{Daddi2005, Trujillo2006, Buitrago2008, Cimatti2008, vanDokkum2008, Cassata2013, Mosleh2017, Suess2019}.
One of the possible drivers of size evolution is that the stellar density of galaxies scales with the density of the universe at its formation epoch, such that galaxies that form later are larger \citep[progenitor bias;][]{Poggianti2013, Carollo2013, Lilly2016}, another possibility is that size increases due to accreted stellar envelopes from dry minor merging \citep{Bezanson2009, Hopkins2009, Naab2009, Trujillo2011, Hilz2012, Oser2012, Bluck2012, Poggianti2013, Ferreras2014, Damjanov2022}. While it is possible local samples are biased against identifying compact galaxies \citep{Taylor2010, Calvi2014, Szomoru2013}, empirical studies appear to favor dry minor merging, which we discuss in further detail in Section \ref{sec:minor_mergers}.

A new avenue for providing constraints on the quenching process involves studying the molecular gas content of quiescent galaxies, because ongoing star formation requires cold gas to collapse into stars. Indeed, this is reflected in global scaling relations such as the correlation between the surface density of star formation and molecular hydrogen \citep{Kennicutt1998}. For galaxies to quench, this cold gas must be unavailable or unusable to form new stars, which can happen because of many different mechanisms such as stabilization against collapse, feedback, or consumption. Several studies have characterized molecular gas in statistical samples of massive quiescent galaxies at $z\approx0$ \citep{Young2011, Saintonge2011, Saintonge2012, Saintonge2017, Davis2016} through ambitious large-scale surveys, investing significant telescope time to detect the faint emission from relatively small molecular gas reservoirs. They found that, after a sustained period of quiescence, the majority of galaxies have low gas fractions, $<1\%$, suggesting that additional gas accretion is prevented (although exceptions with large gas reservoirs exist, up to 24\%, see \citealt{Sansom2019}). Other studies have found that recently quenched galaxies (e.g., post-starburst galaxies) in the local universe have a puzzling diversity of molecular gas, with gas fractions up to $\sim50$\% \citep{Rowlands2015, French2015, French2018, Alatalo2016, Smercina2018, Li2019, Yesuf2020, Smercina2022, French2022}, suggesting that at least some fraction of galaxies that quenched recently have done so by a different mechanism that does not destroy or deplete gas. To better understand the quenching process, it is important to study the gas content of galaxies closer to the peak epoch of quenching at  $z\approx1$.

Beyond the local universe, relatively few studies have made molecular gas mass measurements. Similarly to post-starburst galaxies in the nearby universe, studies at higher redshifts reveal a heterogeneous range in  molecular gas properties. At $z\sim0.7$, there is heterogeneity in the gas content of individual galaxies \citep{Suess2017, Spilker2018, Bezanson2022}, while most of the studies at $z>1$ of individual galaxies find very low gas content, of a few percent or less \citep{Sargent2015, Bezanson2019, Caliendo2021, Williams2021, Whitaker2021a}. This finding could indicate that the quenching process in high-redshift galaxies is both efficient and complete \citep{Bezanson2019, Williams2021}, plausibly due to galaxies running out of gas \citep{Whitaker2021a}. However, average properties measured by stacking dust emission in quiescent galaxies at $z>1$ have, in contrast, shown evidence for large gas reservoirs, with gas fractions an order of magnitude higher \citep{Gobat2018, Magdis2021}. These seemingly conflicting results may result from different methodologies, or they may more broadly reflect gas content in quiescent galaxies that are higher than implied by the small sample of single targets, indicating that the population demonstrates a large diversity of gas content. This diversity could be linked to different quenching processes in galaxies, or it may reflect external factors that could impact their later star formation histories (SFHs), such as recently accreted gas. 

Therefore, understanding the link between cold molecular gas reservoirs and the detailed SFHs of quiescent galaxies would provide important insights into the quenching process. Inferring the SFHs of galaxies requires high-quality spectra and photometry to probe stellar populations for the subtle diagnostic features that imprint the detailed timescales of star formation, so most studies have been limited to the local universe. Recently, large spectroscopic surveys have produced high S/N spectroscopy of galaxies outside of the local universe, enabling the properties of galaxies out to $z\sim1$ to be characterized to similar quality of SDSS at $z\sim0$. The Large Early Galaxy Astrophysics Census (LEGA-C) Survey \citep{vanderWel2016, Straatman2018, vanderWel2021} has produced 20-hour depth spectroscopy of galaxies from $0.6<z<1$ to observed $\lambda \lesssim 8800${\AA}, providing exquisite constraints on rest-frame optical diagnostics that trace ages, SFHs, and velocity dispersions. In addition, a small sample of LEGA-C quiescent galaxies has been observed with the Atacama Large Millimeter/submillimeter Array (ALMA), providing gas mass measurements based on observations of CO(2-1) emission \citep{Spilker2018}, which comprise the sample for this paper.

In this paper, we infer the detailed SFHs of the \citet{Spilker2018} sample while assuming a highly flexible functional form for the SFHs. We are thus able to reconstruct complex and more realistic evolutionary pathways, such as bursts of star formation that cause sharp transitions in the star formation rate, SFR(t), that cannot be captured by traditional parametric models of star formation history. Having both molecular gas measurements and the ability to constrain SFHs, both of which are challenging to measure, opens the door to a better understanding of the quenching process. The only other ALMA survey at comparable redshifts and with similarly excellent quality spectra is SQUIGGLE \citep{Suess2022, Bezanson2022}, where galaxies were selected to be in a post-starburst phase and therefore are a younger population than our sample. Our work thus provides constraints on the mechanisms that cause more early-forming quiescent galaxies to quench their star formation, and it gives new insights into the heterogeneity of their molecular gas mass measurements, complementing studies that target more-recently formed post-starburst galaxies. 

In Section \ref{sec:data}, we present our sample. In Section \ref{sec:SEDfitting}, we present our spectral energy distribution (SED) fitting methods and models. In Section \ref{sec:results}, we discuss our results. In Section \ref{sec:discussion}, we discuss possible causes for our results. We assume a flat $\Lambda$ cold dark matter ($\Lambda$CDM) cosmology with WMAP-9 parameters, $\Omega_m=0.287$, $H_0=69.3\ \mathrm{km\ s^{-1}\ Mpc^{-1}}$ \citep{Hinshaw2013}, and a Chabrier initial mass function (IMF) \citep{Chabrier2003}.

\section{Data} \label{sec:data}
\begin{deluxetable*}{lll}[!htbp]
\tablecaption{\texttt{Prospector} Parameters for the fiducial model \label{tab:priors}}
\tablewidth{0pt}
\tablehead{
\colhead{Parameter} & \colhead{Description} & \colhead{Prior}}
\startdata
log $M/M_{\odot}$ & Total mass formed & Uniform: min=10, max=12.5 \\
$z$ & Redshift & Uniform: min=$z_{spec}-0.02$, max=$z_{spec}+0.02$ \\
log $Z/Z_{\odot}$ & Stellar metallicity & Uniform: min=-0.8, max=0.19 \\
$n$ & \makecell{Power-law modifier to shape of the \citet{Calzetti2000} attenuation \\ curve of the diffuse dust} & Uniform: min=-1.0, max=0.4\\
$\tau_{dust,2}$ & Diffuse dust optical depth & Clipped normal: min=0, max=4, $\mu$=0.3, $\sigma$=1 \\
$\tau_{dust,1}$ & Birth-cloud dust optical depth & Clipped normal in ($\tau_{dust,1}/\tau_{dust,2}$): min=0, max=2, $\mu$=1, $\sigma=0.3$ \\
$\gamma_{e}$ & Mass fraction of dust in high radiation intensity & Log-uniform: min=$10^{-4}$, max=0.1 \\
$U_{min}$ & Minimum starlight intensity to which the dust mass is exposed & Uniform: min=0.1, max=15 \\
$q_{PAH}$ & Percent mass fraction of PAHs in dust & Uniform: min=0, max=7.0 \\
$f_{AGN}$ & AGN luminosity as a fraction of the galaxy bolometric luminosity & Log-uniform: min=$10^{-5}$, max=3 \\
$\tau_{AGN}$ & Optical depth of AGN torus dust & Log-uniform: min=5, max=150 \\
log $\mathrm{Z_{gas}/Z_{\odot}}$ & Gas-phase metallicity & Uniform: min=-1, max=0 \\
log $\mathrm{U_{gas}}$ & Gas Ionization Parameter & Uniform: min=-3.5, max=-2.0 \\
$\sigma$ & Velocity Dispersion & Uniform: min=$\sigma_{LEGA-C}-50$, max=$\sigma_{LEGA-C}+50$ \\
SFR Ratios & \makecell{Ratio of the SFRs in adjacent bins; $N_{SFH}$-1 bins total with  \\ $N_{SFH}$=10} & Student’s t-distribution: $\sigma = 0.3$, $\nu = 2$ \\
$\mathrm{f_{outlier, spec}}$ & Spectra outlier fraction & Uniform: min=$10^{-5}$, max=0.1 \\
spec\_norm & Spectra calibration & Normal: $\mathrm{\sigma}$=0.2, mean=1.0 \\
spec\_jitter & Spectral noise multiplier & Uniform: min=0, max=3.0
\enddata
\tablecomments{$z_{spec}$ and $\sigma$ are the spectroscopic redshift and velocity dispersion from LEGA-C DR3}
\end{deluxetable*}
\subsection{LEGA-C and UltraVISTA}\label{sec:legac_uvista}
LEGA-C \citep{vanderWel2016, Straatman2018, vanderWel2021}) is a 128-night ESO public spectroscopic survey with VIMOS on the Very Large Telescope. The survey includes $\mathrm{\sim3500}$ $\mathrm{K_s}$-band selected galaxies at redshifts $0.6<z<1.0$, stellar masses $\mathrm{M_* > 10^{10} M_{\odot} }$, and observed wavelength range $\mathrm{6300{\AA} \leq \lambda \leq 8800{\AA}}$ in the wide-area COSMOS field \citep{Scoville2007}. The 20 hour long integrations produce continuum spectra with S/N $\mathrm{\sim 20{\AA}^{-1}}$ and high-resolution (R$\mathrm{\approx 3500}$; \citealt{Straatman2018}). We use the spectra from the third data release \citep[DR3:][]{vanderWel2021}.

In addition, the LEGA-C galaxies have excellent ancillary data from the UltraVISTA catalog \citep{Muzzin2013a} that we utilize in our fitting. The LEGA-C galaxies were selected from this catalog, which is a collection of photometric data across 30 passbands from 0.2 to 24$\mathrm{\mu m}$. It includes UV imaging from the GALEX satellite \citep{Martin2005}, optical imaging from the Canada–France–Hawaii Telescope (CFHT) and Subaru telescope \citep{Taniguchi2007, Capak2007}, near-infrared data from the VISTA telescope \citep{McCracken2012}, and mid-infrared data from Spitzer \citep{Sanders2007, Frayer2009}. The IRAC and MIPS photometry from Spitzer was deblended using a well-tested code for many $K_s$ selected catalogs \citep[e.g.][]{Labbe2005, Wuyts2007, Marchesini2009, Williams2009, Whitaker2011, Marchesini2012, Labbe2010, Labbe2013}. See more details from Section 3.5 of \citet{Muzzin2013a}.

\subsection{ALMA}
We use ALMA measurements from \citet{Spilker2018} who observed the CO(2-1) transition, which is a tracer of the molecular ISM. The 8 massive, quiescent galaxies were selected to be log $\mathrm{M_*>10.8\ M_{\odot}}$ and with $\mathrm{SFR_{UV+IR}\ 3-10\times}$ below the star-forming main sequence as defined by \citet{Whitaker2014}. Four of the eight targets have clear detections of CO(2-1) emission. \citet{Spilker2018} use standard assumptions to convert the CO(2-1) luminosities to molecular gas masses. They assume the luminosity ratio between CO(2-1) and CO(1-0) transitions, $r_{21}=0.8$, which is within the plausible expected range, $r_{21}=0.7-1.0$, based on observations of the Milky Way, nearby quiescent and star-forming galaxies, and high-redshift star-forming galaxies \citep{Fixsen1999, Combes2007, Dannerbauer2009, Young2011, Spilker2014, Saintonge2017}. In addition, they assume a "Milky Way-like" conversion factor $\alpha_{CO}=4.4 M_{\odot}\ (K\ km\ s^{-1}\ pc^2)^{-1}$ \citep{Solomon1987, Bolatto2013, Sandstrom2013}. For the 4 CO-undetected galaxies, conservative 3$\sigma$ upper limits are estimated. The noise in 800 km $\mathrm{s^{-1}}$ wide channels is used for this estimation, which would fully include all of the line emission from detected sources. The integrated line flux upper limits are proportional to $\sqrt{\Delta v}$, where $\Delta v$ is the velocity interval over which the spectrum is integrated. The range of molecular gas masses measured for the CO-detected galaxies is $9.82<\log\ (M_{H_2}/M_{\odot})<10.16$ and the limits for the CO-undetected galaxies are $9.72<\log\ (M_{H_2}/M_{\odot})<9.91$. The redshift range for these galaxies is $0.60<z<0.73$. We note that the range of half-light sizes for our sample of galaxies is 0.4-1.6\arcsec, therefore the full extent of the stellar distribution is larger than this. The VIMOS slits widths are 1\arcsec and at $\sim$ 2mm wavelength of the CO(2-1), ALMA has a primary beam Half Power Beam Width of $\sim$38.8\arcsec. Therefore, ALMA is able to probe the entire galaxy and surrounding regions while the optical spectra probe only the inner regions ($<$1\arcsec) of the galaxies.

\section{SED Fitting} \label{sec:SEDfitting}
\begin{figure*}[!htbp]
\includegraphics[trim={10cm 10cm 10cm 10cm}, clip, width=\textwidth]{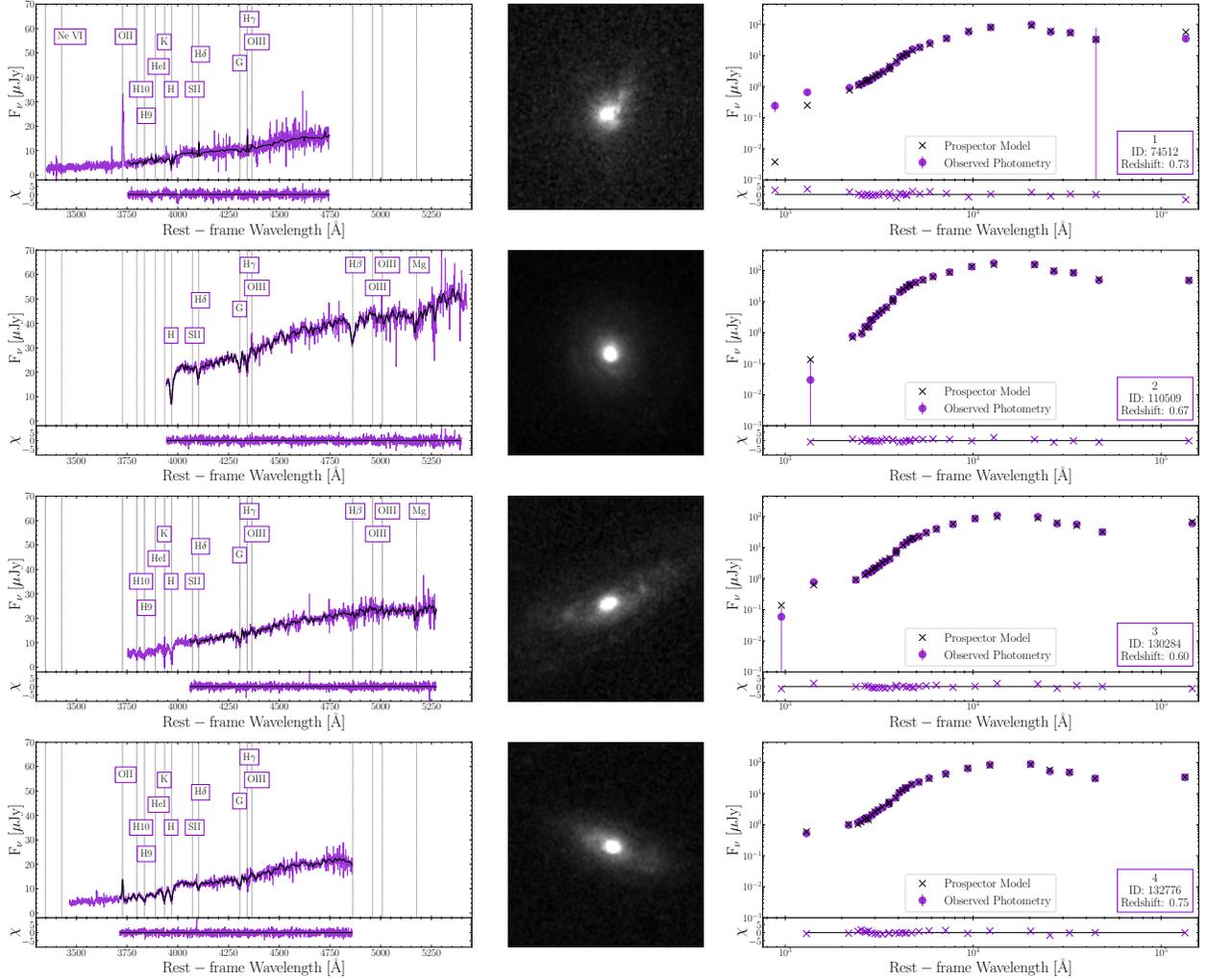}
\caption{Observed data in purple and best fit \texttt{Prospector} model in black for the CO-detected galaxies. A mask was applied to the spectra before fitting to only cover the approximate spectral response limits and to exclude wavelengths below the MILES spectral library limit. The left column shows the spectra while the right column shows the photometry. The smaller panels at the bottom show $\chi$, defined as (data-model)/$\sigma$. The middle columns are \textit{HST/ACS} F814W images of each galaxy, shown as $4\times4 \arcsec$ cutouts. Galaxies are assigned numbers 1-8 and will be labeled as such in all future plots for ease of comparison. The data are well-fit by the model. \label{fig:seds_co}}
\end{figure*}

\renewcommand{\thefigure}{\arabic{figure} (Cont.)}
\addtocounter{figure}{-1}

\begin{figure*}[!htbp]
\includegraphics[trim={10cm 10cm 10cm 10cm}, clip, width=\textwidth]{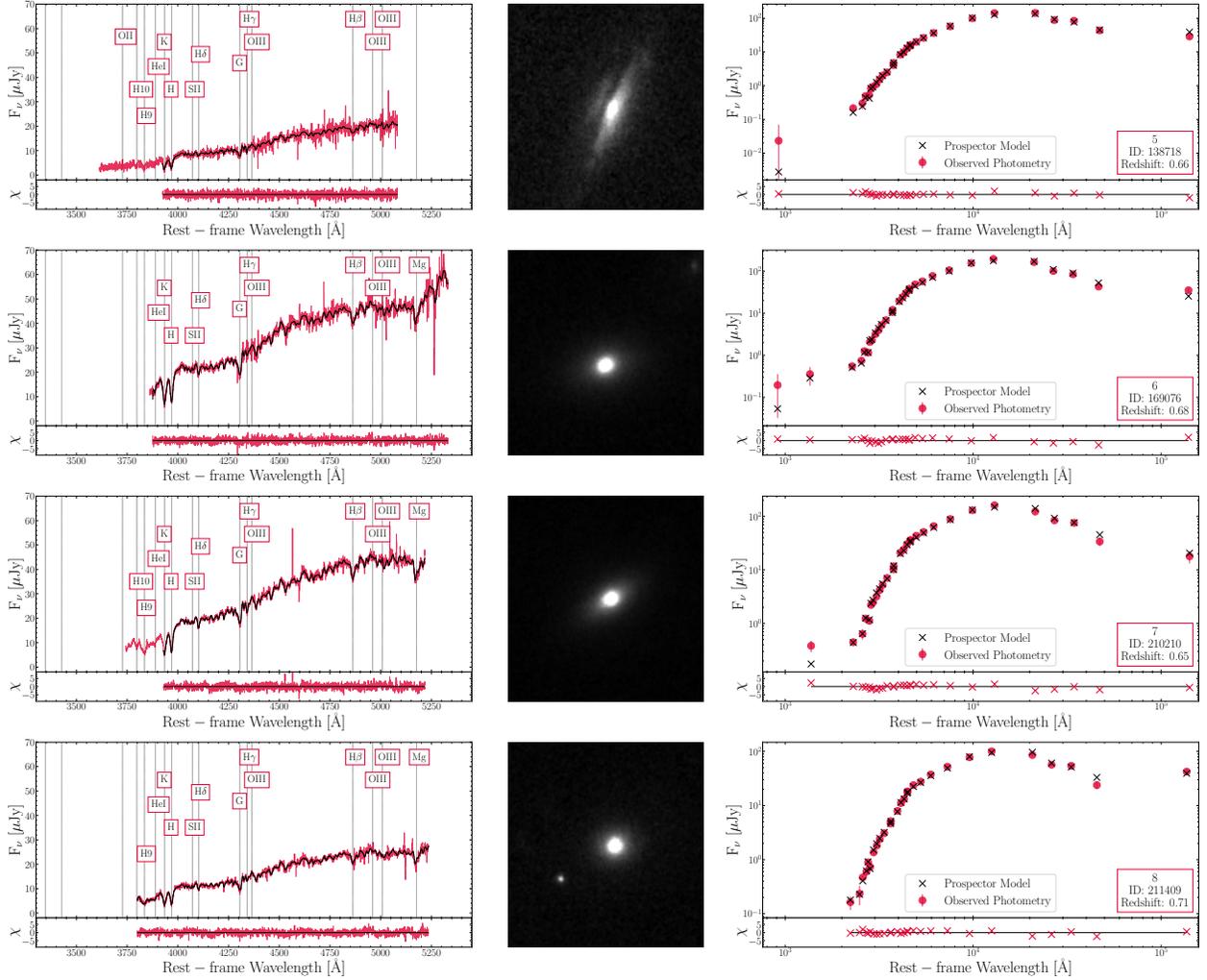}
\caption{Observed data in red and best fit \texttt{Prospector} model in black for the CO-undetected galaxies. \label{fig:seds_noco}}
\end{figure*}
\renewcommand{\thefigure}{\arabic{figure}}

 We use the SED fitting code $\texttt{Prospector}$ (v1.1.0), which uses a Bayesian inference framework \citep{Leja2017, Johnson2021} to infer SFHs. The posterior distributions are sampled using the dynamic nested sampling code $\texttt{dynesty}$ (v1.2.3) \citep{dynesty:2020}. Dynamic nested sampling is preferred over the more popular Markov Chain Monte Carlo (MCMC) methods because it has precise and flexible stopping criteria and improved evaluation of multimodal solutions \citep{Leja2018a}.

The spectra were flux-calibrated by matching the shape of the photometric SED \citep{Straatman2018}. This is beneficial for the simultaneous fitting of both spectra and photometry, as it helps correct any wavelength-dependent normalization offset between the spectra and photometry, from imperfect flux calibration of the spectra. We fit 30 passbands from the photometry, including \texttt{GALEX} (NUV and FUV), CFHT (MegaPrime u*), Subaru Suprime-Cam (IA427, IA464, IA484, IA505, IA527, IA574, IA624, IA679, IA709, IA738, IA767, IA827, B, V, g+, r+, i+, z+), UltraVISTA (Y, J, H, Ks), Spitzer IRAC (ch1, ch2, ch3, and ch4) and MIPS24, see Section \ref{sec:legac_uvista} for more information. We enforce a 5\% error floor for all bands of photometry, to allow for systematic errors in the physical models for stellar, gas, and dust emission. Some galaxies in our sample do not have one or both of the UV data from GALEX, in which case only the remaining 28-29 passbands were fit.

\begin{figure*}[!htbp]
\includegraphics[trim={7cm 10cm 8cm 11cm}, clip, width=\textwidth]{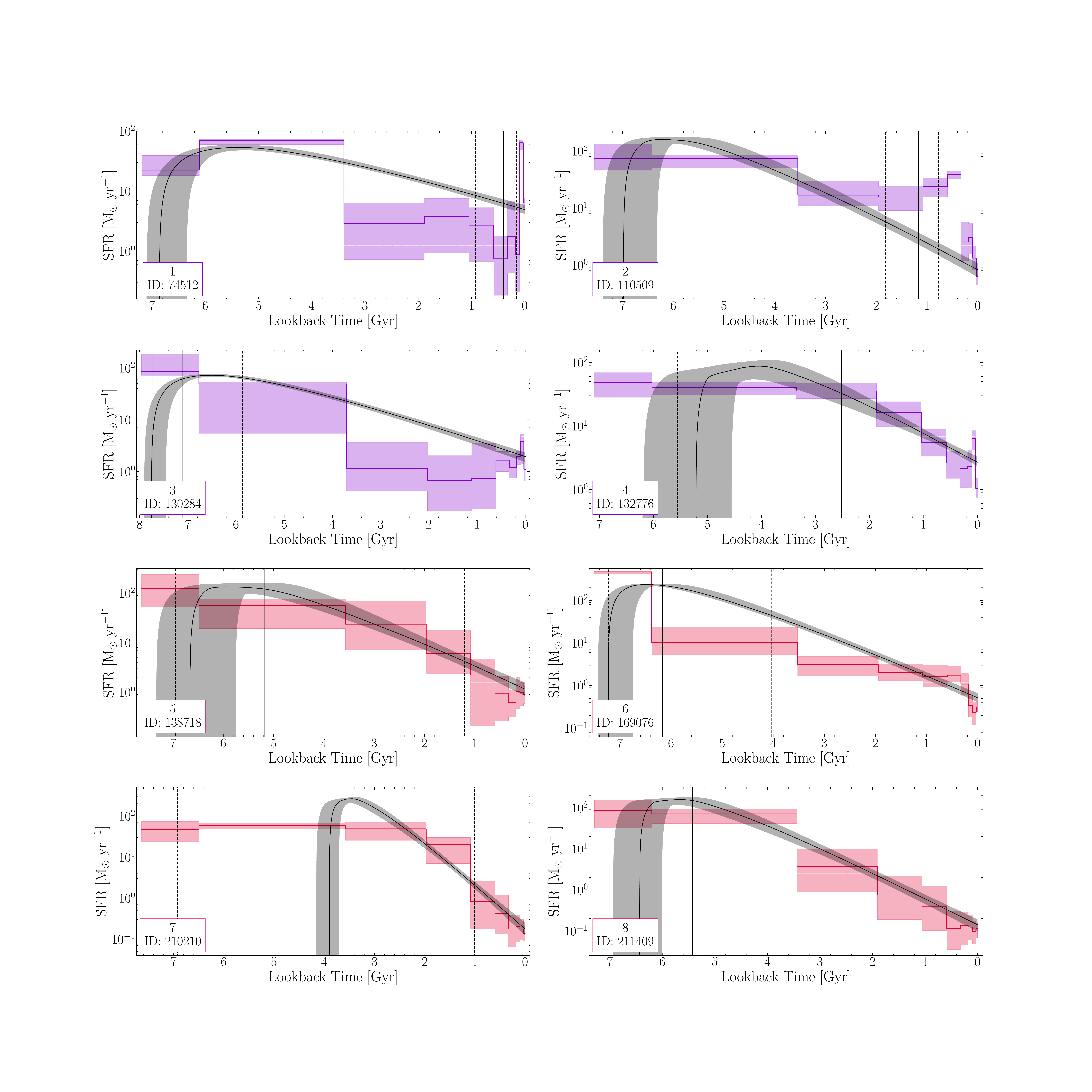}
\caption{A comparison between the parametric and fidudical nonparametric models for individual galaxies. Note that the y-axes are not on the same scale between plots. The parametric SFHs for each galaxy are shown in black. The nonparametric SFHs are shown in purple for the CO-detected galaxies and in red for the CO-undetected galaxies, which are the fiducial models for this study. The shaded regions show the 16th and 84th percentile range of the posterior probability. The time of the burst for the parametric model is shown as a solid black vertical line, with the 16th and 84th percentiles shown as dotted black vertical lines. The fraction of mass formed during the burst for both the parametric and nonparametric models are listed in Table \ref{tab:age}. The CO-detected galaxies show evidence for enhanced late time star-formation (0-1 Gyr in lookback time) and secondary bursts of star formation compared to the CO-undetected galaxies. \label{fig:parametric_SFHs}}
\end{figure*}

\begin{figure*}[!htbp]
\includegraphics[trim={7cm 9cm 8cm 11cm}, clip, width=\textwidth]{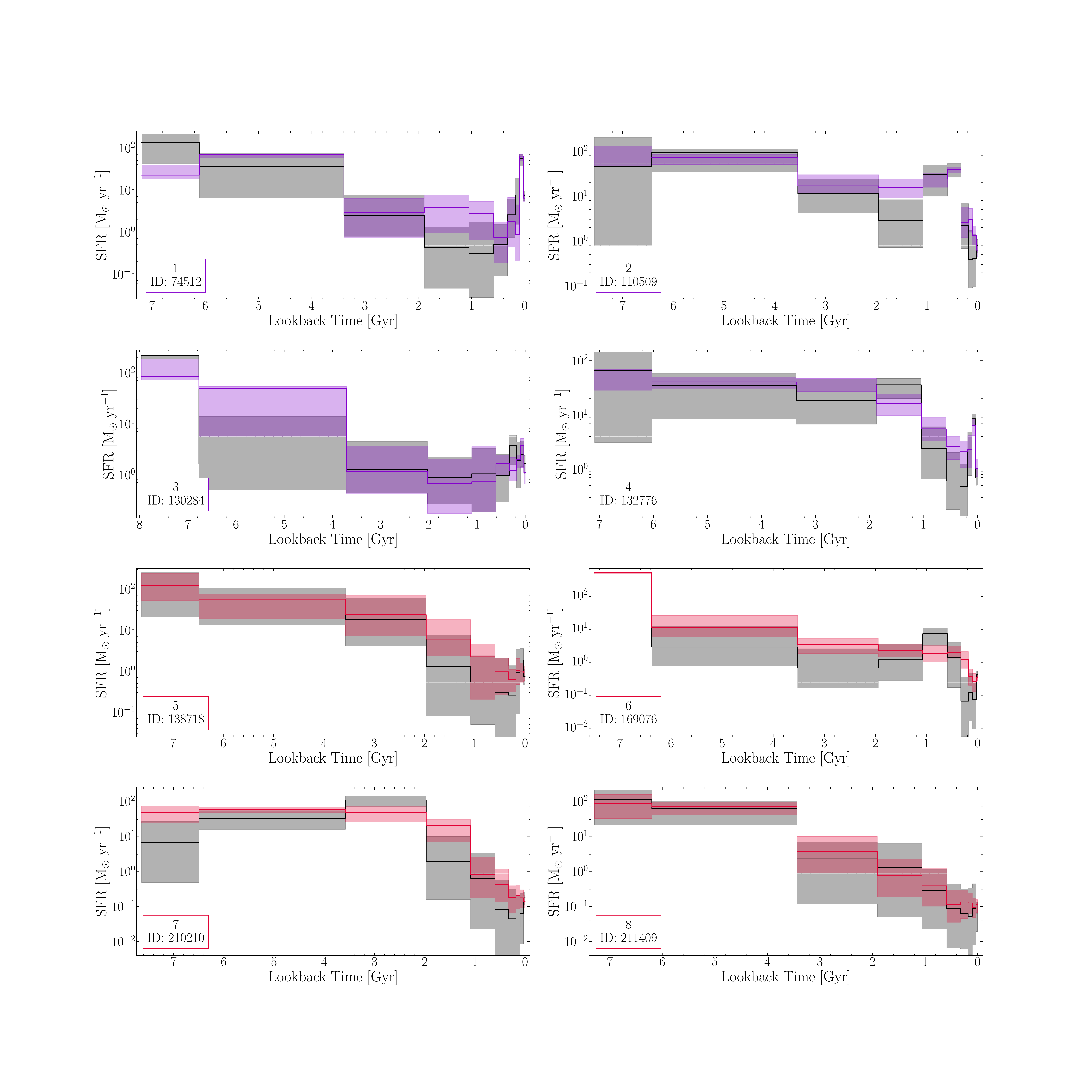}
\caption{The nonparametric SFHs shown in comparison, with the bursty prior in black and the continuity prior in purple for CO-detected galaxies and in red for CO-undetected galaxies. Note that the y-axes are not on the same scale between plots. \label{fig:nonparametric_SFHs}}
\end{figure*}

\subsection{Stellar Populations}
$\texttt{Prospector}$ uses the Flexible Stellar Population Synthesis (\texttt{FSPS}, v3.2) stellar population code \citep{Conroy2009} via \texttt{python-FSPS} (v0.4.1) \citep{ForemanMackey2014}. We employ the MIST stellar evolutionary tracks and isochrones \citep{Choi2016, Dotter2016} which utilizes the MESA stellar evolution package \citep{Paxton2011, Paxton2013, Paxton2015, Paxton2018}. We use MILES for the stellar spectral library \citep{Vazdekis2015, Falcon-Barroso2011} and adopt a Chabrier \citep{Chabrier2003} IMF. The IGM absorption is modeled after \citet{Madau1995}. 

\begin{deluxetable*}{llllllll}[!htbp]
\tablecaption{Star Formation Rates \label{tab:sfrs}}
\tablewidth{0pt}
\tablehead{
\colhead{\#} & \colhead{ID} & \colhead{$\mathrm{SFR_{100Myr}}$} &  \colhead{$\mathrm{SFR_{30Myr}}$} & \colhead{$\mathrm{SFR_{UV+IR}}$} 
& \colhead{$\mathrm{SFR_{radio}}$} & \colhead{$\mathrm{SFR_{peak-burst}}$} & \colhead{$\mathrm{SFR_{\text{[OII]}}}$}}
\startdata
1 & 74512\tablenotemark{a} & $46.6^{+5.2}_{-8.1}$ & $6.4^{+1.4}_{-1.0}$ & 6.3 & $<14$ & $63.9^{+6.3}_{-15.5}$ & $6.7\pm 1.1$ \\
2 & 110509\tablenotemark{b} & $1.1^{+0.6}_{-0.4}$ & $0.6^{+0.2}_{-0.2}$ & 6.5 & $428.4\pm22.2$ \tablenotemark{b} & $39.1^{+5.9}_{-6.5}$ & ...\\
3 & 130284 & $3.0^{+1.0}_{-1.5}$ & $1.1^{+0.6}_{-0.4}$ & 6.8 & $<14$ & $3.8^{+1.4}_{-2.4}$ & ...\\
4 & 132776 & $4.7^{+1.4}_{-1.4}$ & $1.0^{+0.5}_{-0.3}$ & 7.9 & $<14$ & $2.6^{+2.1}_{-2.2}$ & $2.0\pm0.2$ \\
5 & 138718 & $1.0^{+0.5}_{-0.4}$ & $0.9^{+0.4}_{-0.3}$ & 5.6 & $<14$ & ... & $0.6\pm0.4$ \\
6 & 169076\tablenotemark{b} & $0.3^{+0.1}_{-0.1}$ & $0.3^{+0.1}_{-0.1}$ & 5.1 & $126.8\pm7.6$\tablenotemark{b} & ... & ...\\
7 & 210210 & $0.2^{+0.1}_{-0.1}$ & $0.1^{+0.1}_{-0.04}$ & 3.6 & $32.8\pm4.4$ & ... &  ...\\
8 & 211409 & $0.1^{+0.1}_{-0.04}$ & $0.1^{+0.04}_{-0.05}$ & 6.6 & $<14$ & ... &  ...\\
\enddata
\tablenotetext{a}{\citet{Belli2021} found that the stellar absorption lines for this galaxy are offset from the CO and [OII] emission lines by 2400 km $s^{-1}$ and \citet{Spilker2018} note that the gas and stars in this galaxy may be misaligned, both of which may indicate evidence of a merger.} 
\tablenotetext{b}{Radio-loud AGNs with overestimated $\mathrm{SFR_{radio}}$ values \citep{Barisic2017}}
\tablecomments{All values of SFR are in units of $\mathrm{M_{\odot}\ yr^{-1}}$. $\mathrm{SFR_{100Myr}}$ and $\mathrm{SFR_{30Myr}}$ are determined from the SED fitting with \texttt{Prospector}. It is the SFR for the most current 100Myr and 30Myr, respectively. $\mathrm{SFR_{UV+IR}}$ values are from \citet{Spilker2018}. The approximate 3$\sigma$ upper limit SFR from the radio catalog is $\mathrm{14\ M_{\odot}\ yr^{-1}}$. $\mathrm{SFR_{peak-burst}}$ is the SFR at the peak of the potential secondary bursts. \citet{Belli2021} used \texttt{Prospector} to infer SFHs for galaxies 2-8. Their $\mathrm{SFR_{60Myr}}$ values are always consistent within the uncertainties for our $\mathrm{SFR_{30Myr}}$ or $\mathrm{SFR_{100Myr}}$ or have a value that is between the two.}
\end{deluxetable*}

\subsection{Dust Attenuation}
For dust attenuation, we tested different models. We fit with a \citet{Kriek2013} flexible attenuation curve with the UV bump tied to the slope of the curve. We use the \citet{Charlot2000} two-component dust model, which assumes a diffuse dust screen across all stellar light ($\hat{\tau}_{dust, 2}$) and additional attenuation of stellar light for stars that have formed in the past 10 Myr ($\hat{\tau}_{dust, 1}$). The free parameters are the diffuse dust attenuation index ($n$), $\hat{\tau}_{dust, 1}$, and $\hat{\tau}_{dust, 2}$. We use a joint prior on the ratio of the two dust components $0<\hat{\tau}_{dust, 1}/\hat{\tau}_{dust, 2}<2.0$ as in \citet{Leja2018b}. We compared the results obtained using this method to those determined with a \citet{Calzetti2000} attenuation curve and found that all of their inferred masses were consistent within uncertainties. In addition, for 6 of the 8 galaxies, their SFHs were consistent within uncertainties. However, for galaxy 1, five of the agebins were not consistent within uncertainties and for galaxy 2, the three most recent agebins were not consistent within uncertainties. The ratio of the Bayesian evidence can be used to determine which model is better-fit by the photometry and spectra; the evidence increases with more accurate models but is penalized for model complexity. For most galaxies, the Bayesian evidence does not strongly favor either model. However, for galaxies 1 and 2 the evidence strongly supports the more complex dust model. Because the inferred SFHs were not significantly different from each other with the different dust models for all galaxies except galaxy 1 and 2, we use the evidence-supported \citet{Kriek2013} flexible attenuation curve with the \citet{Charlot2000} two-component dust model as our fiducial model for dust attenuation. This is in line with other studies supporting steeper than Calzetti attenuation curves in quiescent galaxies \citep[e.g.][]{Salim2018, Nagaraj2022}.

\subsection{Dust Emission}
Our dust emission model assumes energy balance, where all starlight is attenuated by dust and re-emitted in the IR \citep{daCunha2008}. We use the \citet{Draine2007} dust emission templates, which control the shape of the IR SED with three free parameters. $U_{min}$ and $\gamma_e$ control the shape and location of the thermal dust emission bump and $q_{PAH}$ is the fraction of total dust mass that is in PAHs. These templates are based on the silicate-graphite-PAH model of interstellar dust \citep{Mathis1977, Draine1984}. See Appendix C of \citet{Leja2017} for more information.

\subsection{AGN}
Our AGN model adopts AGN templates from \citet{Nenkova2008a} and \citet{Nenkova2008b}, which are included in \texttt{FSPS}. This model has two free paremeters: $f_{AGN}$, which is the AGN luminosity as a fraction of the galaxy bolometric luminosity, and $\tau_{AGN}$, which is the optical depth of the AGN dust torus. A log-uniform prior describes the observed power-law distribution of black hole accretion rates \citep{Aird2017, Georgakakis2017, Caplar2018}, therefore we adopt a log-uniform prior for $f_{AGN}$. A log-uniform prior is adopted for $\tau_{AGN}$, as the SED response to logarithmic changes in $\tau_{AGN}$ is approximately linear. For more details, see \citet{Leja2018b}. We use nebular marginalization for the AGN emission lines because they are powered by sources not included in our nebular emission model; see next section for more details.

\subsection{Nebular Emission}

The nebular emission is based on CLOUDY model grids \citep{Byler2017}, and includes a nebular continuum component. The ionizing continuum from the stellar populations is modeled to be fully absorbed by the gas and then emitted as both line and continuum emission. Both the gas-phase metallicity and gas ionization are set as free parameters, and the gas-phase metallicity is decoupled from the stellar metallicity. The line emission from quiescent galaxies may be caused by evolved stars, gas shocks, or AGN \citep{Kewley2006}, while the assumption in FSPS is that emission lines are caused by HII regions. Evolved stars, more specifically extreme horizontal branch or post-asymptotic giant branch stars, have been shown to produce ionizing photons in theoretical models \citep{Greggio1990, Binette1994, CidFernandes2011, Belfiore2016, Byler2019}. \citet{Maseda2021} found that 59\% of UVJ-quiescent LEGA-C galaxies have [OII] emission (EW$>1.5${\AA}) and that star formation is unlikely to provide the bulk of the ionizing photons. For this reason, we use the nebular emission-line marginalization option in \texttt{Prospector}. This allows for some flexibility in the emission-line modeling; each emission line is modeled with a Gaussian fit with variable width and amplitude instead of coupling the observed emission line luminosities to stellar physics, see \citet{Johnson2021}.  In other words, this decouples the emission lines from the SFH.

\subsection{Noise and Calibration Models}
Simultaneously fitting high-resolution spectra and photometry can be challenging owing to differing instruments, data processing, and calibration procedures.  Therefore, we include both noise and calibration models in our simultaneous SED fitting. Three parameters are used for noise and calibration including $f_{outlier, spec}$, spec\_norm, and spec\_jitter. $f_{outlier, spec}$ is an outlier model that marginalizes over significant deviations from the model, see \citet{Hogg2010} for more details. For our sample, the median fraction of pixels which are determined to be outliers, or ``bad" spectral data points, from the fits is 0.001 (0.1\% per galaxy). 

To look for spectroscopic uncertainties that are not well-determined, the parameter spec\_jitter is also used. This term essentially inflates the errors such that the reduced $\chi^2 \approx 1$, where when spec\_jitter=1, there is no inflation. The median spec\_jitter value for our sample is 1.5, showing some minor inflation. 

Because we are fitting both spectra and photometry, the overall continuum shape is set using photometry with the continuum of the spectra adjusted to match the shape from the photometry. This is done with the parameter spec\_norm, where the model spectrum is multiplied by a 10th-order maximum-likelihood polynomial using the ratio of the observed and model spectrum before computing the likelihood. The value for spec\_norm is then the ratio between the observed spectrum and model spectrum. The median value for spec\_norm for our sample is 1.03. This value is very close to unity, due to the spectra being flux-calibrated to the photometry before fitting. See Table \ref{tab:priors} for more details, including priors used, and \citet{Johnson2021} for a more in-depth discussion on the noise and calibration models.

\subsection{Star Formation Histories}\label{sec:SFH_methods}
We infer the SFHs with three different models included in \texttt{Prospector} to test whether we find the same results while making different assumptions. We include a parametric model and two nonparametric models with differing priors. 

For the parametric model, we experiment with the commonly used delayed-$\tau$ model, with functional form $SFR(t)=A\ t\ exp(-t/\tau)$, and we include an additional instantaneous random burst (delta-function) of star formation. This SFH model has the following free parameters: age, $\tau$ (timescale of exponentially declining star formation), $t_{burst}$ (time of instantaneous burst), and $f_{burst}$ (fraction of mass formed in instantaneous burst). 

The ratio of the Bayesian evidence (often referred to as Bayes factor) quantifies the relative evidence between two models. \citet{Jeffreys1998} developed a classification scheme based on Bayes factor to interpret the decisiveness of the evidence with the following categories: decisive, very strong, strong, substantial, and barely worth mentioning (weak). Using this classification scheme, we find that the nonparametric model is preferred over the parametric model for six of the galaxies in our sample. The nonparametric model shows decisive evidence for galaxy 1, very strong evidence for galaxy 4, strong evidence for galaxy 6, substantial evidence for galaxies 3 and 5, and weak evidence for galaxy 8. The parametric model shows strong evidence for galaxy 2 and very strong evidence for galaxy 7. The nonparametric model is preferred for most galaxies, and where evidence exists for the parametric model, our results would not change, because the parametric model also recovers the late time burst for galaxy 2 and the shapes of both models agree for galaxy 7.

Nonparametric models assume a highly flexible functional form for SFR(t). \citet{Leja2019a} found that the priors chosen for the nonparametric models are the primary determinants for the size of the posteriors (and therefore the uncertainties). We therefore fit with two different priors for the nonparametric SFHs. We assume the continuity prior using the student's t-distribution with $\sigma=0.3$ and $\nu=2$. This prior fits for smooth transitions in SFR(t) between adjacent time bins and is a flexible model, able to fit both star-forming and maximally old stellar populations, see \citet{Leja2019a}. We also assume a bursty continuity prior \citep{Tacchella2022}, using the student's t-distribution with $\sigma=1.0$ and $\nu=2$, which leads to a more variable SFH by allowing sharper transitions in SFR(t) between adjacent time bins. Both nonparametric SFH models have 10 time bins specified in lookback time: the first two bins are spaced at 0-30 Myr and 30-100 Myr, while the last bin is fixed at $0.15\cdot t_{obs}\ \mbox{before}\ t_{obs}$, where $t_{obs}$ is the age of the universe at the galaxy's observed spectroscopic redshift. The remaining seven bins are spaced evenly in logarithmic time between 100 Myr and $0.85t_{obs}$.

The nonparametric models allow us to fit for complex SFH trends, such as bursts and other sharp transitions in SFR(t), that would not be allowed by commonly used parametric forms such as the delayed tau-model. Parametric functions, such as the exponentially declining $\tau$-models, intrinsically couple the older SFH shape to the recent SFH \citep{Simha2014}, potentially causing an unphysical link between unrelated events in the SFH. \citet{Johnson2021} used mock data based on several galaxies from the large Illustris cosmological hydrodynamic simulation of galaxy formation \citep{Vogelsberger2014, Diemer2017}. They found that the flexibility of a nonparametric SFH model more accurately recovered both the older and more recent SFH shapes of simulated SFHs from Illustris, in comparison to parametric models, which only recovered the recent SFH shape. Similarly, \citet{Lower2020} use model galaxies from the SIMBA cosmological galaxy formation simulations \citep{Dave2019} run with GIZMO’s meshless finite mass hydrodynamics \citep{Hopkins2015, Hopkins2017}. They also find that the nonparametric SFHs were able to match the simulated SFHs from SIMBA across time significantly better than the parametric models. For more discussion comparing parametric models to nonparametric models see \citet{Carnall2019}, \citet{Leja2019}, \citet{Lower2020}, and \citet{Johnson2021}. 

\begin{deluxetable*}{llllllll}[!htbp]
\tablecaption{Ages and $\mathrm{f_{burst}}$ \label{tab:age}}
\tablewidth{0pt}
\tablehead{
\colhead{\#} & \colhead{ID} & \colhead{log $\mathrm{M_*\ [M_{\odot}]}$} & \colhead{log $\mathrm{M_{H2}\ [M_{\odot}]}$} & \colhead{Age [Gyr]} & \colhead{$\mathrm{f_{burst}}$} & \colhead{$\mathrm{f_{burst}}$} & \colhead{$\mathrm{f_{burst}}$}\\
\colhead{} & 
\colhead{} & 
\colhead{} & 
\colhead{} & 
\colhead{} &
\colhead{Nonparametric fiducial} &
\colhead{Nonparametric bursty} & 
\colhead{Parametric} }
\startdata
1 & 74512 & $11.14^{+0.02}_{-0.02}$ & 9.82$\pm0.13$ & 4.7$\pm$0.4 & $0.02^{+0.01}_{-0.01}$ & $0.02^{+0.02}_{-0.01}$ & $0.03^{+0.01}_{-0.01}$\\
2 & 110509 & $11.35^{+0.02}_{-0.02}$ & 9.92$\pm0.07$ & 4.8$\pm$0.3 & $0.06^{+0.02}_{-0.02}$ & $0.07^{+0.03}_{-0.04}$ & $0.06^{+0.01}_{-0.01}$ \\
3 & 130284 & $11.17^{+0.02}_{-0.02}$ & 10.00$\pm0.06$ & 6.0$\pm$0.4 & $0.006^{+0.008}_{-0.003}$ & $0.003^{+0.003}_{-0.002}$ & $0.28^{+0.09}_{-0.17}$ \\
4 & 132776 & $11.15^{+0.02}_{-0.02}$ & 10.16$\pm0.11$ & 4.3$\pm$0.3 & $0.003^{+0.001}_{-0.001}$ & $0.003^{+0.001}_{-0.001}$ & $0.15^{+0.16}_{-0.09}$ \\
5 & 138718 & $11.35^{+0.05}_{-0.05}$ & $<9.84$ & 5.2$\pm$0.7 & ... & ... & $0.18^{+0.14}_{-0.13}$ \\
6 & 169076 & $11.52^{+0.01}_{-0.01}$ & $<9.91$ & 6.7$\pm$0.3 & ... & ... &  $0.19^{+0.15}_{-0.14}$\\
7 & 210210 & $11.29^{+0.02}_{-0.02}$ & $<9.84$ & 4.6$\pm$0.3 & ... & ... &  $0.20^{+0.13}_{-0.14}$ \\
8 & 211409 & $11.25^{+0.02}_{-0.02}$ & $<9.72$ & 5.3$\pm$0.4 & ... & ... & $0.16^{+0.15}_{-0.12}$ \\
\enddata
\tablecomments{Comparisons of ages and fraction of mass formed in secondary bursts of star formation.}
\end{deluxetable*}

\section{Results} \label{sec:results}

\subsection{Inferred SFHs}
In this section, we present the inferred SFHs for our sample of eight massive, quiescent galaxies. In Figure \ref{fig:seds_co} we compare the spectrophotometric data to the fitted \texttt{Prospector} model predictions, finding excellent agreement. We inspect the residuals for both the spectra and photometry and find that $\chi$, defined as (data-model)/$\sigma$, is clustered around 0 for both.

In figure \ref{fig:parametric_SFHs} we show the SFHs of individual galaxies inferred with both the parametric and fiducial nonparametric model. The top four panels show the SFHs of the CO-detected galaxies, while the bottom four panels show the CO-undetected galaxies. There is generally fair agreement in the overall shape for galaxies 5, 7, and 8. However, for the remaining galaxies, the parametric SFH does not agree well with the fiducial model. The fiducial models in these galaxies have sharp transitions in SFR(t) which tend to be poorly reproduced using simple parametric models.

Figure \ref{fig:nonparametric_SFHs} shows the two nonparametric models with differing priors compared with one another. The shaded regions show the 16th and 84th percentile range of the posterior probability. The results using the continuity prior are not significantly different from those using the bursty continuity prior. In addition to being consistent with one another, Figure \ref{fig:nonparametric_SFHs} shows that the overall qualitative shapes of the SFHs are the same for both priors. We adopt the nonparametric star formation history with the continuity prior as our fiducial model.

\subsection{Early time SFH}
To explore differences between the formation pathways of our sample of galaxies, we derive metrics from their fiducial nonparametric SFH models. In particular, we calculate the mass-weighted ages, formation redshifts, and star-formation timescales. The stellar mass-weighted ages ($\mathrm{t_{MW}}$), in units of lookback time, are defined as:

$$\mathrm{t_{MW}}=\frac{\int_{t_{\mathrm{obs}}}^{0} t\ SFR(t) \mathrm{d} t}{\int_{t_{\mathrm{obs}}}^{0} SFR(t) \mathrm{d} t} $$

where $\mathrm{t_{obs}}$ is the age of the Universe at the observed redshift. The formation redshifts, $\mathrm{z_{form}}$, were calculated by converting the mass-weighted ages from units of lookback time into units of cosmic time and finding the corresponding redshift. The star-formation timescale, $\mathrm{\tau_{SF}}$, is measured as the time between when the galaxy formed 20\% and 80\% of its total formed mass, in Gyrs \citep{Tacchella2022a}. 

\begin{figure*}[!htbp]
\includegraphics[trim={1cm 1cm 1cm 1cm}, clip, width=\textwidth]{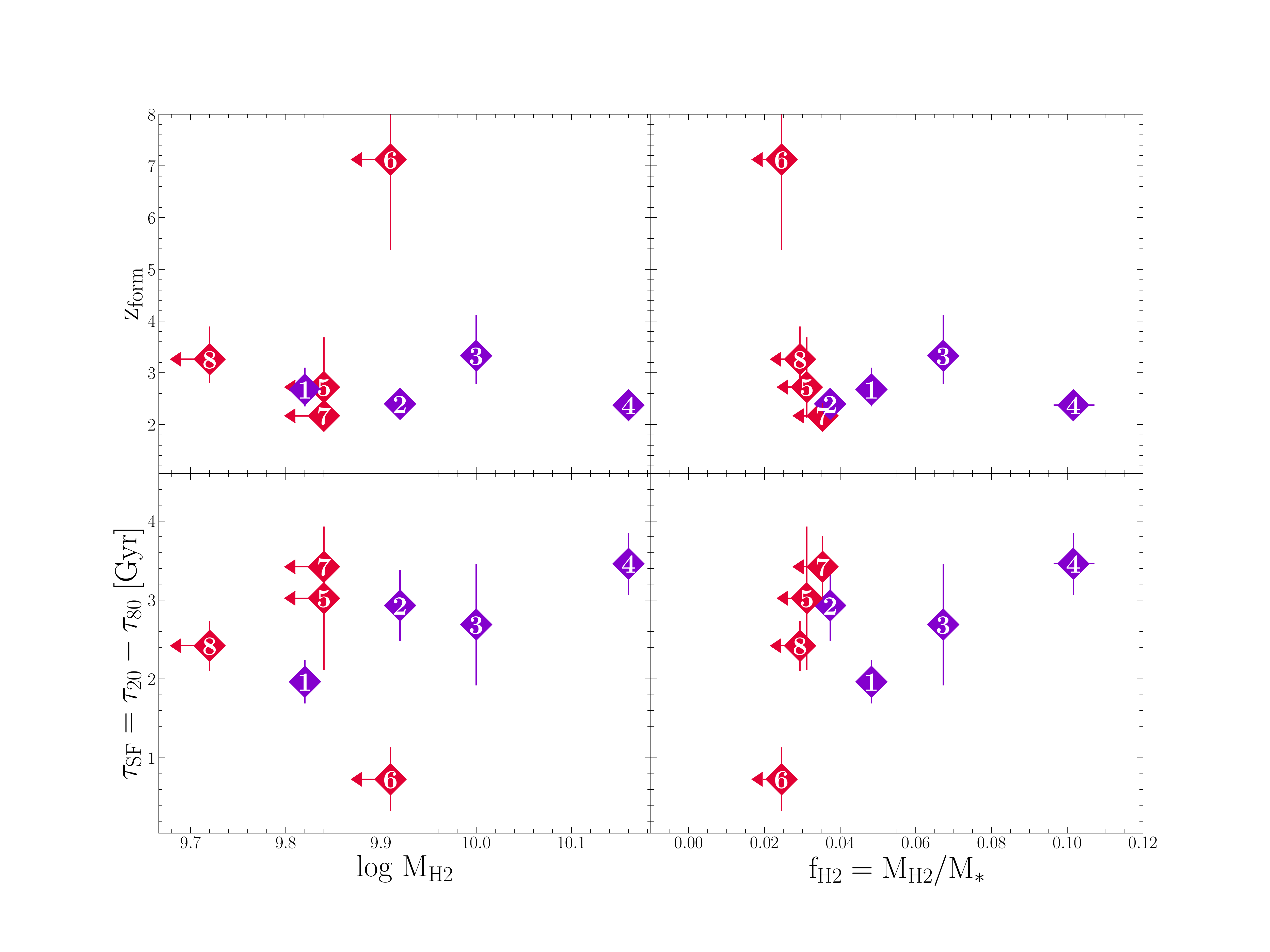}
\caption{The formation redshifts and star-forming timescales as a function of gas mass. There is not a significant difference in the formation redshifts or the star-forming timescales between the CO-undetected galaxies and the CO-detected galaxies.} \label{fig:metrics}
\end{figure*}

Figure \ref{fig:metrics} shows the measurements for the formation redshifts and star-formation timescales versus molecular gas mass. There is no significant difference in the formation redshifts or the star-formation timescales between the CO-undetected galaxies and the CO-detected galaxies, suggesting that the presence of molecular gas does not depend on either metric.

\begin{figure*}[!htbp]
\includegraphics[trim={.1cm .1cm .1cm .1cm}, clip, width=\textwidth]{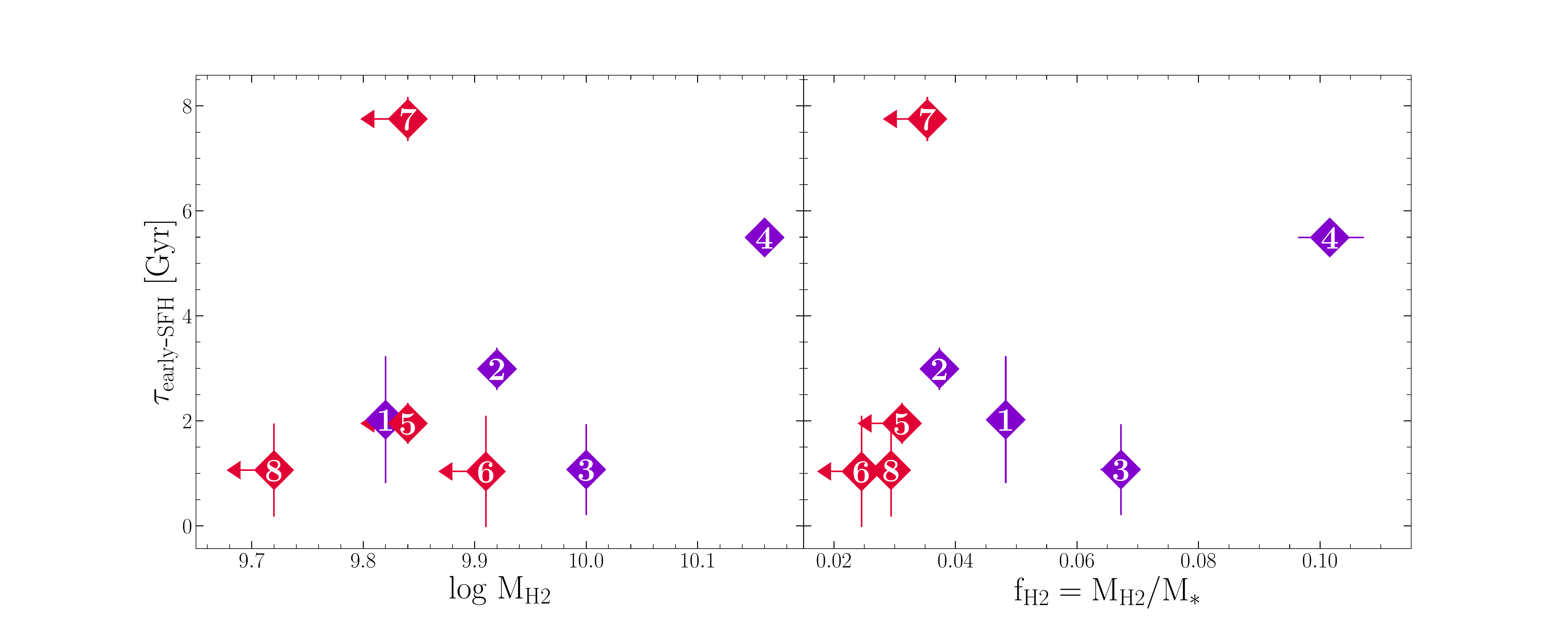}
\caption{Tau-model fits to the early fiducial nonparametric SFH. This measurement shows the rate of star-formation decline in the early SFH. There is no significant difference between the CO-detected and CO-undetected galaxies. Therefore, the detectable gas reservoirs in the CO-detected galaxies are not leftover from their primary epoch of star formation.} \label{fig:SFR_earlySFH_tau}
\end{figure*}

Recent studies have found evidence for both fast and slow paths to quiescence \citep{Carnall2018, Wu2018, Belli2019, Tacchella2022a, Akhshik2022}, suggesting multiple quenching pathways with different associated timescales. This leads to the idea that higher levels of molecular gas in quiescent galaxies could be leftover after a slower quenching timescale. To test if this was the case for our sample, we measured the timescale of decline for the early SFH, which corresponds to lookback times $\gtrsim 1$ Gyr. We fit an exponentially declining $\tau$-model, with functional form SFR(t)=$A\ exp(-t/\tau)$, to the center of the first 4 agebins ($\gtrsim1$ Gyr in lookback time) in the early nonparametric SFH. A high (low) $\tau$ value would indicate a slow (fast) quenching timescale. Six of the eight galaxies have values in the range $\tau=1.0-3.0$ Gyr, while galaxies 4 and 7 have higher values of 5.5 Gyr and 7.8 Gyr respectively, with typical average uncertainties of 0.7 Gyr. There is no correlation between gas mass and $\tau$, as shown in Figure \ref{fig:SFR_earlySFH_tau} and this metric shows no significant difference between the CO-detected and CO-undetected galaxies. Therefore, our data do not show evidence that the gas content is leftover after differing quenching timescales in the early SFHs of the galaxies.

\subsection{Comparison of SFR indicators}

Because the molecular gas content is unrelated to the rate of star formation decline in the early SFHs for our sample of galaxies, we look for differences in their late-time star formation activity. Establishing the level of ongoing star formation in quiescent galaxies is difficult because SFR measurements in quiescent galaxies are known to be uncertain \citep[e.g.][]{Leja2019, Belli2021}. This is particularly true of infrared based SFR, since mid- and far-infrared flux does not always trace ongoing star formation in older galaxies, but instead can come from sources such as AGN, asymptotic giant branch (AGB) stars, and dust heated by older stars \citep{Salim2009, Fumagalli2014, Utomo2014, Hayward2014}. This makes the most common gold standard SFR indicator at higher redshifts, based on UV+IR measurements, unreliable for quiescent galaxies. For this reason, $\mathrm{SFR_{UV+IR}}$ values in particular should be considered an upper limit for quiescent galaxies.

To confirm that our late-time SFH (0-1 Gyr in lookback time) measurements from the full SED-fitting are robust, we present independent SFR indicators based on data that are not used in our SED-fitting. This includes SFRs based on flux from UV+IR, [OII], and radio measurements. $\mathrm{SFR_{UV+IR}}$ is based on a weighted sum of UV and IR (24$\mu$m) fluxes, published in \citet{Spilker2018}. To measure $\mathrm{SFR_{[OII]}}$, we first correct the [OII] flux for dust attenuation using the results from our \texttt{Prospector} fits. We use the dust-corrected fluxes to derive $\mathrm{SFR_{[OII]}}$ with the conversion factor in \citet{Kewley2004}, assuming that the attenuation of gas is the same as the stars. 
Table \ref{tab:sfrs} shows the estimated SFRs from the dust-corrected [OII] flux, assuming that the emission is entirely from star formation \citep{Kewley2004}. This [OII] emission is weak (for galaxies at these redshifts), corresponding to average $<5 M_{\odot}\ yr^{-1}$. 

To calculate $\mathrm{SFR_{radio}}$, we use the 3 GHz flux measurements from \citet{Smolcic2017} and \textit{K}-correct to 1.4 GHz assuming the convention $S_{\nu}\propto \nu^{\alpha}$ with $\alpha=-0.7$ \citep{Condon1992}. We then use the luminosity to derive $\mathrm{SFR_{radio}}$ with the conversion factor from \citet{Murphy2011}. 

We compute $\mathrm{SFR_{radio}}$ for all the galaxies in the sample. We find that galaxies 2 and 6 have high $\mathrm{SFR_{radio}}$ values, in agreement with the results of \citet{Barisic2017} indicating that the radio flux is likely dominated by emission from a radio-loud AGN. This classification was determined using the radio luminosity limit from \citet{Best2005}. We find that they do not show any strong emission lines that would be characteristic of AGN in our spectra. We check if the AGN are obscured by dust in the center, but the IRAC colors for these two galaxies are not near the region populated by AGNs \citep{Kirkpatrick2013}, and are instead near the region for quiescent galaxies, meaning they are not highly dust-obscured AGN. The results from our \texttt{Prospector} fits infer low AGN luminosity fractions with values of $\mathrm{f_{AGN}=0.02^{+0.12}_{-0.02}\%}$ and $\mathrm{f_{AGN}=0.03^{+0.29}_{-0.03}\%}$ for galaxies 2 and 6, respectively. We confirm visually that the optical detections are not offset from their radio detections, meaning they are not contaminated by a nearby companion.

We calculate the radio spectral index for these two galaxies identified as radio loud AGN by \citep{Barisic2017} using their 1.4 GHz flux measurements from \citet{Schinnerer2010} and 3 GHz flux measurements from \citet{Smolcic2017}. The spectral index for galaxy 2 is $-0.1\pm0.1$ and for galaxy 6 is $-0.4\pm0.2$, both of which are consistent with a flat-spectrum AGN \citep{deGasperin2018}. The high $\mathrm{SFR_{radio}}$ values for these galaxies are likely due to added flux from an AGN and the radio flux is not only tracing star formation. We conclude that our measured current SFR (averaged over the past 100 Myr, also presented in Table \ref{tab:sfrs}) we derive from the SED fitting is consistent with the radio measurements, as the high SFR$_{radio}$ is likely contaminated by flux from the AGN. 

Five of the eight galaxies do not have significant 1.4 or 3 GHz emission. Therefore we present the 3$\sigma$ upper limit to $\mathrm{SFR_{radio}}$, 14 $M_{\odot}\ yr^{-1}$, based on the sensitivity of radio imaging of the COSMOS field from \citet{Smolcic2017} (assuming the canonical spectral index for radio emission from star formation). The SFR measurements from the SED fitting are well below the upper limits from the radio imaging, confirming their low SFRs measured by our SED fitting. The remaining galaxy 7 has an $\mathrm{SFR_{radio}}$ value that is higher than the SFR determined from the SED fitting. 

Table \ref{tab:sfrs} shows a comparison of the SFR measurements using all four different techniques for the galaxies in our sample, and we find that there are systematic offsets between them (see also \citealt{Belli2021}). As expected for the reasons discussed above, the $\mathrm{SFR_{UV+IR}}$ values are overestimated, by about a factor of 4, compared to our SED-fitting results ($\mathrm{SFR_{100Myr}}$). For galaxies 4 and 5, the $\mathrm{SFR_{UV+IR}}$ values are 6 times larger than the $\mathrm{SFR_{[OII]}}$ values. Galaxy 1 is an exception, as it has a larger SFR measured from the SED fitting and the [OII] than from the UV+IR. The SFRs determined from the SED-fitting and [OII] are consistent for galaxy 5. However, for galaxy 1 and 4 the SFR from dust-corrected [OII] is smaller than the measurement from the SED-fitting, consistent with similar measurements in \citet{Belli2021}. They discuss possible causes for this disagreement and conclude that it could be due to an inadequate dust correction because of the inability to measure nebular dust attenuation by fitting stellar emission. We therefore determine that the most robust SFR measurements for our sample of quiescent galaxies are based on full SED-fitting, and we adopt these SFRs for the remainder of this work. SED-based SFRs are less likely to be contaminated by other sources in quiescent galaxies and are in line with accumulating evidence that SED-fitting produces the least biased of SFR indicators for quiescent galaxies \citep{Fumagalli2014, Belli2017}, see also Section 6.2.1 of \citet{Leja2022}.

\subsection{Late time SFH}

We look for differences between the late-time SFHs of CO-detected and CO-undetected galaxies. To determine if a galaxy’s SFH shows evidence of a secondary burst, we calculate the difference between the peak in the late-time SFH (most recent 1 Gyr) and the previous minimum in SFR. We consider the galaxy to have a secondary burst if the increase in the SFR at its peak is larger than the previous minimum by a factor of 2.0 or more, and if the resulting difference in SFR between peak and previous minimum is significant at $\mathrm{1\sigma}$ given the uncertainties in the SFRs. Using this criteria, the CO-detected galaxies have possible evidence for secondary bursts of star formation in their late SFH (0-1 Gyr in lookback time), while the CO-undetected galaxies do not. While CO-detected galaxy 3 demonstrates a secondary burst using both priors, the uncertainties are large, and therefore a flat SFH would also be consistent with our measurements. Therefore, for galaxy 3, higher S/N is needed to determine if there is a significant secondary burst of star formation. Galaxies 1, 2 and 4 have statistically significant evidence for secondary bursts of star formation.

We therefore explore whether the evidence for the secondary burst in galaxies 1, 2 and 4 exists under different assumptions, namely our assumed parametric form for the SFH of delayed-tau+burst. The bursts for the parametric and nonparametric models agree quite well for galaxies 1 and 2. The parametric delayed-tau+burst model shows an instantaneous burst at a lookback time of $\lesssim 1$ Gyr that is consistent with the burst found in the nonparametric model for these two galaxies. The fraction of mass formed in these bursts is 2-3\% for galaxy 1 and 6-7\% for galaxy 2, consistent between all 3 models. However, for galaxy 4, the parametric model did not identify a secondary burst with consistent time and mass fraction as the nonparametric model. The fraction of mass formed in this burst is relatively small, only 0.3\%, which could explain why it was not identified by the parametric model. We note that the instantaneous burst time measured by the parametric model for this galaxy is early in its history, aligning with the peak epoch of the SFH (see black lines in Figure \ref{fig:parametric_SFHs}), and forms a relatively minor amount of mass. 

Comparisons of the fraction of mass formed in the secondary bursts, $\mathrm{f_{burst}}$, are shown in Table \ref{tab:age} for all 3 models. To determine the fraction of mass formed in the bursts for the nonparametric models, $\mathrm{f_{burst}}$, we define a burst as an increase in SF. The bin prior to the increase in SF is considered the baseline. The mass formed is then calculated using all the bins following the baseline that have higher SFRs than the baseline.

\begin{figure*}[!htbp]
\includegraphics[trim={2cm 2cm 2cm 2cm}, clip, width=\textwidth]{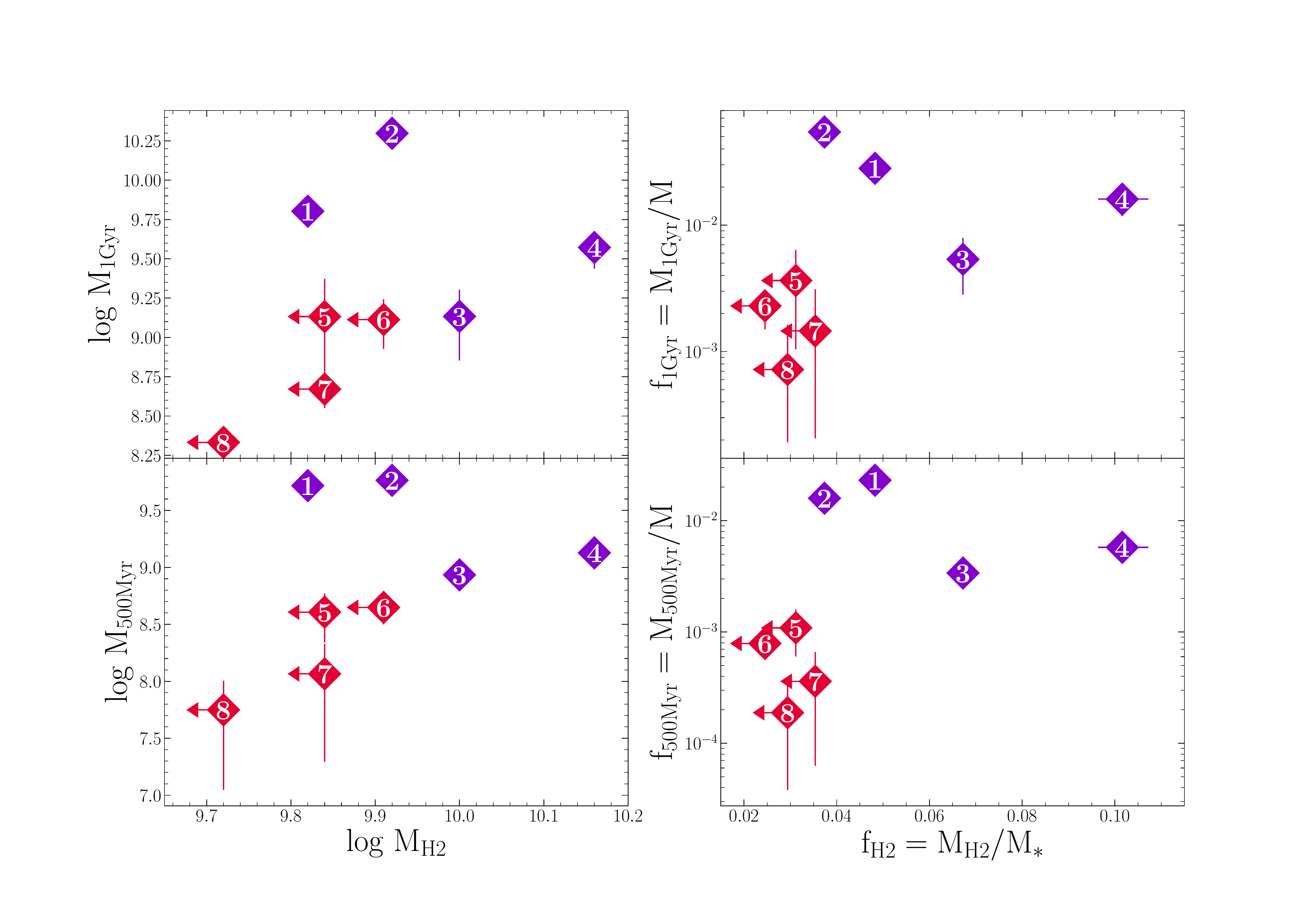}
\caption{$\mathrm{M_{1Gyr}}$ ($\mathrm{M_{500Myr}}$) is the total formed mass in the last Gyr (500 Myr) for each galaxy. The CO-detected galaxies are shown in purple while the CO-undetected galaxies are shown in red. The bootstrapped uncertainties are shown as errorbars. More mass was formed in the last Gyr for galaxies with higher gas masses. This is further evidence that the detectable gas reservoirs are due to an event in their last Gyr, rather than their early SFH.} \label{fig:Gyr_mass}
\end{figure*}

Given that we see evidence in a few, and some tentative in others, for a late time burst, we explore whether this late time SFH is a plausible explanation for higher gas content. Figure \ref{fig:Gyr_mass} shows the total mass formed in the last 1 Gyr (and last 500 Myr) for each galaxy. We find that CO-detected galaxies formed a higher fraction of their mass in the last Gyr than CO-undetected galaxies. The average fraction of mass formed in the last Gyr and the standard deviation for CO-detected galaxies is $\mathrm{2.6\pm1.8\%}$ and for CO-undetected galaxies is $\mathrm{0.2\pm0.1\%}$. 

To summarize, our data are consistent with the idea that detectable gas reservoirs in quiescent galaxies are linked to late time star formation activity, and we do not see evidence that it is leftover from their primary star formation epoch. We discuss possible causes for this finding in the next section.

\section{Discussion} \label{sec:discussion}
The main finding in our study is that quiescent galaxies with detectable gas reservoirs have enhanced late time star formation compared to those without significant gas reservoirs. All four of the CO-detected galaxies have plausible evidence for a secondary burst, although in one case the uncertainties are high enough that we cannot reject a flat level of SF in the last 1 Gyr. In contrast to this, the CO-undetected galaxies have declining SFRs in their most recent 1 Gyr. In addition, the CO-detected galaxies formed a higher fraction of mass in their last Gyr ($f_{M_{1Gyr}}=2.6\pm1.8\%$) than the CO-undetected galaxies ($f_{M_{1Gyr}}=0.2\pm.1\%$). Here, we explore what physical mechanisms could be influencing the late time star formation in galaxies with higher molecular gas content and discuss how this may be a contributing factor to the heterogeneity of molecular gas content observed in quiescent galaxies.

\subsection{Relative timescales of gas depletion and star formation decline}

The relative timescale over which cold molecular gas is depleted, and over which the star formation declines, is a powerful diagnostic of the quenching mechanisms at play. In principal, the CO-detected galaxies in our sample are quenching over a longer timescale than the CO-undetected galaxies (emulating a slow and fast timescale for quenching). However, we find that this is due to a late-time boost in star formation, possibly influenced by external sources such as recently accreted gas, and not necessarily because they are undergoing a different, slower quenching mechanism \citep[e.g.][]{Belli2019}. With a limited sample such as ours, it is difficult to draw broad conclusions, so in this section we compare to other available timescale constraints in the literature for coeval (at similar redshifts) non-starforming galaxies.

We compare our results to the most comprehensive coeval sample, which are post-starburst (PSB) galaxies from the SQUIGGLE survey. We note that none of the galaxies in our sample fall in the region of the Dn4000 vs. $\mathrm{H\delta}$ plane occupied by the SQUIGGLE sample, or by PSBs in general. The $\mathrm{H\delta}$ measurements alone exclude our sample from being classified as PSB galaxies by traditional selections, which require $\mathrm{H\delta>3-5\AA}$, while the galaxies in our sample have values of $\mathrm{H\delta<3\AA}$.  Our sample also have higher Dn4000 than traditional PSBs, with Dn4000 in the range 1.5-1.9\AA, while the PSB galaxies in SQUIGGLE have typical ranges from 1.2-1.5\AA. SQUIGGLE galaxies also have ceased star formation, while exhibiting heterogeneous molecular gas reservoirs \citep{Suess2017, Hunt2018, Suess2022, Bezanson2022}. The SQUIGGLE PSB selection targets strong Balmer breaks and blue slopes redward of the break using rest-frame filters, following \citet{Kriek2010}. The SQUIGGLE sample is selected at $\mathrm{z\approx0.6}$, only slightly lower redshift than this sample. However, their primary burst of SF occurred shortly before observation time, meaning their sample is made up of more-recently quenched galaxies than our sample. Thus, SQUIGGLE PSBs are qualitatively different in that they are ending their primary burst of star formation, while our quiescent galaxies are ending a small secondary burst, having completed their primary growth epoch earlier in cosmic time. Nonetheless, both samples have ceased star formation while retaining signficant gas reservoirs, and together they increase the baseline in galaxy properties for comparing gas and star formation timescales in quenched populations exhibiting heterogeneous gas content. 

\citet{Bezanson2022} find that 6 of the 13 SQUIGGLE galaxies were CO-detected.
Before comparing to our results, we first must convert their molecular gas masses using the same conversion factors assumed in \citet{Spilker2018} with the standard rescaling $M_{H2} \times(0.8/r_{21})(\alpha_{CO}/4.4)$. \citet{Bezanson2022} assume $r_{21}=1.0$ and $\alpha_{CO}=4.0$. We next check for differences in the survey sensitivities. The range of the upper limits on gas masses for the CO-undetected galaxies in \citet{Bezanson2022} using our conversion factors is log $\mathrm{M_{H2}=9.1-9.4 M_{\odot}}$, somewhat deeper than the limits of the \citet{Spilker2018} data with log $\mathrm{M_{H2}=9.7-9.9 M_{\odot}}$. For both studies, the upper limits were determined using the noise in the 800 km/s wide channel at the 3$\mathrm{\sigma}$ level. Therefore, we can safely make this comparison since our entire sample of CO-detected sources from \citep[][]{Spilker2018} would be included as detections by the deeper \citet{Bezanson2022} data.

The main finding in \citet{Bezanson2022} is that the galaxies that quenched $\leq$150 Myr prior to observation have significant gas reservoirs, with an average gas fraction of $f_{H2}\sim 7\%$ (or $f_{H2}\sim 5\%$ using our conversion factors), while older post-starburst galaxies do not. Assuming that these individual galaxies are in different phases of an evolutionary trend, this implies that quenching precedes the disappearance of cold gas in their sample, but that molecular gas must be consumed or destroyed shortly thereafter.

To compare to our sample, we calculate the time between the end of the secondary burst and the time of observation ($t_{end\mbox{-}burst}$). We find for galaxies 1, 3, and 4 that $t_{end\mbox{-}burst}\approx 0-100$Myr, which is consistent with the benchmark suggested by \citet{Bezanson2022} for hosting residual gas. Galaxy 2 (which formed the highest fraction of mass in its secondary burst of star formation among our sample) has a secondary burst $t_{end\mbox{-}burst}\approx$330 Myr prior to observation, despite exceeding the timescale identified by \citet{Bezanson2022} for gas to disappear in their sample of CO-detected galaxies. Based on the overall total fraction of mass formed during the course of the recent bursts (75\% of SQUIGGLE galaxies have burst mass fractions $>$25\% compared to only 0.3-6\% for our sample), the gas content that drove the recent low-level star formation in our sample is likely much smaller than the total amount of gas that drove the large, primary bursts of star formation in the PSBs. It is thus plausible that the gas consumption may not proceed on similar timescales given the dramatically different scales of the growth, perhaps resulting in slower gas consumption than the clearly rapid decline in PSBs.

There are several physical reasons that would explain a slower consumption at low gas content. Firstly, it is possible that small amounts of gas can be dynamically stabilized to prevent star formation without any gas destruction or consumption \citep{Gensior2020}. This simulation predicts that dynamical stabilization of molecular gas can suppress star formation by as much as a factor of 5 in bulge-dominated systems with relatively low gas fraction ($\lesssim$ 5\%) without destroying or consuming gas. In fact, galaxy 2 that exhibits a relatively long timescale since burst is bulge-dominated (see Figure \ref{fig:seds_co}) and has a low enough gas fraction to be in line with the conditions of dynamical stabilization ($f_{gas} = 4\pm0.2\%$). While this may also explain the other 3 CO-detected galaxies that had more recent burst times, this physical reason does not appear to be a universal threshold among quiescent galaxies, given the relatively low $\mathrm{f_{gas}}$ limits among our non-detected sample, and among others across redshifts \citep[e.g.][]{Davis2016,Saintonge2017,Williams2021, Whitaker2021a}. An alternative explanation for unconsumed gas in quiescent galaxies is a decrease in star formation efficiency due to a lower fraction of dense molecular gas compared to the total molecular gas  \citep[e.g.][]{Davis2019, Li2019, French2022}. Unfortunately, our data do not provide insight into the density of molecular gas.

While gas depletion times can in principal measure any difference in gas decline directly, they are difficult to interpret among galaxies below the main sequence because the value depends strongly on which indicator is used to measure the SFR, which can vary by as much as a factor of 6 \citep[see Section \ref{sec:SEDfitting};][]{Belli2021}. We note that both the \citet{Bezanson2022} and \citet{Spilker2018} samples exhibit very long depletion times for gas consumption by the residual star formation ($>2$ Gyr).\footnote{The SFR used to measure the depletion time in \citet{Bezanson2022} comes from SED fitting with a nonparametric star formation history, integrated over a recent time bin that varies by galaxy, but is typically between 100-200 Myr, comparable to our SFR$_{100Myr}$ (K. Suess, personal communication.)} In the case of the PSB sample, such long depletion times are actually inconsistent with the inferred rapid $<150$Myr timescale for gas disappearance. However, the measure of depletion time implicitly assumes that gas is only destroyed by simple consumption by star formation and neglects processes such as heating or removing gas via AGN feedback. It is possible that in such cases that gas is instead actively destroyed or heated by AGN feedback rather than simply consumed by star formation in PSBs \citep[e.g. the PSB sample studied in][]{French2018a}. \citealt{Bezanson2022} also note the possibility that some CO-detected PSBs could appear gas rich because they are caught in a phase immediately prior to rejuvenating star formation (e.g. in a secondary burst such as what we observe). We discuss the possible contribution of rejuvenation to the heterogeneity observed among non-starforming galaxies in the next section.

\subsection{Rejuvenation}
50\% of our sample of 8 quiescent galaxies show plausible evidence for secondary bursts of star formation along with significant molecular gas content. This is an interesting result, in light of the prevailing theoretical picture that quiescent galaxies form early in cosmic time, and lack appreciable new star formation for 10 Gyr since quenching \citep[e.g.][]{Renzini2006, Citro2016}. While this implies rejuvenation is not uncommon around $z\sim0.8$, the observed frequency of rejuvenation in quiescent galaxies remains extremely uncertain, owing to the difficulty of characterizing low-level star formation in quiescent galaxies across cosmic time (see Section \ref{sec:SEDfitting}), the short timescale over which such ``frosting" might be observable, and the difficulty of measuring detailed SFH outside of the nearby Universe.

The majority of studies characterizing the prevalence of rejuvenated star formation in quiescent galaxies are from the local universe. By $z<0.1$ the rejuvenation fraction appears relatively common, with higher rates among quiescent galaxies at lower masses (10$\mathrm{< log\ M_*/M_{\odot}<}$11)  \citep[10-30\%][]{Treu2005,Schawinski2007, Thomas2010, Donas2007, Salim2010, Fang2012,Salvador-Rusinol2020, Werle2020, Jeong2022}, generally in agreement with theoretical predictions \citep[][]{Trayford2016,Pandya2017, Behroozi2019}. Recent advances in higher redshift datasets suggest that the rate that high mass quiescent galaxies such as our sample (log $\mathrm{M_* > 11 M_{\odot}}$) undergo rejuvenation events at earlier times are by comparison quite low \citep[$<$1-16\%, depending on the threshold of star formation used to define a rejuvenation;][]{ Belli2017, Chauke2019, Tacchella2022}. The lower rejuvenation fraction perhaps reflects the shorter available cosmic time, and therefore lower probability, to both quench and experience processes driving rejuvenation such as minor merging. Nonetheless, examples at even higher redshifts $z>1$ are now routinely identified thanks to advances in analyses such as the ones we use here \citep[][]{Jorgensen2014, Williams2021, Akhshik2021, Caliendo2021, Belli2021}.

Generally the criteria used to define rejuvenation identify relatively large secondary bursts: they typically must increase the star formation in quiescent galaxies to within range of the main sequence scatter, with a large range of typical burst masses that can form between $\sim2-25$\% of the galaxy's mass \citep[][]{Chauke2019, Tacchella2022}. We note here that our secondary bursts are small compared to published metrics of rejuvenation in coeval galaxies, with typical mass fractions formed $\mathrm{f_{burst}\sim2\%}$ with a range $0.3-6\%$ (see Table \ref{tab:age}). Based on our inferred SFHs, one of our CO-detected sources (galaxy 1) meet the stringent criteria for identifying rejuvenations at $z\sim0.8$ \citep[e.g.][although we note that these authors may find a low rejuvenating fraction because they specifically select for rejuvenation that is ongoing at the time of observation]{Tacchella2022}. If we exclude the requirement in \citet{Tacchella2022} that the rejuvenation be ongoing at time of observation, galaxy 4 would be identified using the \citet[][]{Tacchella2022} criteria. Our secondary bursts are also more minor than the fiducial rejuvenation measured by \citealt{Chauke2019}, half of which form stellar mass fractions $f_{burst}\gtrsim$8\% in rejuvenation. \citealt{Chauke2019} identify rejuvenations as quiescent galaxies with a prior quiescent phase, with a return to the star forming sequence in between. The star forming and quiescent criteria are based on a sSFR threshold compared to the main sequence ($>$0.3 dex below), that evolves with redshift as defined by \citet[][]{Speagle2014}. Two galaxies (galaxies 1 and 2) meet their criteria using our new SFH measurements (roughly 25\% of our sample, slightly higher than their estimate of 16\% for the fraction of quiescent galaxies with a rejuvenation). These two galaxies also meet their more-relaxed criteria for a rejuvenation that star formation increase above a fixed limit of log sSFR$<-1$ Gyr$^{-1}$ after quiescence, resulting in a consistent fraction of rejuvenation (\citealt{Chauke2019} find 24$\pm$2\%).

We thus find that these lower-level rejuvenation events (that we see in our SFH analysis but are missed by previous rejuvenation criteria) may be more common than stringent criteria imply. This may be in agreement with findings of \citealt{Carnall2018} that up to 60\% of quiescent galaxies may have experienced low level rejuvenation or a merger since $z<1.5$ \citep[see also][]{Mancini2019, Carleton2020}. Some evidence from theory and cosmological simulations imply that a similar fraction of massive galaxies undergo weak rejuvenations (10-40\% of massive halos with $\mathrm{log\ M_{halo} > 13}$), supporting the idea that weak rejuvenations are relatively common \citep{Alarcon2022}.  While the rejuvenations we measured are relatively small, they nonetheless contribute to the surprisingly large number of detectable reservoirs of molecular gas. We explore whether the low level secondary bursts we measure are consistent with common quiescent galaxy evolutionary processes such as minor merging in the next section.

\subsection{Minor Mergers and size evolution}\label{sec:minor_mergers}

While the prevalence of star formation rejuvenation in high-redshift quiescent galaxies still remains uncertain, it is well known that minor merging must be a significant process in their late time evolution after quenching their primary growth phase. At fixed stellar mass, high redshift quiescent galaxies are 3-5 times smaller than in the local universe \citep{Daddi2005, Trujillo2006, Buitrago2008, Cimatti2008, vanDokkum2008}, implying that they have increased significantly in size but not mass over the past 10 Gyr. While gaseous mergers can dissipate energy to collapse to form stars centrally in galaxies (serving to decrease size and increase mass) dry minor merging can match observations by increasing size without adding appreciable mass. Empirical studies find that minor merging is likely the dominant mechanism for size growth over $\mathrm{0<z<1.5}$ \citep{vanDokkum2010, Belli2014, Hamadouche2022} and can account for most or all of the size growth at $\mathrm{z\lesssim1}$ \citep{Newman2012}. Therefore, due to its ability to explain these empirical trends, and the direct evidence that the process occurs \citep[e.g. the presence of tidal debris composed of old stars, and prevalence of quiescent satellites around quiescent massive galaxies][]{vanDokkum2010, Ji2018}, dry minor merging has become the prevailing theoretical picture influencing late time quiescent galaxy evolution \citep[][]{Bezanson2009, Hopkins2009, Naab2009, vandeSande2013, vanDokkum2010, Oser2012}. 

The CO-detected galaxies in our sample tend to have lower stellar densities than the CO-undetected galaxies, which is consistent with having experienced minor merging \citep{Spilker2018}. Given that our data also indicate plausible evidence for accreted gas, perhaps delivered via minor mergers, we explore whether the level of gas measured in the CO-detected sample is consistent with the expectations of dry minor merging. 
Minor mergers are typically defined as mergers between two galaxies with a mass ratio in the range of $10-4:1$ \citep{Kaviraj2009, Kaviraj2012}, and are considered dry if the average gas fraction of both galaxies is gas-poor. To check this possibility, we estimate the limiting quantity of molecular gas allowed in a dry minor merger, assuming our CO-detected galaxies are the larger of the merging pair. For a dry merger, we define a gas-poor galaxy using a very conservative assumption of $\mathrm{f_{gas}<20}$\% (relatively high gas content) following \citet[][see also \citealt{Kaviraj2012}]{Man2012}. For the minor galaxy, assuming a 10:1 mass ratio, this would correspond to the minor galaxy bringing in gas masses of log $\mathrm{M_{gas}\approx 9.4-9.7 M_{\odot}}$ (assuming the range of stellar masses of the CO-detected galaxies as the larger of the merger pair). Assuming a 4:1 ratio predicts a range of log $\mathrm{M_{gas}\approx 9.8-10.1 M_{\odot}}$. Comparatively, the range in measured gas masses in our CO-detected sample is log $\mathrm{M_{gas}=9.8-10.2 M_{\odot}}$.

Thus, we find that a dry minor merger with a 10:1 merger ratio would bring in a lower mass of molecular gas than what is measured for all 4 of our CO-detected galaxies, indicating that our galaxies are inconsistent with the expectations of a dry minor merger with this mass ratio. Assuming a less restrictive 4:1 minor merger ratio, we calculate that a dry minor merger would still be inconsistent with our large measured gas reservoirs in galaxies 3 and 4. However, for galaxies 1 and 2, the limiting gas content calculated for a less restrictive 4:1 dry minor merger ratio are consistent with the gas content that we observe. (We note that since the secondary bursts are nearing their completion at the time of observation, the gas content brought in by the companion galaxy would have to be larger than these estimates). Therefore, we conclude that the detectable gas reservoirs in our CO-detected galaxies are likely too large to result from typical dry minor mergers, favoring instead gas-rich minor mergers as the source of the gas and secondary burst. 

While the accreted gas is plausibly from gas-rich minor mergers, \citet{Spilker2018} argue against the idea of an outside gas source such as gas-rich minor mergers because the rotational axes of the stellar and molecular components of 3 of the 4 CO-detected galaxies are consistent with one another. Several studies have suggested that galaxy mergers would cause star-gas misalignments \citep{Bureau2006, Crocker2009, Martel2020, Khim2021}. \citet{Khim2021} study galaxies from the Horizon-AGN simulation, which is a large-volume cosmological hydrodynamical simulation \citep{Dubois2014} run with the adaptive mesh refinement code RAMSES \citep{Teyssier2002}. They find that major mergers are 2.5$\times$ more effective in causing star-gas misalignments than minor mergers (with the criterion for misalignments being $>$30\textdegree). While the simulations show that the average lifetime of gas misalignment in quiescent galaxies is long ($\sim$2 Gyr), the scatter in lifetimes is quite large, indicating that misalignments indicative of past mergers can decay even on a few hundred Myr timescale \citep[][]{Khim2021}. This is similar to the decline timescale we observe in our secondary bursts. We conclude that a lack of star-gas misalignment does not necessarily rule out the possibility that enhanced late time star formation in our sample could be due to minor mergers that occurred more than a few hundred Myr prior.

Even though we find that any minor merging in our sample is unlikely dry, it is possible they can still contribute to size evolution because the gas mass accreted in the merger is still relatively small. \citet{Sonnenfeld2014} show that purely dry mergers cannot be the only mechanism driving the size and density evolution of massive early type galaxies. They find that dry minor mergers predict a strong decrease in the mass density profile slope, inconsistent with the nearly constant slope inferred from observations over $\mathrm{0<z<1}$. To match the observed evolution in both the size and the mass density profile slope, some dissipation is required and their models prefer a gas fraction of $\mathrm{f_{gas} \approx 8\%}$ for the minor merging companion, higher than what is measured for 3 of the 4 CO-detected galaxies. While we do not have the data to constrain whether the gas at the time of accretion met this criteria, we can use it as an upper limit on the gas fraction that would be observed after the minor merger, since some amount of accreted gas is consumed as the merger completes. \citet{Sonnenfeld2014} indicate that a gas fraction of 8\% in the companion on average results in a 4\% increase in mass of new stars formed. This is within the range of $\mathrm{f_{burst}}$ values that we measure among our CO-detected sample, $\mathrm{f_{burst}=0.3-6\%}$. As such, to the best of the ability of the data, we find that our data are plausibly in line with the theory. Both \citet[][]{Mancini2019} and \citet{Ji2022} find evidence that higher-mass ($\mathrm{log\ M_*>11\ M_{\odot}}$) quiescent galaxies quench and then grow due to merging with satellite star-forming (gas-rich) galaxies based on more extended light profiles in the I814 (bluer) band than in the J125 and H160 (redder) bands. In addition, \citet{Belli2017} find low-level star formation activity based on weak H$\alpha$ emission in nine quiescent galaxies that is likely to be fueled by inflowing gas or gas-rich minor mergers. \citet{Rutkowski2014} find recent bursts of star formation that they conclude could be due to gas-rich minor merger activity, as well. 
Further, late-time star formation from gas-rich minor mergers increases SFR during these secondary bursts, which could slightly decrease the average light-weighted age while also increasing the size. This would be consistent with the size-age trend found in \citet{Williams2017} where younger galaxies are larger, see also \citet{Fagioli2016}. This is often attributed to progenitor bias. With our small sample, it is impossible to draw conclusions about the broader massive galaxy population, but we speculate that part of the size-age trend attributed to progenitor bias could be due to gas-rich minor mergers.

Finally, widespread evidence from quiescent galaxies in the local Universe indicates that minor merging is a significant resource for accreted cold gas, fueling very low levels of star-formation in otherwise inactive systems \citep[][]{Kaviraj2009, Davis2011, Kaviraj2011,Kaviraj2012,Kaviraj2013,vandeVoort2018, DavisYoung2019}. Our data provide evidence that this process also impacts massive galaxies at earlier times, in line with previous findings that gas rich minor merging probably occurred in this population consistently over the past 8 Gyr, and may even be the dominant process for new star formation at $\mathrm{z<1}$ \citep[e.g.][]{Kaviraj2008, Vulcani2016}.  The extent to which gas rich minor mergers impact the quiescent galaxy population at $\mathrm{z>1}$ remains unknown, although possible examples do exist \citep{Williams2021,Akhshik2021, Caliendo2021, Belli2021} and warrants further study.  We posit that gas-rich minor mergers could thus be a contributing factor to the heterogeneity of molecular gas reservoirs observed in quiescent galaxies, a process that increases with frequency with cosmic time since $\mathrm{z<1}$.

\section{Conclusions}\label{sec:conclusions}
We investigate the nonparametric SFHs of 8 quiescent galaxies with existing cold molecular gas mass measurements from \citet{Spilker2018}. Four of the eight galaxies contain signficant gas reservoirs. Our main results are as follows:
\begin{itemize}

\item We find that the molecular gas content is unrelated to the rate of star formation decline in the early SFH. This suggests that the gas reservoirs are not leftover from their primary SF epoch.

\item Quiescent galaxies with detectable gas reservoirs have enhanced late time star formation. We suggest the gas may have accreted via gas-rich minor mergers within the last 1 Gyr based on evidence for a recent low-level rejuvenation ($\mathrm{f_{burst}\approx0.3-6\%}$) in their SFHs. This is also supported by our finding that the galaxies with higher gas masses have formed more mass in the last Gyr of their SFH. Gas-rich minor mergers therefore may contribute to the diversity of molecular gas observed in high-redshift quiescent galaxies.

\end{itemize}

While the current sample is small, increasing the sample size of galaxies with both detailed SFH measurements and cold ISM constraints will increase our understanding of the quenching process. Studying the SFHs of galaxies in upcoming deep photometric and spectroscopic surveys with the James Webb Space Telescope will provide an excellent sample to make follow-up molecular gas observations with ALMA. Finally, we note that although CO observations even with ALMA are relatively expensive, future progress to understand the dust to molecular gas ratios in high-redshift quiescent galaxies \citep[e.g.][]{Whitaker2021b} would more robustly enable efficient constraints on the cold molecular gas in quiescent galaxies based only on dust emission.

\begin{acknowledgments}
We thank the referee for a thorough report that greatly strengthened this work. Based on observations made with ESO Telescopes at the La Silla Paranal Observatory under programme ID 194-A.2005 (The LEGA-C Public Spectroscopy Survey). CW is supported by the National Science Foundation through the Graduate Research Fellowship Program funded by Grant Award No. DGE-1746060. C.C.W. and M.R. were supported by the National Aeronautics and Space Administration (NASA) Contract NAS50210 to the University of Arizona.

This material is based upon High Performance Computing (HPC) resources supported by the University of Arizona TRIF, UITS, and Research, Innovation, and Impact (RII) and maintained by the UArizona Research Technologies department.

We respectfully acknowledge that the University of Arizona is on the land and territories of Indigenous peoples. Today, Arizona is home to 22 federally recognized tribes, with Tucson being home to the O’odham and the Yaqui. 
\end{acknowledgments}

\facilities{HST, ALMA, VLT}

\software{\texttt{Prospector} \citep[v1.1.0,][]{Johnson2021}, \texttt{python-fsps} \citep[v0.4.1,][]{python-fsps}, \texttt{sedpy} \citep[v0.3.0,][]{sedpy}, \texttt{fsps} \citep[v3.2,][]{Conroy2009, Conroy2010}, \texttt{astropy} \citep[v5.0.4,][]{astropy:2013, astropy:2018}, \texttt{matplotlib} \citep[v3.5.1,][]{matplotlib}, \texttt{dynesty} \citep[v1.2.3,][]{dynesty:2020}, \texttt{scipy} \citep[v1.7.3][]{scipy}, \texttt{numpy} \citep[v1.22.3,][]{numpy}, \texttt{corner} \citep[v2.2.1,][]{Foreman-Mackey2016}}

\bibliography{main}{}

\begin{thebibliography}{}
\expandafter\ifx\csname natexlab\endcsname\relax\def\natexlab#1{#1}\fi
\providecommand{\url}[1]{\href{#1}{#1}}
\providecommand{\dodoi}[1]{doi:~\href{http://doi.org/#1}{\nolinkurl{#1}}}
\providecommand{\doeprint}[1]{\href{http://ascl.net/#1}{\nolinkurl{http://ascl.net/#1}}}
\providecommand{\doarXiv}[1]{\href{https://arxiv.org/abs/#1}{\nolinkurl{https://arxiv.org/abs/#1}}}

\bibitem[{{Aird} {et~al.}(2017){Aird}, {Coil}, \& {Georgakakis}}]{Aird2017}
{Aird}, J., {Coil}, A.~L., \& {Georgakakis}, A. 2017, \mnras, 465, 3390,
  \dodoi{10.1093/mnras/stw2932}

\bibitem[{{Akhshik} {et~al.}(2021){Akhshik}, {Whitaker}, {Leja}, {Mahler},
  {Sharon}, {Brammer}, {Toft}, {Bezanson}, {Man}, {Nelson}, {Pacifici},
  {Wellons}, \& {Williams}}]{Akhshik2021}
{Akhshik}, M., {Whitaker}, K.~E., {Leja}, J., {et~al.} 2021, \apjl, 907, L8,
  \dodoi{10.3847/2041-8213/abd416}

\bibitem[{{Akhshik} {et~al.}(2022){Akhshik}, {Whitaker}, {Leja}, {Richard},
  {Spilker}, {Song}, {Brammer}, {Bezanson}, {Ebeling}, {Gallazzi}, {Mahler},
  {Mowla}, {Nelson}, {Pacifici}, {Sharon}, {Toft}, {Williams}, {Wright}, \&
  {Zabl}}]{Akhshik2022}
---. 2022, arXiv e-prints, arXiv:2203.04979.
\newblock \doarXiv{2203.04979}

\bibitem[{{Alarcon} {et~al.}(2022){Alarcon}, {Hearin}, {Becker}, \&
  {Chaves-Montero}}]{Alarcon2022}
{Alarcon}, A., {Hearin}, A.~P., {Becker}, M.~R., \& {Chaves-Montero}, J. 2022,
  arXiv e-prints, arXiv:2205.04273.
\newblock \doarXiv{2205.04273}

\bibitem[{{Alatalo} {et~al.}(2016){Alatalo}, {Lisenfeld}, {Lanz}, {Appleton},
  {Ardila}, {Cales}, {Kewley}, {Lacy}, {Medling}, {Nyland}, {Rich}, \&
  {Urry}}]{Alatalo2016}
{Alatalo}, K., {Lisenfeld}, U., {Lanz}, L., {et~al.} 2016, \apj, 827, 106,
  \dodoi{10.3847/0004-637X/827/2/106}

\bibitem[{{Astropy Collaboration} {et~al.}(2013){Astropy Collaboration},
  {Robitaille}, {Tollerud}, {Greenfield}, {Droettboom}, {Bray}, {Aldcroft},
  {Davis}, {Ginsburg}, {Price-Whelan}, {Kerzendorf}, {Conley}, {Crighton},
  {Barbary}, {Muna}, {Ferguson}, {Grollier}, {Parikh}, {Nair}, {Unther},
  {Deil}, {Woillez}, {Conseil}, {Kramer}, {Turner}, {Singer}, {Fox}, {Weaver},
  {Zabalza}, {Edwards}, {Azalee Bostroem}, {Burke}, {Casey}, {Crawford},
  {Dencheva}, {Ely}, {Jenness}, {Labrie}, {Lim}, {Pierfederici}, {Pontzen},
  {Ptak}, {Refsdal}, {Servillat}, \& {Streicher}}]{astropy:2013}
{Astropy Collaboration}, {Robitaille}, T.~P., {Tollerud}, E.~J., {et~al.} 2013,
  \aap, 558, A33, \dodoi{10.1051/0004-6361/201322068}

\bibitem[{{Astropy Collaboration} {et~al.}(2018){Astropy Collaboration},
  {Price-Whelan}, {Sip{\H{o}}cz}, {G{\"u}nther}, {Lim}, {Crawford}, {Conseil},
  {Shupe}, {Craig}, {Dencheva}, {Ginsburg}, {VanderPlas}, {Bradley},
  {P{\'e}rez-Su{\'a}rez}, {de Val-Borro}, {Paper Contributors}, {Aldcroft},
  {Cruz}, {Robitaille}, {Tollerud}, {Coordination Committee}, {Ardelean},
  {Babej}, {Bach}, {Bachetti}, {Bakanov}, {Bamford}, {Barentsen}, {Barmby},
  {Baumbach}, {Berry}, {Biscani}, {Boquien}, {Bostroem}, {Bouma}, {Brammer},
  {Bray}, {Breytenbach}, {Buddelmeijer}, {Burke}, {Calderone}, {Cano
  Rodr{\'\i}guez}, {Cara}, {Cardoso}, {Cheedella}, {Copin}, {Corrales},
  {Crichton}, {D{\textquoteright}Avella}, {Deil}, {Depagne}, {Dietrich},
  {Donath}, {Droettboom}, {Earl}, {Erben}, {Fabbro}, {Ferreira}, {Finethy},
  {Fox}, {Garrison}, {Gibbons}, {Goldstein}, {Gommers}, {Greco}, {Greenfield},
  {Groener}, {Grollier}, {Hagen}, {Hirst}, {Homeier}, {Horton}, {Hosseinzadeh},
  {Hu}, {Hunkeler}, {Ivezi{\'c}}, {Jain}, {Jenness}, {Kanarek}, {Kendrew},
  {Kern}, {Kerzendorf}, {Khvalko}, {King}, {Kirkby}, {Kulkarni}, {Kumar},
  {Lee}, {Lenz}, {Littlefair}, {Ma}, {Macleod}, {Mastropietro}, {McCully},
  {Montagnac}, {Morris}, {Mueller}, {Mumford}, {Muna}, {Murphy}, {Nelson},
  {Nguyen}, {Ninan}, {N{\"o}the}, {Ogaz}, {Oh}, {Parejko}, {Parley}, {Pascual},
  {Patil}, {Patil}, {Plunkett}, {Prochaska}, {Rastogi}, {Reddy Janga},
  {Sabater}, {Sakurikar}, {Seifert}, {Sherbert}, {Sherwood-Taylor}, {Shih},
  {Sick}, {Silbiger}, {Singanamalla}, {Singer}, {Sladen}, {Sooley},
  {Sornarajah}, {Streicher}, {Teuben}, {Thomas}, {Tremblay}, {Turner},
  {Terr{\'o}n}, {van Kerkwijk}, {de la Vega}, {Watkins}, {Weaver}, {Whitmore},
  {Woillez}, {Zabalza}, \& {Contributors}}]{astropy:2018}
{Astropy Collaboration}, {Price-Whelan}, A.~M., {Sip{\H{o}}cz}, B.~M., {et~al.}
  2018, \aj, 156, 123, \dodoi{10.3847/1538-3881/aabc4f}

\bibitem[{{Bari{\v{s}}i{\'c}} {et~al.}(2017){Bari{\v{s}}i{\'c}}, {van der Wel},
  {Bezanson}, {Pacifici}, {Noeske}, {Mu{\~n}oz-Mateos}, {Franx},
  {Smol{\v{c}}i{\'c}}, {Bell}, {Brammer}, {Calhau}, {Chauk{\'e}}, {van Dokkum},
  {van Houdt}, {Gallazzi}, {Labb{\'e}}, {Maseda}, {Muzzin}, {Sobral},
  {Straatman}, \& {Wu}}]{Barisic2017}
{Bari{\v{s}}i{\'c}}, I., {van der Wel}, A., {Bezanson}, R., {et~al.} 2017,
  \apj, 847, 72, \dodoi{10.3847/1538-4357/aa8768}

\bibitem[{{Behroozi} {et~al.}(2019){Behroozi}, {Wechsler}, {Hearin}, \&
  {Conroy}}]{Behroozi2019}
{Behroozi}, P., {Wechsler}, R.~H., {Hearin}, A.~P., \& {Conroy}, C. 2019,
  \mnras, 488, 3143, \dodoi{10.1093/mnras/stz1182}

\bibitem[{{Belfiore} {et~al.}(2016){Belfiore}, {Maiolino}, {Maraston},
  {Emsellem}, {Bershady}, {Masters}, {Yan}, {Bizyaev}, {Boquien}, {Brownstein},
  {Bundy}, {Drory}, {Heckman}, {Law}, {Roman-Lopes}, {Pan}, {Stanghellini},
  {Thomas}, {Weijmans}, \& {Westfall}}]{Belfiore2016}
{Belfiore}, F., {Maiolino}, R., {Maraston}, C., {et~al.} 2016, \mnras, 461,
  3111, \dodoi{10.1093/mnras/stw1234}

\bibitem[{{Bell} {et~al.}(2004){Bell}, {Wolf}, {Meisenheimer}, {Rix}, {Borch},
  {Dye}, {Kleinheinrich}, {Wisotzki}, \& {McIntosh}}]{Bell2004}
{Bell}, E.~F., {Wolf}, C., {Meisenheimer}, K., {et~al.} 2004, \apj, 608, 752,
  \dodoi{10.1086/420778}

\bibitem[{{Belli} {et~al.}(2014){Belli}, {Newman}, \& {Ellis}}]{Belli2014}
{Belli}, S., {Newman}, A.~B., \& {Ellis}, R.~S. 2014, \apj, 783, 117,
  \dodoi{10.1088/0004-637X/783/2/117}

\bibitem[{{Belli} {et~al.}(2019){Belli}, {Newman}, \& {Ellis}}]{Belli2019}
---. 2019, \apj, 874, 17, \dodoi{10.3847/1538-4357/ab07af}

\bibitem[{{Belli} {et~al.}(2017){Belli}, {Genzel}, {F{\"o}rster Schreiber},
  {Wisnioski}, {Wilman}, {Wuyts}, {Mendel}, {Beifiori}, {Bender}, {Brammer},
  {Burkert}, {Chan}, {Davies}, {Davies}, {Fabricius}, {Fossati}, {Galametz},
  {Lang}, {Lutz}, {Momcheva}, {Nelson}, {Saglia}, {Tacconi}, {Tadaki},
  {{\"U}bler}, \& {van Dokkum}}]{Belli2017}
{Belli}, S., {Genzel}, R., {F{\"o}rster Schreiber}, N.~M., {et~al.} 2017,
  \apjl, 841, L6, \dodoi{10.3847/2041-8213/aa70e5}

\bibitem[{{Belli} {et~al.}(2021){Belli}, {Contursi}, {Genzel}, {Tacconi},
  {F{\"o}rster-Schreiber}, {Lutz}, {Combes}, {Neri}, {Garc{\'\i}a-Burillo},
  {Schuster}, {Herrera-Camus}, {Tadaki}, {Davies}, {Davies}, {Johnson}, {Lee},
  {Leja}, {Nelson}, {Price}, {Shangguan}, {Shimizu}, {Tacchella}, \&
  {{\"U}bler}}]{Belli2021}
{Belli}, S., {Contursi}, A., {Genzel}, R., {et~al.} 2021, \apjl, 909, L11,
  \dodoi{10.3847/2041-8213/abe6a6}

\bibitem[{{Best} {et~al.}(2005){Best}, {Kauffmann}, {Heckman}, {Brinchmann},
  {Charlot}, {Ivezi{\'c}}, \& {White}}]{Best2005}
{Best}, P.~N., {Kauffmann}, G., {Heckman}, T.~M., {et~al.} 2005, \mnras, 362,
  25, \dodoi{10.1111/j.1365-2966.2005.09192.x}

\bibitem[{{Bezanson} {et~al.}(2019){Bezanson}, {Spilker}, {Williams},
  {Whitaker}, {Narayanan}, {Weiner}, \& {Franx}}]{Bezanson2019}
{Bezanson}, R., {Spilker}, J., {Williams}, C.~C., {et~al.} 2019, \apjl, 873,
  L19, \dodoi{10.3847/2041-8213/ab0c9c}

\bibitem[{{Bezanson} {et~al.}(2009){Bezanson}, {van Dokkum}, {Tal},
  {Marchesini}, {Kriek}, {Franx}, \& {Coppi}}]{Bezanson2009}
{Bezanson}, R., {van Dokkum}, P.~G., {Tal}, T., {et~al.} 2009, \apj, 697, 1290,
  \dodoi{10.1088/0004-637X/697/2/1290}

\bibitem[{{Bezanson} {et~al.}(2022){Bezanson}, {Spilker}, {Suess}, {Setton},
  {Feldmann}, {Greene}, {Kriek}, {Narayanan}, \& {Verrico}}]{Bezanson2022}
{Bezanson}, R., {Spilker}, J.~S., {Suess}, K.~A., {et~al.} 2022, \apj, 925,
  153, \dodoi{10.3847/1538-4357/ac3dfa}

\bibitem[{{Binette} {et~al.}(1994){Binette}, {Magris}, {Stasi{\'n}ska}, \&
  {Bruzual}}]{Binette1994}
{Binette}, L., {Magris}, C.~G., {Stasi{\'n}ska}, G., \& {Bruzual}, A.~G. 1994,
  \aap, 292, 13

\bibitem[{{Birnboim} \& {Dekel}(2003)}]{Birnboim2003}
{Birnboim}, Y., \& {Dekel}, A. 2003, \mnras, 345, 349,
  \dodoi{10.1046/j.1365-8711.2003.06955.x}

\bibitem[{{Bluck} {et~al.}(2012){Bluck}, {Conselice}, {Buitrago},
  {Gr{\"u}tzbauch}, {Hoyos}, {Mortlock}, \& {Bauer}}]{Bluck2012}
{Bluck}, A. F.~L., {Conselice}, C.~J., {Buitrago}, F., {et~al.} 2012, \apj,
  747, 34, \dodoi{10.1088/0004-637X/747/1/34}

\bibitem[{{Bolatto} {et~al.}(2013){Bolatto}, {Wolfire}, \&
  {Leroy}}]{Bolatto2013}
{Bolatto}, A.~D., {Wolfire}, M., \& {Leroy}, A.~K. 2013, \araa, 51, 207,
  \dodoi{10.1146/annurev-astro-082812-140944}

\bibitem[{{Buitrago} {et~al.}(2008){Buitrago}, {Trujillo}, {Conselice},
  {Bouwens}, {Dickinson}, \& {Yan}}]{Buitrago2008}
{Buitrago}, F., {Trujillo}, I., {Conselice}, C.~J., {et~al.} 2008, \apjl, 687,
  L61, \dodoi{10.1086/592836}

\bibitem[{{Bureau} \& {Chung}(2006)}]{Bureau2006}
{Bureau}, M., \& {Chung}, A. 2006, \mnras, 366, 182,
  \dodoi{10.1111/j.1365-2966.2005.09840.x}

\bibitem[{{Byler} {et~al.}(2017){Byler}, {Dalcanton}, {Conroy}, \&
  {Johnson}}]{Byler2017}
{Byler}, N., {Dalcanton}, J.~J., {Conroy}, C., \& {Johnson}, B.~D. 2017, \apj,
  840, 44, \dodoi{10.3847/1538-4357/aa6c66}

\bibitem[{{Byler} {et~al.}(2019){Byler}, {Dalcanton}, {Conroy}, {Johnson},
  {Choi}, {Dotter}, \& {Rosenfield}}]{Byler2019}
{Byler}, N., {Dalcanton}, J.~J., {Conroy}, C., {et~al.} 2019, \aj, 158, 2,
  \dodoi{10.3847/1538-3881/ab1b70}

\bibitem[{{Caliendo} {et~al.}(2021){Caliendo}, {Whitaker}, {Akhshik}, {Wilson},
  {Williams}, {Spilker}, {Mahler}, {Pope}, {Sharon}, {Aguilar}, {Bezanson},
  {Chavez Dagostino}, {G{\'o}mez-Ruiz}, {Monta{\~n}a}, {Toft}, {Velazquez de la
  Rosa}, \& {Zeballos}}]{Caliendo2021}
{Caliendo}, J.~N., {Whitaker}, K.~E., {Akhshik}, M., {et~al.} 2021, \apjl, 910,
  L7, \dodoi{10.3847/2041-8213/abe132}

\bibitem[{{Calvi} {et~al.}(2014){Calvi}, {Stiavelli}, {Bradley}, {Pizzella}, \&
  {Kim}}]{Calvi2014}
{Calvi}, V., {Stiavelli}, M., {Bradley}, L., {Pizzella}, A., \& {Kim}, S. 2014,
  \apj, 796, 102, \dodoi{10.1088/0004-637X/796/2/102}

\bibitem[{{Calzetti} {et~al.}(2000){Calzetti}, {Armus}, {Bohlin}, {Kinney},
  {Koornneef}, \& {Storchi-Bergmann}}]{Calzetti2000}
{Calzetti}, D., {Armus}, L., {Bohlin}, R.~C., {et~al.} 2000, \apj, 533, 682,
  \dodoi{10.1086/308692}

\bibitem[{{Capak} {et~al.}(2007){Capak}, {Aussel}, {Ajiki}, {McCracken},
  {Mobasher}, {Scoville}, {Shopbell}, {Taniguchi}, {Thompson}, {Tribiano},
  {Sasaki}, {Blain}, {Brusa}, {Carilli}, {Comastri}, {Carollo}, {Cassata},
  {Colbert}, {Ellis}, {Elvis}, {Giavalisco}, {Green}, {Guzzo}, {Hasinger},
  {Ilbert}, {Impey}, {Jahnke}, {Kartaltepe}, {Kneib}, {Koda}, {Koekemoer},
  {Komiyama}, {Leauthaud}, {Le Fevre}, {Lilly}, {Liu}, {Massey}, {Miyazaki},
  {Murayama}, {Nagao}, {Peacock}, {Pickles}, {Porciani}, {Renzini}, {Rhodes},
  {Rich}, {Salvato}, {Sanders}, {Scarlata}, {Schiminovich}, {Schinnerer},
  {Scodeggio}, {Sheth}, {Shioya}, {Tasca}, {Taylor}, {Yan}, \&
  {Zamorani}}]{Capak2007}
{Capak}, P., {Aussel}, H., {Ajiki}, M., {et~al.} 2007, \apjs, 172, 99,
  \dodoi{10.1086/519081}

\bibitem[{{Caplar} {et~al.}(2018){Caplar}, {Lilly}, \&
  {Trakhtenbrot}}]{Caplar2018}
{Caplar}, N., {Lilly}, S.~J., \& {Trakhtenbrot}, B. 2018, \apj, 867, 148,
  \dodoi{10.3847/1538-4357/aae691}

\bibitem[{{Carleton} {et~al.}(2020){Carleton}, {Guo}, {Nayyeri}, {Cooper},
  {Rudnick}, \& {Whitaker}}]{Carleton2020}
{Carleton}, T., {Guo}, Y., {Nayyeri}, H., {et~al.} 2020, \mnras, 491, 2822,
  \dodoi{10.1093/mnras/stz3216}

\bibitem[{Carnall {et~al.}(2019)Carnall, Leja, Johnson, Conroy, \&
  Speagle}]{Carnall2019}
Carnall, A.~C., Leja, J., Johnson, B.~D., Conroy, C., \& Speagle, J.~S. 2019,
  The Astrophysical Journal, 876, 3, \dodoi{10.3847/1538-4357/ab133c}

\bibitem[{{Carnall} {et~al.}(2018){Carnall}, {McLure}, {Dunlop}, \&
  {Dav{\'e}}}]{Carnall2018}
{Carnall}, A.~C., {McLure}, R.~J., {Dunlop}, J.~S., \& {Dav{\'e}}, R. 2018,
  \mnras, 480, 4379, \dodoi{10.1093/mnras/sty2169}

\bibitem[{{Carollo} {et~al.}(2013){Carollo}, {Bschorr}, {Renzini}, {Lilly},
  {Capak}, {Cibinel}, {Ilbert}, {Onodera}, {Scoville}, {Cameron}, {Mobasher},
  {Sanders}, \& {Taniguchi}}]{Carollo2013}
{Carollo}, C.~M., {Bschorr}, T.~J., {Renzini}, A., {et~al.} 2013, \apj, 773,
  112, \dodoi{10.1088/0004-637X/773/2/112}

\bibitem[{{Cassata} {et~al.}(2013){Cassata}, {Giavalisco}, {Williams}, {Guo},
  {Lee}, {Renzini}, {Ferguson}, {Faber}, {Barro}, {McIntosh}, {Lu}, {Bell},
  {Koo}, {Papovich}, {Ryan}, {Conselice}, {Grogin}, {Koekemoer}, \&
  {Hathi}}]{Cassata2013}
{Cassata}, P., {Giavalisco}, M., {Williams}, C.~C., {et~al.} 2013, \apj, 775,
  106, \dodoi{10.1088/0004-637X/775/2/106}

\bibitem[{{Chabrier}(2003)}]{Chabrier2003}
{Chabrier}, G. 2003, \pasp, 115, 763, \dodoi{10.1086/376392}

\bibitem[{{Charlot} \& {Fall}(2000)}]{Charlot2000}
{Charlot}, S., \& {Fall}, S.~M. 2000, \apj, 539, 718, \dodoi{10.1086/309250}

\bibitem[{{Chauke} {et~al.}(2019){Chauke}, {van der Wel}, {Pacifici},
  {Bezanson}, {Wu}, {Gallazzi}, {Straatman}, {Franx}, {Bari{\v{s}}i{\'c}},
  {Bell}, {van Houdt}, {Maseda}, {Muzzin}, {Sobral}, \& {Spilker}}]{Chauke2019}
{Chauke}, P., {van der Wel}, A., {Pacifici}, C., {et~al.} 2019, \apj, 877, 48,
  \dodoi{10.3847/1538-4357/ab164d}

\bibitem[{Choi {et~al.}(2016)Choi, Dotter, Conroy, Cantiello, Paxton, \&
  Johnson}]{Choi2016}
Choi, J., Dotter, A., Conroy, C., {et~al.} 2016, The Astrophysical Journal,
  823, 102, \dodoi{10.3847/0004-637x/823/2/102}

\bibitem[{{Cid Fernandes} {et~al.}(2011){Cid Fernandes}, {Stasi{\'n}ska},
  {Mateus}, \& {Vale Asari}}]{CidFernandes2011}
{Cid Fernandes}, R., {Stasi{\'n}ska}, G., {Mateus}, A., \& {Vale Asari}, N.
  2011, \mnras, 413, 1687, \dodoi{10.1111/j.1365-2966.2011.18244.x}

\bibitem[{{Cimatti} {et~al.}(2008){Cimatti}, {Cassata}, {Pozzetti}, {Kurk},
  {Mignoli}, {Renzini}, {Daddi}, {Bolzonella}, {Brusa}, {Rodighiero},
  {Dickinson}, {Franceschini}, {Zamorani}, {Berta}, {Rosati}, \&
  {Halliday}}]{Cimatti2008}
{Cimatti}, A., {Cassata}, P., {Pozzetti}, L., {et~al.} 2008, \aap, 482, 21,
  \dodoi{10.1051/0004-6361:20078739}

\bibitem[{{Citro} {et~al.}(2016){Citro}, {Pozzetti}, {Moresco}, \&
  {Cimatti}}]{Citro2016}
{Citro}, A., {Pozzetti}, L., {Moresco}, M., \& {Cimatti}, A. 2016, \aap, 592,
  A19, \dodoi{10.1051/0004-6361/201527772}

\bibitem[{{Combes} {et~al.}(2007){Combes}, {Young}, \& {Bureau}}]{Combes2007}
{Combes}, F., {Young}, L.~M., \& {Bureau}, M. 2007, \mnras, 377, 1795,
  \dodoi{10.1111/j.1365-2966.2007.11759.x}

\bibitem[{{Condon}(1992)}]{Condon1992}
{Condon}, J.~J. 1992, \araa, 30, 575,
  \dodoi{10.1146/annurev.aa.30.090192.003043}

\bibitem[{{Conroy} \& {Gunn}(2010)}]{Conroy2010}
{Conroy}, C., \& {Gunn}, J.~E. 2010, \apj, 712, 833,
  \dodoi{10.1088/0004-637X/712/2/833}

\bibitem[{{Conroy} {et~al.}(2009){Conroy}, {Gunn}, \& {White}}]{Conroy2009}
{Conroy}, C., {Gunn}, J.~E., \& {White}, M. 2009, \apj, 699, 486,
  \dodoi{10.1088/0004-637X/699/1/486}

\bibitem[{{Crocker} {et~al.}(2009){Crocker}, {Jeong}, {Komugi}, {Combes},
  {Bureau}, {Young}, \& {Yi}}]{Crocker2009}
{Crocker}, A.~F., {Jeong}, H., {Komugi}, S., {et~al.} 2009, \mnras, 393, 1255,
  \dodoi{10.1111/j.1365-2966.2008.14295.x}

\bibitem[{{da Cunha} {et~al.}(2008){da Cunha}, {Charlot}, \&
  {Elbaz}}]{daCunha2008}
{da Cunha}, E., {Charlot}, S., \& {Elbaz}, D. 2008, \mnras, 388, 1595,
  \dodoi{10.1111/j.1365-2966.2008.13535.x}

\bibitem[{{Daddi} {et~al.}(2005){Daddi}, {Renzini}, {Pirzkal}, {Cimatti},
  {Malhotra}, {Stiavelli}, {Xu}, {Pasquali}, {Rhoads}, {Brusa}, {di Serego
  Alighieri}, {Ferguson}, {Koekemoer}, {Moustakas}, {Panagia}, \&
  {Windhorst}}]{Daddi2005}
{Daddi}, E., {Renzini}, A., {Pirzkal}, N., {et~al.} 2005, \apj, 626, 680,
  \dodoi{10.1086/430104}

\bibitem[{{Damjanov} {et~al.}(2022){Damjanov}, {Sohn}, {Geller}, {Utsumi}, \&
  {Dell'Antonio}}]{Damjanov2022}
{Damjanov}, I., {Sohn}, J., {Geller}, M.~J., {Utsumi}, Y., \& {Dell'Antonio},
  I. 2022, arXiv e-prints, arXiv:2210.01129.
\newblock \doarXiv{2210.01129}

\bibitem[{{Dannerbauer} {et~al.}(2009){Dannerbauer}, {Daddi}, {Riechers},
  {Walter}, {Carilli}, {Dickinson}, {Elbaz}, \& {Morrison}}]{Dannerbauer2009}
{Dannerbauer}, H., {Daddi}, E., {Riechers}, D.~A., {et~al.} 2009, \apjl, 698,
  L178, \dodoi{10.1088/0004-637X/698/2/L178}

\bibitem[{{Dav{\'e}} {et~al.}(2019){Dav{\'e}}, {Angl{\'e}s-Alc{\'a}zar},
  {Narayanan}, {Li}, {Rafieferantsoa}, \& {Appleby}}]{Dave2019}
{Dav{\'e}}, R., {Angl{\'e}s-Alc{\'a}zar}, D., {Narayanan}, D., {et~al.} 2019,
  \mnras, 486, 2827, \dodoi{10.1093/mnras/stz937}

\bibitem[{{Davidzon} {et~al.}(2017){Davidzon}, {Ilbert}, {Laigle}, {Coupon},
  {McCracken}, {Delvecchio}, {Masters}, {Capak}, {Hsieh}, {Le F{\`e}vre},
  {Tresse}, {Bethermin}, {Chang}, {Faisst}, {Le Floc'h}, {Steinhardt}, {Toft},
  {Aussel}, {Dubois}, {Hasinger}, {Salvato}, {Sanders}, {Scoville}, \&
  {Silverman}}]{Davidzon2017}
{Davidzon}, I., {Ilbert}, O., {Laigle}, C., {et~al.} 2017, \aap, 605, A70,
  \dodoi{10.1051/0004-6361/201730419}

\bibitem[{{Davis} {et~al.}(2016){Davis}, {Greene}, {Ma}, {Pandya}, {Blakeslee},
  {McConnell}, \& {Thomas}}]{Davis2016}
{Davis}, T.~A., {Greene}, J., {Ma}, C.-P., {et~al.} 2016, \mnras, 455, 214,
  \dodoi{10.1093/mnras/stv2313}

\bibitem[{{Davis} {et~al.}(2019){Davis}, {van de Voort}, {Rowlands},
  {McAlpine}, {Wild}, \& {Crain}}]{Davis2019}
{Davis}, T.~A., {van de Voort}, F., {Rowlands}, K., {et~al.} 2019, \mnras, 484,
  2447, \dodoi{10.1093/mnras/stz180}

\bibitem[{{Davis} \& {Young}(2019)}]{DavisYoung2019}
{Davis}, T.~A., \& {Young}, L.~M. 2019, \mnras, 489, L108,
  \dodoi{10.1093/mnrasl/slz138}

\bibitem[{{Davis} {et~al.}(2011){Davis}, {Alatalo}, {Sarzi}, {Bureau}, {Young},
  {Blitz}, {Serra}, {Crocker}, {Krajnovi{\'c}}, {McDermid}, {Bois}, {Bournaud},
  {Cappellari}, {Davies}, {Duc}, {de Zeeuw}, {Emsellem}, {Khochfar},
  {Kuntschner}, {Lablanche}, {Morganti}, {Naab}, {Oosterloo}, {Scott}, \&
  {Weijmans}}]{Davis2011}
{Davis}, T.~A., {Alatalo}, K., {Sarzi}, M., {et~al.} 2011, \mnras, 417, 882,
  \dodoi{10.1111/j.1365-2966.2011.19355.x}

\bibitem[{{de Gasperin} {et~al.}(2018){de Gasperin}, {Intema}, \&
  {Frail}}]{deGasperin2018}
{de Gasperin}, F., {Intema}, H.~T., \& {Frail}, D.~A. 2018, \mnras, 474, 5008,
  \dodoi{10.1093/mnras/stx3125}

\bibitem[{{Dekel} \& {Birnboim}(2006)}]{Dekel2006}
{Dekel}, A., \& {Birnboim}, Y. 2006, \mnras, 368, 2,
  \dodoi{10.1111/j.1365-2966.2006.10145.x}

\bibitem[{{Dekel} \& {Burkert}(2014)}]{Dekel2014}
{Dekel}, A., \& {Burkert}, A. 2014, \mnras, 438, 1870,
  \dodoi{10.1093/mnras/stt2331}

\bibitem[{{Dekel} {et~al.}(2009){Dekel}, {Birnboim}, {Engel}, {Freundlich},
  {Goerdt}, {Mumcuoglu}, {Neistein}, {Pichon}, {Teyssier}, \&
  {Zinger}}]{Dekel2009}
{Dekel}, A., {Birnboim}, Y., {Engel}, G., {et~al.} 2009, \nat, 457, 451,
  \dodoi{10.1038/nature07648}

\bibitem[{{Diemer} {et~al.}(2017){Diemer}, {Sparre}, {Abramson}, \&
  {Torrey}}]{Diemer2017}
{Diemer}, B., {Sparre}, M., {Abramson}, L.~E., \& {Torrey}, P. 2017, \apj, 839,
  26, \dodoi{10.3847/1538-4357/aa68e5}

\bibitem[{{Donas} {et~al.}(2007){Donas}, {Deharveng}, {Rich}, {Yi}, {Lee},
  {Boselli}, {Gil de Paz}, {Boissier}, {Charlot}, {Salim}, {Bianchi}, {Barlow},
  {Forster}, {Friedman}, {Heckman}, {Madore}, {Martin}, {Milliard},
  {Morrissey}, {Neff}, {Schiminovich}, {Seibert}, {Small}, {Szalay}, {Welsh},
  \& {Wyder}}]{Donas2007}
{Donas}, J., {Deharveng}, J.-M., {Rich}, R.~M., {et~al.} 2007, \apjs, 173, 597,
  \dodoi{10.1086/516643}

\bibitem[{{Dotter}(2016)}]{Dotter2016}
{Dotter}, A. 2016, \apjs, 222, 8, \dodoi{10.3847/0067-0049/222/1/8}

\bibitem[{{Draine} \& {Lee}(1984)}]{Draine1984}
{Draine}, B.~T., \& {Lee}, H.~M. 1984, \apj, 285, 89, \dodoi{10.1086/162480}

\bibitem[{{Draine} \& {Li}(2007)}]{Draine2007}
{Draine}, B.~T., \& {Li}, A. 2007, \apj, 657, 810, \dodoi{10.1086/511055}

\bibitem[{{Dubois} {et~al.}(2014){Dubois}, {Pichon}, {Welker}, {Le Borgne},
  {Devriendt}, {Laigle}, {Codis}, {Pogosyan}, {Arnouts}, {Benabed}, {Bertin},
  {Blaizot}, {Bouchet}, {Cardoso}, {Colombi}, {de Lapparent}, {Desjacques},
  {Gavazzi}, {Kassin}, {Kimm}, {McCracken}, {Milliard}, {Peirani}, {Prunet},
  {Rouberol}, {Silk}, {Slyz}, {Sousbie}, {Teyssier}, {Tresse}, {Treyer},
  {Vibert}, \& {Volonteri}}]{Dubois2014}
{Dubois}, Y., {Pichon}, C., {Welker}, C., {et~al.} 2014, \mnras, 444, 1453,
  \dodoi{10.1093/mnras/stu1227}

\bibitem[{{Estrada-Carpenter} {et~al.}(2020){Estrada-Carpenter}, {Papovich},
  {Momcheva}, {Brammer}, {Simons}, {Bridge}, {Cleri}, {Ferguson},
  {Finkelstein}, {Giavalisco}, {Jung}, {Matharu}, {Trump}, \&
  {Weiner}}]{Estrada-Carpenter2020}
{Estrada-Carpenter}, V., {Papovich}, C., {Momcheva}, I., {et~al.} 2020, \apj,
  898, 171, \dodoi{10.3847/1538-4357/aba004}

\bibitem[{{Faber} {et~al.}(2007){Faber}, {Willmer}, {Wolf}, {Koo}, {Weiner},
  {Newman}, {Im}, {Coil}, {Conroy}, {Cooper}, {Davis}, {Finkbeiner}, {Gerke},
  {Gebhardt}, {Groth}, {Guhathakurta}, {Harker}, {Kaiser}, {Kassin},
  {Kleinheinrich}, {Konidaris}, {Kron}, {Lin}, {Luppino}, {Madgwick},
  {Meisenheimer}, {Noeske}, {Phillips}, {Sarajedini}, {Schiavon}, {Simard},
  {Szalay}, {Vogt}, \& {Yan}}]{Faber2007}
{Faber}, S.~M., {Willmer}, C.~N.~A., {Wolf}, C., {et~al.} 2007, \apj, 665, 265,
  \dodoi{10.1086/519294}

\bibitem[{{Fagioli} {et~al.}(2016){Fagioli}, {Carollo}, {Renzini}, {Lilly},
  {Onodera}, \& {Tacchella}}]{Fagioli2016}
{Fagioli}, M., {Carollo}, C.~M., {Renzini}, A., {et~al.} 2016, \apj, 831, 173,
  \dodoi{10.3847/0004-637X/831/2/173}

\bibitem[{{Falc{\'o}n-Barroso} {et~al.}(2011){Falc{\'o}n-Barroso},
  {S{\'a}nchez-Bl{\'a}zquez}, {Vazdekis}, {Ricciardelli}, {Cardiel}, {Cenarro},
  {Gorgas}, \& {Peletier}}]{Falcon-Barroso2011}
{Falc{\'o}n-Barroso}, J., {S{\'a}nchez-Bl{\'a}zquez}, P., {Vazdekis}, A.,
  {et~al.} 2011, \aap, 532, A95, \dodoi{10.1051/0004-6361/201116842}

\bibitem[{{Fang} {et~al.}(2012){Fang}, {Faber}, {Salim}, {Graves}, \&
  {Rich}}]{Fang2012}
{Fang}, J.~J., {Faber}, S.~M., {Salim}, S., {Graves}, G.~J., \& {Rich}, R.~M.
  2012, \apj, 761, 23, \dodoi{10.1088/0004-637X/761/1/23}

\bibitem[{{Faucher-Gigu{\`e}re} {et~al.}(2011){Faucher-Gigu{\`e}re},
  {Kere{\v{s}}}, \& {Ma}}]{Faucher2011}
{Faucher-Gigu{\`e}re}, C.-A., {Kere{\v{s}}}, D., \& {Ma}, C.-P. 2011, \mnras,
  417, 2982, \dodoi{10.1111/j.1365-2966.2011.19457.x}

\bibitem[{{Feldmann} \& {Mayer}(2015)}]{Feldmann2015}
{Feldmann}, R., \& {Mayer}, L. 2015, \mnras, 446, 1939,
  \dodoi{10.1093/mnras/stu2207}

\bibitem[{{Ferreras} {et~al.}(2014){Ferreras}, {Trujillo},
  {M{\'a}rmol-Queralt{\'o}}, {P{\'e}rez-Gonz{\'a}lez}, {Cava}, {Barro},
  {Cenarro}, {Hern{\'a}n-Caballero}, {Cardiel}, {Rodr{\'\i}guez-Zaur{\'\i}n},
  \& {Cebri{\'a}n}}]{Ferreras2014}
{Ferreras}, I., {Trujillo}, I., {M{\'a}rmol-Queralt{\'o}}, E., {et~al.} 2014,
  \mnras, 444, 906, \dodoi{10.1093/mnras/stu1425}

\bibitem[{{Fixsen} {et~al.}(1999){Fixsen}, {Bennett}, \& {Mather}}]{Fixsen1999}
{Fixsen}, D.~J., {Bennett}, C.~L., \& {Mather}, J.~C. 1999, \apj, 526, 207,
  \dodoi{10.1086/307962}

\bibitem[{Foreman-Mackey(2016)}]{Foreman-Mackey2016}
Foreman-Mackey, D. 2016, Journal of Open Source Software, 1, 24,
  \dodoi{10.21105/joss.00024}

\bibitem[{Foreman-Mackey {et~al.}(2014)Foreman-Mackey, Sick, \&
  Johnson}]{ForemanMackey2014}
Foreman-Mackey, D., Sick, J., \& Johnson, B. 2014, python-fsps: Python bindings
  to FSPS (v0.1.1), v0.1.1,  Zenodo, \dodoi{10.5281/zenodo.12157}

\bibitem[{{Frayer} {et~al.}(2009){Frayer}, {Sanders}, {Surace}, {Aussel},
  {Salvato}, {Le Floc'h}, {Huynh}, {Scoville}, {Afonso-Luis}, {Bhattacharya},
  {Capak}, {Fadda}, {Fu}, {Helou}, {Ilbert}, {Kartaltepe}, {Koekemoer}, {Lee},
  {Murphy}, {Sargent}, {Schinnerer}, {Sheth}, {Shopbell}, {Shupe}, \&
  {Yan}}]{Frayer2009}
{Frayer}, D.~T., {Sanders}, D.~B., {Surace}, J.~A., {et~al.} 2009, \aj, 138,
  1261, \dodoi{10.1088/0004-6256/138/5/1261}

\bibitem[{{French} {et~al.}(2015){French}, {Yang}, {Zabludoff}, {Narayanan},
  {Shirley}, {Walter}, {Smith}, \& {Tremonti}}]{French2015}
{French}, K.~D., {Yang}, Y., {Zabludoff}, A., {et~al.} 2015, \apj, 801, 1,
  \dodoi{10.1088/0004-637X/801/1/1}

\bibitem[{{French} {et~al.}(2018{\natexlab{a}}){French}, {Yang}, {Zabludoff},
  \& {Tremonti}}]{French2018a}
{French}, K.~D., {Yang}, Y., {Zabludoff}, A.~I., \& {Tremonti}, C.~A.
  2018{\natexlab{a}}, \apj, 862, 2, \dodoi{10.3847/1538-4357/aacb2d}

\bibitem[{{French} {et~al.}(2018{\natexlab{b}}){French}, {Zabludoff}, {Yoon},
  {Shirley}, {Yang}, {Smercina}, {Smith}, \& {Narayanan}}]{French2018}
{French}, K.~D., {Zabludoff}, A.~I., {Yoon}, I., {et~al.} 2018{\natexlab{b}},
  \apj, 861, 123, \dodoi{10.3847/1538-4357/aac8de}

\bibitem[{{French} {et~al.}(2022){French}, {Smercina}, {Rowlands}, {Tripathi},
  {Zabludoff}, {Smith}, {Narayanan}, {Yang}, {Shirley}, \&
  {Alatalo}}]{French2022}
{French}, K.~D., {Smercina}, A., {Rowlands}, K., {et~al.} 2022, arXiv e-prints,
  arXiv:2204.07465.
\newblock \doarXiv{2204.07465}

\bibitem[{{Fumagalli} {et~al.}(2014){Fumagalli}, {Labb{\'e}}, {Patel}, {Franx},
  {van Dokkum}, {Brammer}, {da Cunha}, {F{\"o}rster Schreiber}, {Kriek},
  {Quadri}, {Rix}, {Wake}, {Whitaker}, {Lundgren}, {Marchesini}, {Maseda},
  {Momcheva}, {Nelson}, {Pacifici}, \& {Skelton}}]{Fumagalli2014}
{Fumagalli}, M., {Labb{\'e}}, I., {Patel}, S.~G., {et~al.} 2014, \apj, 796, 35,
  \dodoi{10.1088/0004-637X/796/1/35}

\bibitem[{{Gensior} {et~al.}(2020){Gensior}, {Kruijssen}, \&
  {Keller}}]{Gensior2020}
{Gensior}, J., {Kruijssen}, J.~M.~D., \& {Keller}, B.~W. 2020, \mnras, 495,
  199, \dodoi{10.1093/mnras/staa1184}

\bibitem[{{Georgakakis} {et~al.}(2017){Georgakakis}, {Aird}, {Schulze},
  {Dwelly}, {Salvato}, {Nandra}, {Merloni}, \& {Schneider}}]{Georgakakis2017}
{Georgakakis}, A., {Aird}, J., {Schulze}, A., {et~al.} 2017, \mnras, 471, 1976,
  \dodoi{10.1093/mnras/stx1602}

\bibitem[{{Gobat} {et~al.}(2018){Gobat}, {Daddi}, {Magdis}, {Bournaud},
  {Sargent}, {Martig}, {Jin}, {Finoguenov}, {B{\'e}thermin}, {Hwang},
  {Renzini}, {Wilson}, {Aretxaga}, {Yun}, {Strazzullo}, \&
  {Valentino}}]{Gobat2018}
{Gobat}, R., {Daddi}, E., {Magdis}, G., {et~al.} 2018, Nature Astronomy, 2,
  239, \dodoi{10.1038/s41550-017-0352-5}

\bibitem[{{Greggio} \& {Renzini}(1990)}]{Greggio1990}
{Greggio}, L., \& {Renzini}, A. 1990, \apj, 364, 35, \dodoi{10.1086/169384}

\bibitem[{{Hamadouche} {et~al.}(2022){Hamadouche}, {Carnall}, {McLure},
  {Dunlop}, {McLeod}, {Cullen}, {Begley}, {Bolzonella}, {Buitrago},
  {Castellano}, {Cucciati}, {Fontana}, {Gargiulo}, {Moresco}, {Pozzetti}, \&
  {Zamorani}}]{Hamadouche2022}
{Hamadouche}, M.~L., {Carnall}, A.~C., {McLure}, R.~J., {et~al.} 2022, \mnras,
  512, 1262, \dodoi{10.1093/mnras/stac535}

\bibitem[{{Harris} {et~al.}(2020){Harris}, {Millman}, {van der Walt},
  {Gommers}, {Virtanen}, {Cournapeau}, {Wieser}, {Taylor}, {Berg}, {Smith},
  {Kern}, {Picus}, {Hoyer}, {van Kerkwijk}, {Brett}, {Haldane}, {del R{\'\i}o},
  {Wiebe}, {Peterson}, {G{\'e}rard-Marchant}, {Sheppard}, {Reddy}, {Weckesser},
  {Abbasi}, {Gohlke}, \& {Oliphant}}]{numpy}
{Harris}, C.~R., {Millman}, K.~J., {van der Walt}, S.~J., {et~al.} 2020, \nat,
  585, 357, \dodoi{10.1038/s41586-020-2649-2}

\bibitem[{{Hayward} {et~al.}(2014){Hayward}, {Lanz}, {Ashby}, {Fazio},
  {Hernquist}, {Mart{\'\i}nez-Galarza}, {Noeske}, {Smith}, {Wuyts}, \&
  {Zezas}}]{Hayward2014}
{Hayward}, C.~C., {Lanz}, L., {Ashby}, M. L.~N., {et~al.} 2014, \mnras, 445,
  1598, \dodoi{10.1093/mnras/stu1843}

\bibitem[{{Hilz} {et~al.}(2012){Hilz}, {Naab}, {Ostriker}, {Thomas}, {Burkert},
  \& {Jesseit}}]{Hilz2012}
{Hilz}, M., {Naab}, T., {Ostriker}, J.~P., {et~al.} 2012, \mnras, 425, 3119,
  \dodoi{10.1111/j.1365-2966.2012.21541.x}

\bibitem[{{Hinshaw} {et~al.}(2013){Hinshaw}, {Larson}, {Komatsu}, {Spergel},
  {Bennett}, {Dunkley}, {Nolta}, {Halpern}, {Hill}, {Odegard}, {Page}, {Smith},
  {Weiland}, {Gold}, {Jarosik}, {Kogut}, {Limon}, {Meyer}, {Tucker}, {Wollack},
  \& {Wright}}]{Hinshaw2013}
{Hinshaw}, G., {Larson}, D., {Komatsu}, E., {et~al.} 2013, \apjs, 208, 19,
  \dodoi{10.1088/0067-0049/208/2/19}

\bibitem[{{Hogg} {et~al.}(2010){Hogg}, {Bovy}, \& {Lang}}]{Hogg2010}
{Hogg}, D.~W., {Bovy}, J., \& {Lang}, D. 2010, arXiv e-prints, arXiv:1008.4686.
\newblock \doarXiv{1008.4686}

\bibitem[{{Hopkins}(2015)}]{Hopkins2015}
{Hopkins}, P.~F. 2015, \mnras, 450, 53, \dodoi{10.1093/mnras/stv195}

\bibitem[{{Hopkins}(2017)}]{Hopkins2017}
---. 2017, arXiv e-prints, arXiv:1712.01294.
\newblock \doarXiv{1712.01294}

\bibitem[{{Hopkins} {et~al.}(2009){Hopkins}, {Bundy}, {Murray}, {Quataert},
  {Lauer}, \& {Ma}}]{Hopkins2009}
{Hopkins}, P.~F., {Bundy}, K., {Murray}, N., {et~al.} 2009, \mnras, 398, 898,
  \dodoi{10.1111/j.1365-2966.2009.15062.x}

\bibitem[{{Hopkins} {et~al.}(2006){Hopkins}, {Hernquist}, {Cox}, {Di Matteo},
  {Robertson}, \& {Springel}}]{Hopkins2006}
{Hopkins}, P.~F., {Hernquist}, L., {Cox}, T.~J., {et~al.} 2006, \apjs, 163, 1,
  \dodoi{10.1086/499298}

\bibitem[{{Hunt} {et~al.}(2018){Hunt}, {Bezanson}, {Greene}, {Spilker},
  {Suess}, {Kriek}, {Narayanan}, {Feldmann}, {van der Wel}, \&
  {Pattarakijwanich}}]{Hunt2018}
{Hunt}, Q., {Bezanson}, R., {Greene}, J.~E., {et~al.} 2018, \apjl, 860, L18,
  \dodoi{10.3847/2041-8213/aaca9a}

\bibitem[{{Hunter}(2007)}]{matplotlib}
{Hunter}, J.~D. 2007, Computing in Science and Engineering, 9, 90,
  \dodoi{10.1109/MCSE.2007.55}

\bibitem[{Jeffreys(1998)}]{Jeffreys1998}
Jeffreys, H. 1998, The Theory of Probability, Oxford Classic Texts in the
  Physical Sciences (OUP Oxford).
\newblock \url{https://books.google.com/books?id=vh9Act9rtzQC}

\bibitem[{{Jeong} {et~al.}(2022){Jeong}, {Oh}, {Joo}, \& {Yi}}]{Jeong2022}
{Jeong}, H., {Oh}, K., {Joo}, S.-J., \& {Yi}, S.~K. 2022, \mnras, 509, 550,
  \dodoi{10.1093/mnras/stab2990}

\bibitem[{{Ji} \& {Giavalisco}(2022)}]{Ji2022}
{Ji}, Z., \& {Giavalisco}, M. 2022, arXiv e-prints, arXiv:2204.02414.
\newblock \doarXiv{2204.02414}

\bibitem[{{Ji} {et~al.}(2018){Ji}, {Giavalisco}, {Williams}, {Faber},
  {Ferguson}, {Guo}, {Liu}, \& {Lee}}]{Ji2018}
{Ji}, Z., {Giavalisco}, M., {Williams}, C.~C., {et~al.} 2018, \apj, 862, 135,
  \dodoi{10.3847/1538-4357/aacc2c}

\bibitem[{{Johnson} {et~al.}(2021{\natexlab{a}}){Johnson}, {Foreman-Mackey},
  {Sick}, {Leja}, {Byler}, {Walmsley}, {Tollerud}, {Leung}, \&
  {Scott}}]{python-fsps}
{Johnson}, B., {Foreman-Mackey}, D., {Sick}, J., {et~al.} 2021{\natexlab{a}},
  {dfm/python-fsps: python-fsps v0.4.1rc1}, v0.4.1rc1, Zenodo,  Zenodo,
  \dodoi{10.5281/zenodo.4737461}

\bibitem[{{Johnson}(2021)}]{sedpy}
{Johnson}, B.~D. 2021, {bd-j/sedpy: sedpy v0.2.0}, v0.2.0, Zenodo,  Zenodo,
  \dodoi{10.5281/zenodo.4582723}

\bibitem[{{Johnson} {et~al.}(2021{\natexlab{b}}){Johnson}, {Leja}, {Conroy}, \&
  {Speagle}}]{Johnson2021}
{Johnson}, B.~D., {Leja}, J., {Conroy}, C., \& {Speagle}, J.~S.
  2021{\natexlab{b}}, \apjs, 254, 22, \dodoi{10.3847/1538-4365/abef67}

\bibitem[{{J{\o}rgensen} {et~al.}(2014){J{\o}rgensen}, {Chiboucas}, {Toft},
  {Bergmann}, {Zirm}, {Schiavon}, \& {Gr{\"u}tzbauch}}]{Jorgensen2014}
{J{\o}rgensen}, I., {Chiboucas}, K., {Toft}, S., {et~al.} 2014, \aj, 148, 117,
  \dodoi{10.1088/0004-6256/148/6/117}

\bibitem[{{Kaviraj} {et~al.}(2009){Kaviraj}, {Peirani}, {Khochfar}, {Silk}, \&
  {Kay}}]{Kaviraj2009}
{Kaviraj}, S., {Peirani}, S., {Khochfar}, S., {Silk}, J., \& {Kay}, S. 2009,
  \mnras, 394, 1713, \dodoi{10.1111/j.1365-2966.2009.14403.x}

\bibitem[{{Kaviraj} {et~al.}(2011){Kaviraj}, {Tan}, {Ellis}, \&
  {Silk}}]{Kaviraj2011}
{Kaviraj}, S., {Tan}, K.-M., {Ellis}, R.~S., \& {Silk}, J. 2011, \mnras, 411,
  2148, \dodoi{10.1111/j.1365-2966.2010.17754.x}

\bibitem[{{Kaviraj} {et~al.}(2008){Kaviraj}, {Khochfar}, {Schawinski}, {Yi},
  {Gawiser}, {Silk}, {Virani}, {Cardamone}, {van Dokkum}, \&
  {Urry}}]{Kaviraj2008}
{Kaviraj}, S., {Khochfar}, S., {Schawinski}, K., {et~al.} 2008, \mnras, 388,
  67, \dodoi{10.1111/j.1365-2966.2008.13392.x}

\bibitem[{{Kaviraj} {et~al.}(2012){Kaviraj}, {Ting}, {Bureau}, {Shabala},
  {Crockett}, {Silk}, {Lintott}, {Smith}, {Keel}, {Masters}, {Schawinski}, \&
  {Bamford}}]{Kaviraj2012}
{Kaviraj}, S., {Ting}, Y.-S., {Bureau}, M., {et~al.} 2012, \mnras, 423, 49,
  \dodoi{10.1111/j.1365-2966.2012.20957.x}

\bibitem[{{Kaviraj} {et~al.}(2013){Kaviraj}, {Rowlands}, {Alpaslan}, {Dunne},
  {Ting}, {Bureau}, {Shabala}, {Lintott}, {Smith}, {Agius}, {Auld}, {Baes},
  {Bourne}, {Cava}, {Clements}, {Cooray}, {Dariush}, {De Zotti}, {Driver},
  {Eales}, {Hopwood}, {Hoyos}, {Ibar}, {Maddox}, {Micha{\l}owski}, {Sansom},
  {Smith}, \& {Valiante}}]{Kaviraj2013}
{Kaviraj}, S., {Rowlands}, K., {Alpaslan}, M., {et~al.} 2013, \mnras, 435,
  1463, \dodoi{10.1093/mnras/stt1629}

\bibitem[{{Kennicutt}(1998)}]{Kennicutt1998}
{Kennicutt}, Robert~C., J. 1998, \araa, 36, 189,
  \dodoi{10.1146/annurev.astro.36.1.189}

\bibitem[{{Kere{\v{s}}} {et~al.}(2005){Kere{\v{s}}}, {Katz}, {Weinberg}, \&
  {Dav{\'e}}}]{Keres2005}
{Kere{\v{s}}}, D., {Katz}, N., {Weinberg}, D.~H., \& {Dav{\'e}}, R. 2005,
  \mnras, 363, 2, \dodoi{10.1111/j.1365-2966.2005.09451.x}

\bibitem[{{Kewley} {et~al.}(2004){Kewley}, {Geller}, \& {Jansen}}]{Kewley2004}
{Kewley}, L.~J., {Geller}, M.~J., \& {Jansen}, R.~A. 2004, \aj, 127, 2002,
  \dodoi{10.1086/382723}

\bibitem[{{Kewley} {et~al.}(2006){Kewley}, {Groves}, {Kauffmann}, \&
  {Heckman}}]{Kewley2006}
{Kewley}, L.~J., {Groves}, B., {Kauffmann}, G., \& {Heckman}, T. 2006, \mnras,
  372, 961, \dodoi{10.1111/j.1365-2966.2006.10859.x}

\bibitem[{{Khim} {et~al.}(2021){Khim}, {Yi}, {Pichon}, {Dubois}, {Devriendt},
  {Choi}, {Bryant}, \& {Croom}}]{Khim2021}
{Khim}, D.~J., {Yi}, S.~K., {Pichon}, C., {et~al.} 2021, \apjs, 254, 27,
  \dodoi{10.3847/1538-4365/abf043}

\bibitem[{{Kirkpatrick} {et~al.}(2013){Kirkpatrick}, {Pope}, {Charmandaris},
  {Daddi}, {Elbaz}, {Hwang}, {Pannella}, {Scott}, {Altieri}, {Aussel}, {Coia},
  {Dannerbauer}, {Dasyra}, {Dickinson}, {Kartaltepe}, {Leiton}, {Magdis},
  {Magnelli}, {Popesso}, \& {Valtchanov}}]{Kirkpatrick2013}
{Kirkpatrick}, A., {Pope}, A., {Charmandaris}, V., {et~al.} 2013, \apj, 763,
  123, \dodoi{10.1088/0004-637X/763/2/123}

\bibitem[{{Kriek} \& {Conroy}(2013)}]{Kriek2013}
{Kriek}, M., \& {Conroy}, C. 2013, \apjl, 775, L16,
  \dodoi{10.1088/2041-8205/775/1/L16}

\bibitem[{{Kriek} {et~al.}(2010){Kriek}, {Labb{\'e}}, {Conroy}, {Whitaker},
  {van Dokkum}, {Brammer}, {Franx}, {Illingworth}, {Marchesini}, {Muzzin},
  {Quadri}, \& {Rudnick}}]{Kriek2010}
{Kriek}, M., {Labb{\'e}}, I., {Conroy}, C., {et~al.} 2010, \apjl, 722, L64,
  \dodoi{10.1088/2041-8205/722/1/L64}

\bibitem[{{Labb{\'e}} {et~al.}(2005){Labb{\'e}}, {Huang}, {Franx}, {Rudnick},
  {Barmby}, {Daddi}, {van Dokkum}, {Fazio}, {F{\"o}rster Schreiber},
  {Moorwood}, {Rix}, {R{\"o}ttgering}, {Trujillo}, \& {van der
  Werf}}]{Labbe2005}
{Labb{\'e}}, I., {Huang}, J., {Franx}, M., {et~al.} 2005, \apjl, 624, L81,
  \dodoi{10.1086/430700}

\bibitem[{{Labb{\'e}} {et~al.}(2010){Labb{\'e}}, {Gonz{\'a}lez}, {Bouwens},
  {Illingworth}, {Franx}, {Trenti}, {Oesch}, {van Dokkum}, {Stiavelli},
  {Carollo}, {Kriek}, \& {Magee}}]{Labbe2010}
{Labb{\'e}}, I., {Gonz{\'a}lez}, V., {Bouwens}, R.~J., {et~al.} 2010, \apjl,
  716, L103, \dodoi{10.1088/2041-8205/716/2/L103}

\bibitem[{{Labb{\'e}} {et~al.}(2013){Labb{\'e}}, {Oesch}, {Bouwens},
  {Illingworth}, {Magee}, {Gonz{\'a}lez}, {Carollo}, {Franx}, {Trenti}, {van
  Dokkum}, \& {Stiavelli}}]{Labbe2013}
{Labb{\'e}}, I., {Oesch}, P.~A., {Bouwens}, R.~J., {et~al.} 2013, \apjl, 777,
  L19, \dodoi{10.1088/2041-8205/777/2/L19}

\bibitem[{{Lee} {et~al.}(2018){Lee}, {Giavalisco}, {Whitaker}, {Williams},
  {Ferguson}, {Acquaviva}, {Koekemoer}, {Straughn}, {Guo}, {Kartaltepe},
  {Lotz}, {Pacifici}, {Croton}, {Somerville}, \& {Lu}}]{Lee2018}
{Lee}, B., {Giavalisco}, M., {Whitaker}, K., {et~al.} 2018, \apj, 853, 131,
  \dodoi{10.3847/1538-4357/aaa40f}

\bibitem[{{Leja} {et~al.}(2019{\natexlab{a}}){Leja}, {Carnall}, {Johnson},
  {Conroy}, \& {Speagle}}]{Leja2019a}
{Leja}, J., {Carnall}, A.~C., {Johnson}, B.~D., {Conroy}, C., \& {Speagle},
  J.~S. 2019{\natexlab{a}}, \apj, 876, 3, \dodoi{10.3847/1538-4357/ab133c}

\bibitem[{{Leja} {et~al.}(2018){Leja}, {Johnson}, {Conroy}, \& {van
  Dokkum}}]{Leja2018b}
{Leja}, J., {Johnson}, B.~D., {Conroy}, C., \& {van Dokkum}, P. 2018, \apj,
  854, 62, \dodoi{10.3847/1538-4357/aaa8db}

\bibitem[{Leja {et~al.}(2017)Leja, Johnson, Conroy, van Dokkum, \&
  Byler}]{Leja2017}
Leja, J., Johnson, B.~D., Conroy, C., van Dokkum, P.~G., \& Byler, N. 2017, The
  Astrophysical Journal, 837, 170, \dodoi{10.3847/1538-4357/aa5ffe}

\bibitem[{Leja {et~al.}(2018)Leja, Johnson, Conroy, van Dokkum, Speagle,
  Brammer, Momcheva, Skelton, Whitaker, Franx, \& Nelson}]{Leja2018a}
Leja, J., Johnson, B.~D., Conroy, C., {et~al.} 2018, The Astrophysical Journal,
  877, 140, \dodoi{10.3847/1538-4357/ab1d5a}

\bibitem[{{Leja} {et~al.}(2019{\natexlab{b}}){Leja}, {Johnson}, {Conroy}, {van
  Dokkum}, {Speagle}, {Brammer}, {Momcheva}, {Skelton}, {Whitaker}, {Franx}, \&
  {Nelson}}]{Leja2019}
{Leja}, J., {Johnson}, B.~D., {Conroy}, C., {et~al.} 2019{\natexlab{b}}, \apj,
  877, 140, \dodoi{10.3847/1538-4357/ab1d5a}

\bibitem[{{Leja} {et~al.}(2021){Leja}, {Speagle}, {Ting}, {Johnson}, {Conroy},
  {Whitaker}, {Nelson}, {van Dokkum}, \& {Franx}}]{Leja2022}
{Leja}, J., {Speagle}, J.~S., {Ting}, Y.-S., {et~al.} 2021, arXiv e-prints,
  arXiv:2110.04314.
\newblock \doarXiv{2110.04314}

\bibitem[{{Li} {et~al.}(2019){Li}, {French}, {Zabludoff}, \& {Ho}}]{Li2019}
{Li}, Z., {French}, K.~D., {Zabludoff}, A.~I., \& {Ho}, L.~C. 2019, \apj, 879,
  131, \dodoi{10.3847/1538-4357/ab1f68}

\bibitem[{{Lilly} \& {Carollo}(2016)}]{Lilly2016}
{Lilly}, S.~J., \& {Carollo}, C.~M. 2016, \apj, 833, 1,
  \dodoi{10.3847/0004-637X/833/1/1}

\bibitem[{{Lower} {et~al.}(2020){Lower}, {Narayanan}, {Leja}, {Johnson},
  {Conroy}, \& {Dav{\'e}}}]{Lower2020}
{Lower}, S., {Narayanan}, D., {Leja}, J., {et~al.} 2020, \apj, 904, 33,
  \dodoi{10.3847/1538-4357/abbfa7}

\bibitem[{{Madau}(1995)}]{Madau1995}
{Madau}, P. 1995, \apj, 441, 18, \dodoi{10.1086/175332}

\bibitem[{{Magdis} {et~al.}(2021){Magdis}, {Gobat}, {Valentino}, {Daddi},
  {Zanella}, {Kokorev}, {Toft}, {Jin}, \& {Whitaker}}]{Magdis2021}
{Magdis}, G.~E., {Gobat}, R., {Valentino}, F., {et~al.} 2021, \aap, 647, A33,
  \dodoi{10.1051/0004-6361/202039280}

\bibitem[{{Man} {et~al.}(2012){Man}, {Toft}, {Zirm}, {Wuyts}, \& {van der
  Wel}}]{Man2012}
{Man}, A. W.~S., {Toft}, S., {Zirm}, A.~W., {Wuyts}, S., \& {van der Wel}, A.
  2012, \apj, 744, 85, \dodoi{10.1088/0004-637X/744/2/85}

\bibitem[{{Mancini} {et~al.}(2019){Mancini}, {Daddi}, {Juneau}, {Renzini},
  {Rodighiero}, {Cappellari}, {Rodr{\'\i}guez-Mu{\~n}oz}, {Liu}, {Pannella},
  {Baronchelli}, {Franceschini}, {Bergamini}, {D'Eugenio}, \&
  {Puglisi}}]{Mancini2019}
{Mancini}, C., {Daddi}, E., {Juneau}, S., {et~al.} 2019, \mnras, 489, 1265,
  \dodoi{10.1093/mnras/stz2130}

\bibitem[{{Marchesini} {et~al.}(2012){Marchesini}, {Stefanon}, {Brammer}, \&
  {Whitaker}}]{Marchesini2012}
{Marchesini}, D., {Stefanon}, M., {Brammer}, G.~B., \& {Whitaker}, K.~E. 2012,
  \apj, 748, 126, \dodoi{10.1088/0004-637X/748/2/126}

\bibitem[{{Marchesini} {et~al.}(2009){Marchesini}, {van Dokkum}, {F{\"o}rster
  Schreiber}, {Franx}, {Labb{\'e}}, \& {Wuyts}}]{Marchesini2009}
{Marchesini}, D., {van Dokkum}, P.~G., {F{\"o}rster Schreiber}, N.~M., {et~al.}
  2009, \apj, 701, 1765, \dodoi{10.1088/0004-637X/701/2/1765}

\bibitem[{{Martel} \& {Richard}(2020)}]{Martel2020}
{Martel}, H., \& {Richard}, S. 2020, \mnras, 498, 940,
  \dodoi{10.1093/mnras/staa2122}

\bibitem[{{Martig} {et~al.}(2009){Martig}, {Bournaud}, {Teyssier}, \&
  {Dekel}}]{Martig2009}
{Martig}, M., {Bournaud}, F., {Teyssier}, R., \& {Dekel}, A. 2009, \apj, 707,
  250, \dodoi{10.1088/0004-637X/707/1/250}

\bibitem[{{Martin} {et~al.}(2005){Martin}, {Fanson}, {Schiminovich},
  {Morrissey}, {Friedman}, {Barlow}, {Conrow}, {Grange}, {Jelinsky},
  {Milliard}, {Siegmund}, {Bianchi}, {Byun}, {Donas}, {Forster}, {Heckman},
  {Lee}, {Madore}, {Malina}, {Neff}, {Rich}, {Small}, {Surber}, {Szalay},
  {Welsh}, \& {Wyder}}]{Martin2005}
{Martin}, D.~C., {Fanson}, J., {Schiminovich}, D., {et~al.} 2005, \apjl, 619,
  L1, \dodoi{10.1086/426387}

\bibitem[{{Maseda} {et~al.}(2021){Maseda}, {van der Wel}, {Franx}, {Bell},
  {Bezanson}, {Muzzin}, {Sobral}, {D'Eugenio}, {Gallazzi}, {de Graaff}, {Leja},
  {Straatman}, {Whitaker}, {Williams}, \& {Wu}}]{Maseda2021}
{Maseda}, M.~V., {van der Wel}, A., {Franx}, M., {et~al.} 2021, \apj, 923, 18,
  \dodoi{10.3847/1538-4357/ac2bfe}

\bibitem[{{Mathis} {et~al.}(1977){Mathis}, {Rumpl}, \&
  {Nordsieck}}]{Mathis1977}
{Mathis}, J.~S., {Rumpl}, W., \& {Nordsieck}, K.~H. 1977, \apj, 217, 425,
  \dodoi{10.1086/155591}

\bibitem[{{McCracken} {et~al.}(2012){McCracken}, {Milvang-Jensen}, {Dunlop},
  {Franx}, {Fynbo}, {Le F{\`e}vre}, {Holt}, {Caputi}, {Goranova}, {Buitrago},
  {Emerson}, {Freudling}, {Hudelot}, {L{\'o}pez-Sanjuan}, {Magnard}, {Mellier},
  {M{\o}ller}, {Nilsson}, {Sutherland}, {Tasca}, \& {Zabl}}]{McCracken2012}
{McCracken}, H.~J., {Milvang-Jensen}, B., {Dunlop}, J., {et~al.} 2012, \aap,
  544, A156, \dodoi{10.1051/0004-6361/201219507}

\bibitem[{{Mosleh} {et~al.}(2017){Mosleh}, {Tacchella}, {Renzini}, {Carollo},
  {Molaeinezhad}, {Onodera}, {Khosroshahi}, \& {Lilly}}]{Mosleh2017}
{Mosleh}, M., {Tacchella}, S., {Renzini}, A., {et~al.} 2017, \apj, 837, 2,
  \dodoi{10.3847/1538-4357/aa5f14}

\bibitem[{{Murphy} {et~al.}(2011){Murphy}, {Condon}, {Schinnerer}, {Kennicutt},
  {Calzetti}, {Armus}, {Helou}, {Turner}, {Aniano}, {Beir{\~a}o}, {Bolatto},
  {Brandl}, {Croxall}, {Dale}, {Donovan Meyer}, {Draine}, {Engelbracht},
  {Hunt}, {Hao}, {Koda}, {Roussel}, {Skibba}, \& {Smith}}]{Murphy2011}
{Murphy}, E.~J., {Condon}, J.~J., {Schinnerer}, E., {et~al.} 2011, \apj, 737,
  67, \dodoi{10.1088/0004-637X/737/2/67}

\bibitem[{{Muzzin} {et~al.}(2013{\natexlab{a}}){Muzzin}, {Marchesini},
  {Stefanon}, {Franx}, {McCracken}, {Milvang-Jensen}, {Dunlop}, {Fynbo},
  {Brammer}, {Labb{\'e}}, \& {van Dokkum}}]{Muzzin2013b}
{Muzzin}, A., {Marchesini}, D., {Stefanon}, M., {et~al.} 2013{\natexlab{a}},
  \apj, 777, 18, \dodoi{10.1088/0004-637X/777/1/18}

\bibitem[{{Muzzin} {et~al.}(2013{\natexlab{b}}){Muzzin}, {Marchesini},
  {Stefanon}, {Franx}, {Milvang-Jensen}, {Dunlop}, {Fynbo}, {Brammer},
  {Labb{\'e}}, \& {van Dokkum}}]{Muzzin2013a}
---. 2013{\natexlab{b}}, \apjs, 206, 8, \dodoi{10.1088/0067-0049/206/1/8}

\bibitem[{{Naab} {et~al.}(2009){Naab}, {Johansson}, \& {Ostriker}}]{Naab2009}
{Naab}, T., {Johansson}, P.~H., \& {Ostriker}, J.~P. 2009, \apjl, 699, L178,
  \dodoi{10.1088/0004-637X/699/2/L178}

\bibitem[{{Nagaraj} {et~al.}(2022){Nagaraj}, {Forbes}, {Leja},
  {Foreman-Mackey}, \& {Hayward}}]{Nagaraj2022}
{Nagaraj}, G., {Forbes}, J.~C., {Leja}, J., {Foreman-Mackey}, D., \& {Hayward},
  C.~C. 2022, \apj, 932, 54, \dodoi{10.3847/1538-4357/ac6c80}

\bibitem[{{Nenkova} {et~al.}(2008{\natexlab{a}}){Nenkova}, {Sirocky},
  {Ivezi{\'c}}, \& {Elitzur}}]{Nenkova2008a}
{Nenkova}, M., {Sirocky}, M.~M., {Ivezi{\'c}}, {\v{Z}}., \& {Elitzur}, M.
  2008{\natexlab{a}}, \apj, 685, 147, \dodoi{10.1086/590482}

\bibitem[{{Nenkova} {et~al.}(2008{\natexlab{b}}){Nenkova}, {Sirocky},
  {Nikutta}, {Ivezi{\'c}}, \& {Elitzur}}]{Nenkova2008b}
{Nenkova}, M., {Sirocky}, M.~M., {Nikutta}, R., {Ivezi{\'c}}, {\v{Z}}., \&
  {Elitzur}, M. 2008{\natexlab{b}}, \apj, 685, 160, \dodoi{10.1086/590483}

\bibitem[{{Newman} {et~al.}(2012){Newman}, {Ellis}, {Bundy}, \&
  {Treu}}]{Newman2012}
{Newman}, A.~B., {Ellis}, R.~S., {Bundy}, K., \& {Treu}, T. 2012, \apj, 746,
  162, \dodoi{10.1088/0004-637X/746/2/162}

\bibitem[{{Oser} {et~al.}(2012){Oser}, {Naab}, {Ostriker}, \&
  {Johansson}}]{Oser2012}
{Oser}, L., {Naab}, T., {Ostriker}, J.~P., \& {Johansson}, P.~H. 2012, \apj,
  744, 63, \dodoi{10.1088/0004-637X/744/1/63}

\bibitem[{{Pandya} {et~al.}(2017){Pandya}, {Brennan}, {Somerville}, {Choi},
  {Barro}, {Wuyts}, {Taylor}, {Behroozi}, {Kirkpatrick}, {Faber}, {Primack},
  {Koo}, {McIntosh}, {Kocevski}, {Bell}, {Dekel}, {Fang}, {Ferguson}, {Grogin},
  {Koekemoer}, {Lu}, {Mantha}, {Mobasher}, {Newman}, {Pacifici}, {Papovich},
  {van der Wel}, \& {Yesuf}}]{Pandya2017}
{Pandya}, V., {Brennan}, R., {Somerville}, R.~S., {et~al.} 2017, \mnras, 472,
  2054, \dodoi{10.1093/mnras/stx2027}

\bibitem[{{Paxton} {et~al.}(2011){Paxton}, {Bildsten}, {Dotter}, {Herwig},
  {Lesaffre}, \& {Timmes}}]{Paxton2011}
{Paxton}, B., {Bildsten}, L., {Dotter}, A., {et~al.} 2011, \apjs, 192, 3,
  \dodoi{10.1088/0067-0049/192/1/3}

\bibitem[{{Paxton} {et~al.}(2013){Paxton}, {Cantiello}, {Arras}, {Bildsten},
  {Brown}, {Dotter}, {Mankovich}, {Montgomery}, {Stello}, {Timmes}, \&
  {Townsend}}]{Paxton2013}
{Paxton}, B., {Cantiello}, M., {Arras}, P., {et~al.} 2013, \apjs, 208, 4,
  \dodoi{10.1088/0067-0049/208/1/4}

\bibitem[{{Paxton} {et~al.}(2015){Paxton}, {Marchant}, {Schwab}, {Bauer},
  {Bildsten}, {Cantiello}, {Dessart}, {Farmer}, {Hu}, {Langer}, {Townsend},
  {Townsley}, \& {Timmes}}]{Paxton2015}
{Paxton}, B., {Marchant}, P., {Schwab}, J., {et~al.} 2015, \apjs, 220, 15,
  \dodoi{10.1088/0067-0049/220/1/15}

\bibitem[{{Paxton} {et~al.}(2018){Paxton}, {Schwab}, {Bauer}, {Bildsten},
  {Blinnikov}, {Duffell}, {Farmer}, {Goldberg}, {Marchant}, {Sorokina},
  {Thoul}, {Townsend}, \& {Timmes}}]{Paxton2018}
{Paxton}, B., {Schwab}, J., {Bauer}, E.~B., {et~al.} 2018, \apjs, 234, 34,
  \dodoi{10.3847/1538-4365/aaa5a8}

\bibitem[{{Peng} {et~al.}(2010){Peng}, {Lilly}, {Kova{\v{c}}}, {Bolzonella},
  {Pozzetti}, {Renzini}, {Zamorani}, {Ilbert}, {Knobel}, {Iovino}, {Maier},
  {Cucciati}, {Tasca}, {Carollo}, {Silverman}, {Kampczyk}, {de Ravel},
  {Sanders}, {Scoville}, {Contini}, {Mainieri}, {Scodeggio}, {Kneib}, {Le
  F{\`e}vre}, {Bardelli}, {Bongiorno}, {Caputi}, {Coppa}, {de la Torre},
  {Franzetti}, {Garilli}, {Lamareille}, {Le Borgne}, {Le Brun}, {Mignoli},
  {Perez Montero}, {Pello}, {Ricciardelli}, {Tanaka}, {Tresse}, {Vergani},
  {Welikala}, {Zucca}, {Oesch}, {Abbas}, {Barnes}, {Bordoloi}, {Bottini},
  {Cappi}, {Cassata}, {Cimatti}, {Fumana}, {Hasinger}, {Koekemoer},
  {Leauthaud}, {Maccagni}, {Marinoni}, {McCracken}, {Memeo}, {Meneux}, {Nair},
  {Porciani}, {Presotto}, \& {Scaramella}}]{Peng2010}
{Peng}, Y.-j., {Lilly}, S.~J., {Kova{\v{c}}}, K., {et~al.} 2010, \apj, 721,
  193, \dodoi{10.1088/0004-637X/721/1/193}

\bibitem[{{Poggianti} {et~al.}(2013){Poggianti}, {Calvi}, {Bindoni},
  {D'Onofrio}, {Moretti}, {Valentinuzzi}, {Fasano}, {Fritz}, {De Lucia},
  {Vulcani}, {Bettoni}, {Gullieuszik}, \& {Omizzolo}}]{Poggianti2013}
{Poggianti}, B.~M., {Calvi}, R., {Bindoni}, D., {et~al.} 2013, \apj, 762, 77,
  \dodoi{10.1088/0004-637X/762/2/77}

\bibitem[{Renzini(2006)}]{Renzini2006}
Renzini, A. 2006, Annual Review of Astronomy and Astrophysics, 44, 92,
  \dodoi{10.1146/annurev.astro.44.051905.092450}

\bibitem[{{Rowlands} {et~al.}(2015){Rowlands}, {Wild}, {Nesvadba}, {Sibthorpe},
  {Mortier}, {Lehnert}, \& {da Cunha}}]{Rowlands2015}
{Rowlands}, K., {Wild}, V., {Nesvadba}, N., {et~al.} 2015, \mnras, 448, 258,
  \dodoi{10.1093/mnras/stu2714}

\bibitem[{{Rutkowski} {et~al.}(2014){Rutkowski}, {Jeong}, {Cohen}, {Kaviraj},
  {Windhorst}, {Ryan}, {Koekemoer}, {Yi}, {Hathi}, \& {Dopita}}]{Rutkowski2014}
{Rutkowski}, M.~J., {Jeong}, H., {Cohen}, S.~H., {et~al.} 2014, \apj, 796, 101,
  \dodoi{10.1088/0004-637X/796/2/101}

\bibitem[{{Saintonge} {et~al.}(2011){Saintonge}, {Kauffmann}, {Kramer},
  {Tacconi}, {Buchbender}, {Catinella}, {Fabello}, {Graci{\'a}-Carpio}, {Wang},
  {Cortese}, {Fu}, {Genzel}, {Giovanelli}, {Guo}, {Haynes}, {Heckman},
  {Krumholz}, {Lemonias}, {Li}, {Moran}, {Rodriguez-Fernandez}, {Schiminovich},
  {Schuster}, \& {Sievers}}]{Saintonge2011}
{Saintonge}, A., {Kauffmann}, G., {Kramer}, C., {et~al.} 2011, \mnras, 415, 32,
  \dodoi{10.1111/j.1365-2966.2011.18677.x}

\bibitem[{{Saintonge} {et~al.}(2012){Saintonge}, {Tacconi}, {Fabello}, {Wang},
  {Catinella}, {Genzel}, {Graci{\'a}-Carpio}, {Kramer}, {Moran}, {Heckman},
  {Schiminovich}, {Schuster}, \& {Wuyts}}]{Saintonge2012}
{Saintonge}, A., {Tacconi}, L.~J., {Fabello}, S., {et~al.} 2012, \apj, 758, 73,
  \dodoi{10.1088/0004-637X/758/2/73}

\bibitem[{{Saintonge} {et~al.}(2017){Saintonge}, {Catinella}, {Tacconi},
  {Kauffmann}, {Genzel}, {Cortese}, {Dav{\'e}}, {Fletcher},
  {Graci{\'a}-Carpio}, {Kramer}, {Heckman}, {Janowiecki}, {Lutz}, {Rosario},
  {Schiminovich}, {Schuster}, {Wang}, {Wuyts}, {Borthakur}, {Lamperti}, \&
  {Roberts-Borsani}}]{Saintonge2017}
{Saintonge}, A., {Catinella}, B., {Tacconi}, L.~J., {et~al.} 2017, \apjs, 233,
  22, \dodoi{10.3847/1538-4365/aa97e0}

\bibitem[{{Salim} {et~al.}(2018){Salim}, {Boquien}, \& {Lee}}]{Salim2018}
{Salim}, S., {Boquien}, M., \& {Lee}, J.~C. 2018, \apj, 859, 11,
  \dodoi{10.3847/1538-4357/aabf3c}

\bibitem[{{Salim} \& {Rich}(2010)}]{Salim2010}
{Salim}, S., \& {Rich}, R.~M. 2010, \apjl, 714, L290,
  \dodoi{10.1088/2041-8205/714/2/L290}

\bibitem[{{Salim} {et~al.}(2009){Salim}, {Dickinson}, {Michael Rich},
  {Charlot}, {Lee}, {Schiminovich}, {P{\'e}rez-Gonz{\'a}lez}, {Ashby},
  {Papovich}, {Faber}, {Ivison}, {Frayer}, {Walton}, {Weiner}, {Chary},
  {Bundy}, {Noeske}, \& {Koekemoer}}]{Salim2009}
{Salim}, S., {Dickinson}, M., {Michael Rich}, R., {et~al.} 2009, \apj, 700,
  161, \dodoi{10.1088/0004-637X/700/1/161}

\bibitem[{{Salvador-Rusi{\~n}ol} {et~al.}(2020){Salvador-Rusi{\~n}ol},
  {Vazdekis}, {La Barbera}, {Beasley}, {Ferreras}, {Negri}, \& {Dalla
  Vecchia}}]{Salvador-Rusinol2020}
{Salvador-Rusi{\~n}ol}, N., {Vazdekis}, A., {La Barbera}, F., {et~al.} 2020,
  Nature Astronomy, 4, 252, \dodoi{10.1038/s41550-019-0955-0}

\bibitem[{{Sanders} {et~al.}(2007){Sanders}, {Salvato}, {Aussel}, {Ilbert},
  {Scoville}, {Surace}, {Frayer}, {Sheth}, {Helou}, {Brooke}, {Bhattacharya},
  {Yan}, {Kartaltepe}, {Barnes}, {Blain}, {Calzetti}, {Capak}, {Carilli},
  {Carollo}, {Comastri}, {Daddi}, {Ellis}, {Elvis}, {Fall}, {Franceschini},
  {Giavalisco}, {Hasinger}, {Impey}, {Koekemoer}, {Le F{\`e}vre}, {Lilly},
  {Liu}, {McCracken}, {Mobasher}, {Renzini}, {Rich}, {Schinnerer}, {Shopbell},
  {Taniguchi}, {Thompson}, {Urry}, \& {Williams}}]{Sanders2007}
{Sanders}, D.~B., {Salvato}, M., {Aussel}, H., {et~al.} 2007, \apjs, 172, 86,
  \dodoi{10.1086/517885}

\bibitem[{{Sandstrom} {et~al.}(2013){Sandstrom}, {Leroy}, {Walter}, {Bolatto},
  {Croxall}, {Draine}, {Wilson}, {Wolfire}, {Calzetti}, {Kennicutt}, {Aniano},
  {Donovan Meyer}, {Usero}, {Bigiel}, {Brinks}, {de Blok}, {Crocker}, {Dale},
  {Engelbracht}, {Galametz}, {Groves}, {Hunt}, {Koda}, {Kreckel}, {Linz},
  {Meidt}, {Pellegrini}, {Rix}, {Roussel}, {Schinnerer}, {Schruba}, {Schuster},
  {Skibba}, {van der Laan}, {Appleton}, {Armus}, {Brandl}, {Gordon}, {Hinz},
  {Krause}, {Montiel}, {Sauvage}, {Schmiedeke}, {Smith}, \&
  {Vigroux}}]{Sandstrom2013}
{Sandstrom}, K.~M., {Leroy}, A.~K., {Walter}, F., {et~al.} 2013, \apj, 777, 5,
  \dodoi{10.1088/0004-637X/777/1/5}

\bibitem[{{Sansom} {et~al.}(2019){Sansom}, {Glass}, {Bendo}, {Davis},
  {Rowlands}, {Bourne}, {Dunne}, {Eales}, {Kaviraj}, {Popescu}, {Smith}, \&
  {Viaene}}]{Sansom2019}
{Sansom}, A.~E., {Glass}, D.~H.~W., {Bendo}, G.~J., {et~al.} 2019, \mnras, 482,
  4617, \dodoi{10.1093/mnras/sty3021}

\bibitem[{{Sargent} {et~al.}(2015){Sargent}, {Daddi}, {Bournaud}, {Onodera},
  {Feruglio}, {Martig}, {Gobat}, {Dannerbauer}, \& {Schinnerer}}]{Sargent2015}
{Sargent}, M.~T., {Daddi}, E., {Bournaud}, F., {et~al.} 2015, \apjl, 806, L20,
  \dodoi{10.1088/2041-8205/806/1/L20}

\bibitem[{{Schawinski} {et~al.}(2007){Schawinski}, {Kaviraj}, {Khochfar},
  {Yoon}, {Yi}, {Deharveng}, {Boselli}, {Barlow}, {Conrow}, {Forster},
  {Friedman}, {Martin}, {Morrissey}, {Neff}, {Schiminovich}, {Seibert},
  {Small}, {Wyder}, {Bianchi}, {Donas}, {Heckman}, {Lee}, {Madore}, {Milliard},
  {Rich}, \& {Szalay}}]{Schawinski2007}
{Schawinski}, K., {Kaviraj}, S., {Khochfar}, S., {et~al.} 2007, \apjs, 173,
  512, \dodoi{10.1086/516631}

\bibitem[{{Schinnerer} {et~al.}(2010){Schinnerer}, {Sargent}, {Bondi},
  {Smol{\v{c}}i{\'c}}, {Datta}, {Carilli}, {Bertoldi}, {Blain}, {Ciliegi},
  {Koekemoer}, \& {Scoville}}]{Schinnerer2010}
{Schinnerer}, E., {Sargent}, M.~T., {Bondi}, M., {et~al.} 2010, \apjs, 188,
  384, \dodoi{10.1088/0067-0049/188/2/384}

\bibitem[{Scoville {et~al.}(2007)Scoville, Abraham, Aussel, Barnes, Benson,
  Blain, Calzetti, Comastri, Capak, Carilli, Carlstrom, Carollo, Colbert,
  Daddi, Ellis, Elvis, Ewald, Fall, Franceschini, Giavalisco, Green, Griffiths,
  Guzzo, Hasinger, Impey, Kneib, Koda, Koekemoer, Lefevre, Lilly, Liu,
  McCracken, Massey, Mellier, Miyazaki, Mobasher, Mould, Norman, Refregier,
  Renzini, Rhodes, Rich, Sanders, Schiminovich, Schinnerer, Scodeggio, Sheth,
  Shopbell, Taniguchi, Tyson, Urry, Waerbeke, Vettolani, White, \&
  Yan}]{Scoville2007}
Scoville, N., Abraham, R.~G., Aussel, H., {et~al.} 2007, The Astrophysical
  Journal Supplement Series, 172, 38, \dodoi{10.1086/516580}

\bibitem[{{Simha} {et~al.}(2014){Simha}, {Weinberg}, {Conroy}, {Dave},
  {Fardal}, {Katz}, \& {Oppenheimer}}]{Simha2014}
{Simha}, V., {Weinberg}, D.~H., {Conroy}, C., {et~al.} 2014, arXiv e-prints,
  arXiv:1404.0402.
\newblock \doarXiv{1404.0402}

\bibitem[{{Smercina} {et~al.}(2018){Smercina}, {Smith}, {Dale}, {French},
  {Croxall}, {Zhukovska}, {Togi}, {Bell}, {Crocker}, {Draine}, {Jarrett},
  {Tremonti}, {Yang}, \& {Zabludoff}}]{Smercina2018}
{Smercina}, A., {Smith}, J.~D.~T., {Dale}, D.~A., {et~al.} 2018, \apj, 855, 51,
  \dodoi{10.3847/1538-4357/aaafcd}

\bibitem[{{Smercina} {et~al.}(2022){Smercina}, {Smith}, {French}, {Bell},
  {Dale}, {Medling}, {Nyland}, {Privon}, {Rowlands}, {Walter}, \&
  {Zabludoff}}]{Smercina2022}
{Smercina}, A., {Smith}, J.-D.~T., {French}, K.~D., {et~al.} 2022, \apj, 929,
  154, \dodoi{10.3847/1538-4357/ac5d5f}

\bibitem[{{Smol{\v{c}}i{\'c}} {et~al.}(2017){Smol{\v{c}}i{\'c}}, {Novak},
  {Bondi}, {Ciliegi}, {Mooley}, {Schinnerer}, {Zamorani}, {Navarrete},
  {Bourke}, {Karim}, {Vardoulaki}, {Leslie}, {Delhaize}, {Carilli}, {Myers},
  {Baran}, {Delvecchio}, {Miettinen}, {Banfield}, {Balokovi{\'c}}, {Bertoldi},
  {Capak}, {Frail}, {Hallinan}, {Hao}, {Herrera Ruiz}, {Horesh}, {Ilbert},
  {Intema}, {Jeli{\'c}}, {Kl{\"o}ckner}, {Krpan}, {Kulkarni}, {McCracken},
  {Laigle}, {Middleberg}, {Murphy}, {Sargent}, {Scoville}, \&
  {Sheth}}]{Smolcic2017}
{Smol{\v{c}}i{\'c}}, V., {Novak}, M., {Bondi}, M., {et~al.} 2017, \aap, 602,
  A1, \dodoi{10.1051/0004-6361/201628704}

\bibitem[{{Solomon} {et~al.}(1987){Solomon}, {Rivolo}, {Barrett}, \&
  {Yahil}}]{Solomon1987}
{Solomon}, P.~M., {Rivolo}, A.~R., {Barrett}, J., \& {Yahil}, A. 1987, \apj,
  319, 730, \dodoi{10.1086/165493}

\bibitem[{{Sonnenfeld} {et~al.}(2014){Sonnenfeld}, {Nipoti}, \&
  {Treu}}]{Sonnenfeld2014}
{Sonnenfeld}, A., {Nipoti}, C., \& {Treu}, T. 2014, \apj, 786, 89,
  \dodoi{10.1088/0004-637X/786/2/89}

\bibitem[{{Speagle}(2020)}]{dynesty:2020}
{Speagle}, J.~S. 2020, \mnras, \dodoi{10.1093/mnras/staa278}

\bibitem[{{Speagle} {et~al.}(2014){Speagle}, {Steinhardt}, {Capak}, \&
  {Silverman}}]{Speagle2014}
{Speagle}, J.~S., {Steinhardt}, C.~L., {Capak}, P.~L., \& {Silverman}, J.~D.
  2014, \apjs, 214, 15, \dodoi{10.1088/0067-0049/214/2/15}

\bibitem[{{Spilker} {et~al.}(2018){Spilker}, {Bezanson}, {Bari{\v{s}}i{\'c}},
  {Bell}, {Lagos}, {Maseda}, {Muzzin}, {Pacifici}, {Sobral}, {Straatman}, {van
  der Wel}, {van Dokkum}, {Weiner}, {Whitaker}, {Williams}, \&
  {Wu}}]{Spilker2018}
{Spilker}, J., {Bezanson}, R., {Bari{\v{s}}i{\'c}}, I., {et~al.} 2018, \apj,
  860, 103, \dodoi{10.3847/1538-4357/aac438}

\bibitem[{{Spilker} {et~al.}(2014){Spilker}, {Marrone}, {Aguirre}, {Aravena},
  {Ashby}, {B{\'e}thermin}, {Bradford}, {Bothwell}, {Brodwin}, {Carlstrom},
  {Chapman}, {Crawford}, {de Breuck}, {Fassnacht}, {Gonzalez}, {Greve},
  {Gullberg}, {Hezaveh}, {Holzapfel}, {Husband}, {Ma}, {Malkan}, {Murphy},
  {Reichardt}, {Rotermund}, {Stalder}, {Stark}, {Strandet}, {Vieira},
  {Wei{\ss}}, \& {Welikala}}]{Spilker2014}
{Spilker}, J.~S., {Marrone}, D.~P., {Aguirre}, J.~E., {et~al.} 2014, \apj, 785,
  149, \dodoi{10.1088/0004-637X/785/2/149}

\bibitem[{{Straatman} {et~al.}(2018){Straatman}, {van der Wel}, {Bezanson},
  {Pacifici}, {Gallazzi}, {Wu}, {Noeske}, {Bari{\v{s}}i{\'c}}, {Bell},
  {Brammer}, {Calhau}, {Chauke}, {Franx}, {van Houdt}, {Labb{\'e}}, {Maseda},
  {Mu{\~n}oz-Mateos}, {Muzzin}, {van de Sande}, {Sobral}, \&
  {Spilker}}]{Straatman2018}
{Straatman}, C. M.~S., {van der Wel}, A., {Bezanson}, R., {et~al.} 2018, \apjs,
  239, 27, \dodoi{10.3847/1538-4365/aae37a}

\bibitem[{{Suess} {et~al.}(2017){Suess}, {Bezanson}, {Spilker}, {Kriek},
  {Greene}, {Feldmann}, {Hunt}, \& {Narayanan}}]{Suess2017}
{Suess}, K.~A., {Bezanson}, R., {Spilker}, J.~S., {et~al.} 2017, \apjl, 846,
  L14, \dodoi{10.3847/2041-8213/aa85dc}

\bibitem[{{Suess} {et~al.}(2019){Suess}, {Kriek}, {Price}, \&
  {Barro}}]{Suess2019}
{Suess}, K.~A., {Kriek}, M., {Price}, S.~H., \& {Barro}, G. 2019, \apjl, 885,
  L22, \dodoi{10.3847/2041-8213/ab4db3}

\bibitem[{{Suess} {et~al.}(2022){Suess}, {Kriek}, {Bezanson}, {Greene},
  {Setton}, {Spilker}, {Feldmann}, {Goulding}, {Johnson}, {Leja}, {Narayanan},
  {Hall-Hooper}, {Hunt}, {Lower}, \& {Verrico}}]{Suess2022}
{Suess}, K.~A., {Kriek}, M., {Bezanson}, R., {et~al.} 2022, \apj, 926, 89,
  \dodoi{10.3847/1538-4357/ac404a}

\bibitem[{{Szomoru} {et~al.}(2013){Szomoru}, {Franx}, {van Dokkum}, {Trenti},
  {Illingworth}, {Labb{\'e}}, \& {Oesch}}]{Szomoru2013}
{Szomoru}, D., {Franx}, M., {van Dokkum}, P.~G., {et~al.} 2013, \apj, 763, 73,
  \dodoi{10.1088/0004-637X/763/2/73}

\bibitem[{{Tacchella} {et~al.}(2017){Tacchella}, {Carollo}, {Faber}, {Cibinel},
  {Dekel}, {Koo}, {Renzini}, \& {Woo}}]{Tacchella2017}
{Tacchella}, S., {Carollo}, C.~M., {Faber}, S.~M., {et~al.} 2017, \apjl, 844,
  L1, \dodoi{10.3847/2041-8213/aa7cfb}

\bibitem[{{Tacchella} {et~al.}(2022{\natexlab{a}}){Tacchella}, {Conroy},
  {Faber}, {Johnson}, {Leja}, {Barro}, {Cunningham}, {Deason}, {Guhathakurta},
  {Guo}, {Hernquist}, {Koo}, {McKinnon}, {Rockosi}, {Speagle}, {van Dokkum}, \&
  {Yesuf}}]{Tacchella2022a}
{Tacchella}, S., {Conroy}, C., {Faber}, S.~M., {et~al.} 2022{\natexlab{a}},
  \apj, 926, 134, \dodoi{10.3847/1538-4357/ac449b}

\bibitem[{{Tacchella} {et~al.}(2022{\natexlab{b}}){Tacchella}, {Finkelstein},
  {Bagley}, {Dickinson}, {Ferguson}, {Giavalisco}, {Graziani}, {Grogin},
  {Hathi}, {Hutchison}, {Jung}, {Koekemoer}, {Larson}, {Papovich}, {Pirzkal},
  {Rojas-Ruiz}, {Song}, {Schneider}, {Somerville}, {Wilkins}, \&
  {Yung}}]{Tacchella2022}
{Tacchella}, S., {Finkelstein}, S.~L., {Bagley}, M., {et~al.}
  2022{\natexlab{b}}, \apj, 927, 170, \dodoi{10.3847/1538-4357/ac4cad}

\bibitem[{{Taniguchi} {et~al.}(2007){Taniguchi}, {Scoville}, {Murayama},
  {Sanders}, {Mobasher}, {Aussel}, {Capak}, {Ajiki}, {Miyazaki}, {Komiyama},
  {Shioya}, {Nagao}, {Sasaki}, {Koda}, {Carilli}, {Giavalisco}, {Guzzo},
  {Hasinger}, {Impey}, {LeFevre}, {Lilly}, {Renzini}, {Rich}, {Schinnerer},
  {Shopbell}, {Kaifu}, {Karoji}, {Arimoto}, {Okamura}, \&
  {Ohta}}]{Taniguchi2007}
{Taniguchi}, Y., {Scoville}, N., {Murayama}, T., {et~al.} 2007, \apjs, 172, 9,
  \dodoi{10.1086/516596}

\bibitem[{{Taylor} {et~al.}(2010){Taylor}, {Franx}, {Glazebrook}, {Brinchmann},
  {van der Wel}, \& {van Dokkum}}]{Taylor2010}
{Taylor}, E.~N., {Franx}, M., {Glazebrook}, K., {et~al.} 2010, \apj, 720, 723,
  \dodoi{10.1088/0004-637X/720/1/723}

\bibitem[{{Teyssier}(2002)}]{Teyssier2002}
{Teyssier}, R. 2002, \aap, 385, 337, \dodoi{10.1051/0004-6361:20011817}

\bibitem[{{Thomas} {et~al.}(2010){Thomas}, {Maraston}, {Schawinski}, {Sarzi},
  \& {Silk}}]{Thomas2010}
{Thomas}, D., {Maraston}, C., {Schawinski}, K., {Sarzi}, M., \& {Silk}, J.
  2010, \mnras, 404, 1775, \dodoi{10.1111/j.1365-2966.2010.16427.x}

\bibitem[{{Tomczak} {et~al.}(2014){Tomczak}, {Quadri}, {Tran}, {Labb{\'e}},
  {Straatman}, {Papovich}, {Glazebrook}, {Allen}, {Brammer}, {Kacprzak},
  {Kawinwanichakij}, {Kelson}, {McCarthy}, {Mehrtens}, {Monson}, {Persson},
  {Spitler}, {Tilvi}, \& {van Dokkum}}]{Tomczak2014}
{Tomczak}, A.~R., {Quadri}, R.~F., {Tran}, K.-V.~H., {et~al.} 2014, \apj, 783,
  85, \dodoi{10.1088/0004-637X/783/2/85}

\bibitem[{{Trayford} {et~al.}(2016){Trayford}, {Theuns}, {Bower}, {Crain},
  {Lagos}, {Schaller}, \& {Schaye}}]{Trayford2016}
{Trayford}, J.~W., {Theuns}, T., {Bower}, R.~G., {et~al.} 2016, \mnras, 460,
  3925, \dodoi{10.1093/mnras/stw1230}

\bibitem[{Treu {et~al.}(2005)Treu, Ellis, Liao, van Dokkum, Tozzi, Coil,
  Newman, Cooper, \& Davis}]{Treu2005}
Treu, T., Ellis, R.~S., Liao, T.~X., {et~al.} 2005, The Astrophysical Journal,
  633, 174, \dodoi{10.1086/444585}

\bibitem[{{Trujillo} {et~al.}(2011){Trujillo}, {Ferreras}, \& {de La
  Rosa}}]{Trujillo2011}
{Trujillo}, I., {Ferreras}, I., \& {de La Rosa}, I.~G. 2011, \mnras, 415, 3903,
  \dodoi{10.1111/j.1365-2966.2011.19017.x}

\bibitem[{{Trujillo} {et~al.}(2006){Trujillo}, {F{\"o}rster Schreiber},
  {Rudnick}, {Barden}, {Franx}, {Rix}, {Caldwell}, {McIntosh}, {Toft},
  {H{\"a}ussler}, {Zirm}, {van Dokkum}, {Labb{\'e}}, {Moorwood},
  {R{\"o}ttgering}, {van der Wel}, {van der Werf}, \& {van
  Starkenburg}}]{Trujillo2006}
{Trujillo}, I., {F{\"o}rster Schreiber}, N.~M., {Rudnick}, G., {et~al.} 2006,
  \apj, 650, 18, \dodoi{10.1086/506464}

\bibitem[{{Utomo} {et~al.}(2014){Utomo}, {Kriek}, {Labb{\'e}}, {Conroy}, \&
  {Fumagalli}}]{Utomo2014}
{Utomo}, D., {Kriek}, M., {Labb{\'e}}, I., {Conroy}, C., \& {Fumagalli}, M.
  2014, \apjl, 783, L30, \dodoi{10.1088/2041-8205/783/2/L30}

\bibitem[{{van de Sande} {et~al.}(2013){van de Sande}, {Kriek}, {Franx}, {van
  Dokkum}, {Bezanson}, {Bouwens}, {Quadri}, {Rix}, \&
  {Skelton}}]{vandeSande2013}
{van de Sande}, J., {Kriek}, M., {Franx}, M., {et~al.} 2013, \apj, 771, 85,
  \dodoi{10.1088/0004-637X/771/2/85}

\bibitem[{{van de Voort} {et~al.}(2018){van de Voort}, {Davis}, {Matsushita},
  {Rowlands}, {Shabala}, {Allison}, {Ting}, {Sansom}, \& {van der
  Werf}}]{vandeVoort2018}
{van de Voort}, F., {Davis}, T.~A., {Matsushita}, S., {et~al.} 2018, \mnras,
  476, 122, \dodoi{10.1093/mnras/sty228}

\bibitem[{{van der Wel} {et~al.}(2016){van der Wel}, {Noeske}, {Bezanson},
  {Pacifici}, {Gallazzi}, {Franx}, {Mu{\~n}oz-Mateos}, {Bell}, {Brammer},
  {Charlot}, {Chauk{\'e}}, {Labb{\'e}}, {Maseda}, {Muzzin}, {Rix}, {Sobral},
  {van de Sande}, {van Dokkum}, {Wild}, \& {Wolf}}]{vanderWel2016}
{van der Wel}, A., {Noeske}, K., {Bezanson}, R., {et~al.} 2016, \apjs, 223, 29,
  \dodoi{10.3847/0067-0049/223/2/29}

\bibitem[{{van der Wel} {et~al.}(2021){van der Wel}, {Bezanson}, {D'Eugenio},
  {Straatman}, {Franx}, {van Houdt}, {Maseda}, {Gallazzi}, {Wu}, {Pacifici},
  {Barisic}, {Brammer}, {Munoz-Mateos}, {Vervalcke}, {Zibetti}, {Sobral}, {de
  Graaff}, {Calhau}, {Kaushal}, {Muzzin}, {Bell}, \& {van
  Dokkum}}]{vanderWel2021}
{van der Wel}, A., {Bezanson}, R., {D'Eugenio}, F., {et~al.} 2021, \apjs, 256,
  44, \dodoi{10.3847/1538-4365/ac1356}

\bibitem[{{van Dokkum} {et~al.}(2008){van Dokkum}, {Franx}, {Kriek}, {Holden},
  {Illingworth}, {Magee}, {Bouwens}, {Marchesini}, {Quadri}, {Rudnick},
  {Taylor}, \& {Toft}}]{vanDokkum2008}
{van Dokkum}, P.~G., {Franx}, M., {Kriek}, M., {et~al.} 2008, \apjl, 677, L5,
  \dodoi{10.1086/587874}

\bibitem[{{van Dokkum} {et~al.}(2010){van Dokkum}, {Whitaker}, {Brammer},
  {Franx}, {Kriek}, {Labb{\'e}}, {Marchesini}, {Quadri}, {Bezanson},
  {Illingworth}, {Muzzin}, {Rudnick}, {Tal}, \& {Wake}}]{vanDokkum2010}
{van Dokkum}, P.~G., {Whitaker}, K.~E., {Brammer}, G., {et~al.} 2010, \apj,
  709, 1018, \dodoi{10.1088/0004-637X/709/2/1018}

\bibitem[{Vazdekis {et~al.}(2015)Vazdekis, Coelho, Cassisi, Ricciardelli,
  Falcon-Barroso, Sanchez-Blazquez, Barbera, Beasley, \&
  Pietrinferni}]{Vazdekis2015}
Vazdekis, A., Coelho, P., Cassisi, S., {et~al.} 2015, Monthly Notices of the
  Royal Astronomical Society, 449, 1177, \dodoi{10.1093/mnras/stv151}

\bibitem[{{Virtanen} {et~al.}(2020){Virtanen}, {Gommers}, {Oliphant},
  {Haberland}, {Reddy}, {Cournapeau}, {Burovski}, {Peterson}, {Weckesser},
  {Bright}, {van der Walt}, {Brett}, {Wilson}, {Jarrod Millman}, {Mayorov},
  {Nelson}, {Jones}, {Kern}, {Larson}, {Carey}, {Polat}, {Feng}, {Moore}, {Vand
  erPlas}, {Laxalde}, {Perktold}, {Cimrman}, {Henriksen}, {Quintero}, {Harris},
  {Archibald}, {Ribeiro}, {Pedregosa}, {van Mulbregt}, \&
  {Contributors}}]{scipy}
{Virtanen}, P., {Gommers}, R., {Oliphant}, T.~E., {et~al.} 2020, Nature
  Methods, \dodoi{https://doi.org/10.1038/s41592-019-0686-2}

\bibitem[{{Vogelsberger} {et~al.}(2014){Vogelsberger}, {Genel}, {Springel},
  {Torrey}, {Sijacki}, {Xu}, {Snyder}, {Nelson}, \&
  {Hernquist}}]{Vogelsberger2014}
{Vogelsberger}, M., {Genel}, S., {Springel}, V., {et~al.} 2014, \mnras, 444,
  1518, \dodoi{10.1093/mnras/stu1536}

\bibitem[{{Vulcani} {et~al.}(2016){Vulcani}, {Marchesini}, {De Lucia},
  {Muzzin}, {Stefanon}, {Brammer}, {Labb{\'e}}, {Le F{\`e}vre}, \&
  {Milvang-Jensen}}]{Vulcani2016}
{Vulcani}, B., {Marchesini}, D., {De Lucia}, G., {et~al.} 2016, \apj, 816, 86,
  \dodoi{10.3847/0004-637X/816/2/86}

\bibitem[{{Werle} {et~al.}(2020){Werle}, {Cid Fernandes}, {Vale Asari},
  {Coelho}, {Bruzual}, {Charlot}, {de Carvalho}, {Herpich}, {Mendes de
  Oliveira}, {Sodr{\'e}}, {Ruschel-Dutra}, {de Amorim}, \&
  {Sampaio}}]{Werle2020}
{Werle}, A., {Cid Fernandes}, R., {Vale Asari}, N., {et~al.} 2020, \mnras, 497,
  3251, \dodoi{10.1093/mnras/staa2217}

\bibitem[{{Whitaker} {et~al.}(2011){Whitaker}, {Labb{\'e}}, {van Dokkum},
  {Brammer}, {Kriek}, {Marchesini}, {Quadri}, {Franx}, {Muzzin}, {Williams},
  {Bezanson}, {Illingworth}, {Lee}, {Lundgren}, {Nelson}, {Rudnick}, {Tal}, \&
  {Wake}}]{Whitaker2011}
{Whitaker}, K.~E., {Labb{\'e}}, I., {van Dokkum}, P.~G., {et~al.} 2011, \apj,
  735, 86, \dodoi{10.1088/0004-637X/735/2/86}

\bibitem[{{Whitaker} {et~al.}(2014){Whitaker}, {Franx}, {Leja}, {van Dokkum},
  {Henry}, {Skelton}, {Fumagalli}, {Momcheva}, {Brammer}, {Labb{\'e}},
  {Nelson}, \& {Rigby}}]{Whitaker2014}
{Whitaker}, K.~E., {Franx}, M., {Leja}, J., {et~al.} 2014, \apj, 795, 104,
  \dodoi{10.1088/0004-637X/795/2/104}

\bibitem[{{Whitaker} {et~al.}(2021{\natexlab{a}}){Whitaker}, {Williams},
  {Mowla}, {Spilker}, {Toft}, {Narayanan}, {Pope}, {Magdis}, {van Dokkum},
  {Akhshik}, {Bezanson}, {Brammer}, {Leja}, {Man}, {Nelson}, {Richard},
  {Pacifici}, {Sharon}, \& {Valentino}}]{Whitaker2021a}
{Whitaker}, K.~E., {Williams}, C.~C., {Mowla}, L., {et~al.} 2021{\natexlab{a}},
  \nat, 597, 485, \dodoi{10.1038/s41586-021-03806-7}

\bibitem[{{Whitaker} {et~al.}(2021{\natexlab{b}}){Whitaker}, {Narayanan},
  {Williams}, {Li}, {Spilker}, {Dav{\'e}}, {Akhshik}, {Akins}, {Bezanson},
  {Katz}, {Leja}, {Magdis}, {Mowla}, {Nelson}, {Pope}, {Privon}, {Toft}, \&
  {Valentino}}]{Whitaker2021b}
{Whitaker}, K.~E., {Narayanan}, D., {Williams}, C.~C., {et~al.}
  2021{\natexlab{b}}, \apjl, 922, L30, \dodoi{10.3847/2041-8213/ac399f}

\bibitem[{Williams {et~al.}(2017)Williams, Giavalisco, Bezanson, Cappelluti,
  Cassata, Liu, Lee, Tundo, \& Vanzella}]{Williams2017}
Williams, C.~C., Giavalisco, M., Bezanson, R., {et~al.} 2017, The Astrophysical
  Journal, \dodoi{10.3847/1538-4357/aa662f}

\bibitem[{{Williams} {et~al.}(2021){Williams}, {Spilker}, {Whitaker},
  {Dav{\'e}}, {Woodrum}, {Brammer}, {Bezanson}, {Narayanan}, \&
  {Weiner}}]{Williams2021}
{Williams}, C.~C., {Spilker}, J.~S., {Whitaker}, K.~E., {et~al.} 2021, \apj,
  908, 54, \dodoi{10.3847/1538-4357/abcbf6}

\bibitem[{{Williams} {et~al.}(2009){Williams}, {Quadri}, {Franx}, {van Dokkum},
  \& {Labb{\'e}}}]{Williams2009}
{Williams}, R.~J., {Quadri}, R.~F., {Franx}, M., {van Dokkum}, P., \&
  {Labb{\'e}}, I. 2009, \apj, 691, 1879, \dodoi{10.1088/0004-637X/691/2/1879}

\bibitem[{{Wu} {et~al.}(2018){Wu}, {van der Wel}, {Bezanson}, {Gallazzi},
  {Pacifici}, {Straatman}, {Bari{\v{s}}i{\'c}}, {Bell}, {Chauke}, {van Houdt},
  {Franx}, {Muzzin}, {Sobral}, \& {Wild}}]{Wu2018}
{Wu}, P.-F., {van der Wel}, A., {Bezanson}, R., {et~al.} 2018, \apj, 868, 37,
  \dodoi{10.3847/1538-4357/aae822}

\bibitem[{{Wuyts} {et~al.}(2007){Wuyts}, {Labb{\'e}}, {Franx}, {Rudnick}, {van
  Dokkum}, {Fazio}, {F{\"o}rster Schreiber}, {Huang}, {Moorwood}, {Rix},
  {R{\"o}ttgering}, \& {van der Werf}}]{Wuyts2007}
{Wuyts}, S., {Labb{\'e}}, I., {Franx}, M., {et~al.} 2007, \apj, 655, 51,
  \dodoi{10.1086/509708}

\bibitem[{{Yang} {et~al.}(2008){Yang}, {Zabludoff}, {Zaritsky}, \&
  {Mihos}}]{Yang2008}
{Yang}, Y., {Zabludoff}, A.~I., {Zaritsky}, D., \& {Mihos}, J.~C. 2008, \apj,
  688, 945, \dodoi{10.1086/591656}

\bibitem[{{Yesuf} \& {Ho}(2020)}]{Yesuf2020}
{Yesuf}, H.~M., \& {Ho}, L.~C. 2020, \apj, 900, 107,
  \dodoi{10.3847/1538-4357/abaa43}

\bibitem[{{Young} {et~al.}(2011){Young}, {Bureau}, {Davis}, {Combes},
  {McDermid}, {Alatalo}, {Blitz}, {Bois}, {Bournaud}, {Cappellari}, {Davies},
  {de Zeeuw}, {Emsellem}, {Khochfar}, {Krajnovi{\'c}}, {Kuntschner},
  {Lablanche}, {Morganti}, {Naab}, {Oosterloo}, {Sarzi}, {Scott}, {Serra}, \&
  {Weijmans}}]{Young2011}
{Young}, L.~M., {Bureau}, M., {Davis}, T.~A., {et~al.} 2011, \mnras, 414, 940,
  \dodoi{10.1111/j.1365-2966.2011.18561.x}

\bibitem[{{Zolotov} {et~al.}(2015){Zolotov}, {Dekel}, {Mandelker}, {Tweed},
  {Inoue}, {DeGraf}, {Ceverino}, {Primack}, {Barro}, \& {Faber}}]{Zolotov2015}
{Zolotov}, A., {Dekel}, A., {Mandelker}, N., {et~al.} 2015, \mnras, 450, 2327,
  \dodoi{10.1093/mnras/stv740}

\end{thebibliography}
\bibliographystyle{aasjournal}

\end{document}